\documentclass[fleqn,usenatbib]{mnras}

\usepackage{newtxtext,newtxmath}
\usepackage[T1]{fontenc}
\usepackage{ae,aecompl}

\DeclareRobustCommand{\VAN}[3]{#2}
\let\VANthebibliography\thebibliography
\def\thebibliography{\DeclareRobustCommand{\VAN}[3]{##3}\VANthebibliography}

\usepackage[normalem]{ulem} 
\usepackage{graphicx}	% Including figure files
\usepackage{amsmath}	% Advanced maths commands
\usepackage{upgreek}
\usepackage{stfloats}
\usepackage{mathrsfs,bm}
\usepackage[mathscr]{eucal}

\newcommand{\bnabla}{\bm{\nabla}}
\newcommand{\btimes}{\bm{\times}}

\newcommand{\bs}[1]{\boldsymbol{#1}}
\newcommand{\bcdot}{\bs{\cdot}}
\newcommand{\dd}{\mathrm{d}}
\newcommand{\e}{\mathrm{e}}
\newcommand{\p}{\mathrm{p}}
\newcommand{\eps}{\varepsilon}

\newcommand{\B}{{\mathcal B}}
\newcommand{\C}{{\mathcal C}}
\newcommand{\x}{\mathbfit{x}}
\renewcommand{\vec}{\mathbfit}
\newcommand{\rvec}{\mathbfit{r}}
\newcommand{\kvec}{\mathbfit{k}}
\newcommand{\msun}{\mathrm{M}_\odot}

\title[Simulating radio synchrotron emission in galaxies]{Simulating radio synchrotron emission in star-forming galaxies: small-scale magnetic dynamo and the origin of the far infrared--radio correlation}

\author[Pfrommer et al.]
       {Christoph Pfrommer$^{1}\thanks{E-mail: cpfrommer@aip.de}$,
         Maria Werhahn$^{1,2}$,
         R{\"u}diger Pakmor$^{3}$,
         Philipp Girichidis$^{1}$,\newauthor
         Christine M.~Simpson$^{4,5}$\\
$^{1}$Leibniz-Institute for Astrophysics Potsdam (AIP), An der Sternwarte 16, 14482 Potsdam, Germany\\
$^2$Institut f\"ur Physik und Astronomie, Universit\"at Potsdam, Karl-Liebknecht-Str.\,24/25, 14476 Golm, Germany\\
$^{3}$Max Planck Institute for Astrophysics, Karl-Schwarzschild-Str. 1, 85741 Garching, Germany\\
$^{4}$Enrico Fermi Institute, The University of Chicago, Chicago, IL 60637, USA\\
$^{5}$Department of Astronomy \& Astrophysics, The University of Chicago, Chicago, IL 60637, USA
}

\date{Accepted XXX. Received YYY; in original form ZZZ}

\pubyear{2022}

\begin{document}
\label{firstpage}
\pagerange{\pageref{firstpage}--\pageref{lastpage}}
\maketitle

\begin{abstract}
In star-forming galaxies, the far-infrared (FIR) and radio-continuum
luminosities obey a tight empirical relation over a large range of
star-formation rates (SFR). To understand the physics, we examine
magneto-hydrodynamic galaxy simulations, which follow the genesis of cosmic ray
(CR) protons at supernovae and their advective and anisotropic diffusive
transport. We show that gravitational collapse of the proto-galaxy generates a
corrugated accretion shock, which injects turbulence and drives a small-scale
magnetic dynamo. As the shock propagates outwards and the associated turbulence
decays, the large velocity shear between the supersonically rotating cool disc
with respect to the (partially) pressure-supported hot circumgalactic medium
excites Kelvin-Helmholtz surface and body modes. Those interact non-linearly,
inject additional turbulence and continuously drive multiple small-scale
dynamos, which exponentially amplify weak seed magnetic fields.  After
saturation at small scales, they grow in scale to reach equipartition with
thermal and CR energies in Milky Way-mass galaxies. In small galaxies, the
magnetic energy saturates at the turbulent energy while it fails to reach
equipartition with thermal and CR energies. We solve for steady-state spectra of
CR protons, secondary electrons/positrons from hadronic CR-proton interactions
with the interstellar medium, and primary shock-accelerated electrons at
supernovae. The radio-synchrotron emission is dominated by primary electrons,
irradiates the magnetised disc and bulge of our simulated Milky Way-mass galaxy
and weakly traces bubble-shaped magnetically-loaded outflows.  Our star-forming
and star-bursting galaxies with saturated magnetic fields match the global
FIR-radio correlation (FRC) across four orders of magnitude. Its intrinsic
scatter arises due to (i) different magnetic saturation levels that result from
different seed magnetic fields, (ii) different radio synchrotron luminosities
for different specific SFRs at fixed SFR and (iii) a varying radio intensity
with galactic inclination. In agreement with observations, several 100-pc-sized
regions within star-forming galaxies also obey the FRC, while the centres of
starbursts substantially exceed the FRC.
\end{abstract}


\section{Introduction}
\label{sec:intro}

The FIR emission of star-forming and -bursting galaxies tightly correlates with
their radio continuum luminosities, forming the (nearly) linear ``FIR-radio
correlation''
\citep[FRC,][]{1971A&A....15..110V,1973A&A....29..263V,1985A&A...147L...6D,
  1985ApJ...298L...7H,1992ARA&A..30..575C,2001ApJ...554..803Y,
  2003ApJ...586..794B,2021MNRAS.504..118M,2021ApJ...914..126M}, which extends
over five decades in luminosity. It not only applies to entire galaxies, but
also holds on small scales down to a few~100~pc within local star-forming
galaxies (M31, M33, M101 and IC 342: \citealt{1988A&A...191L...9B}, M31:
\citealt{1998A&A...334...57H}, M33: \citealt{2003A&A...407..137H,
  2007A&A...466..509T}, M51: \citealt{2011AJ....141...41D}, LMC:
\citealt{2006MNRAS.370..363H}, samples of twenty to thirty star-forming galaxies
at GHz radio frequencies:
\citealt{1990ApJ...362...59B,2008ApJ...678..828M,2014AJ....147..103H}, as well
as at 140 MHz: \citealt{2019A&A...622A...8H}), thus providing important insight
into the star formation process in galaxies.

The birth and death of massive stars shape the FRC. Young massive stars
predominantly emit ultra-violet (UV) photons that are absorbed by their
dust-enshrouded environments, and subsequently re-emitted in the FIR. Provided
that dust is optically thick to UV photons, the emitted FIR radiation is
proportional to the SFR. When massive stars explode as supernovae, their remnant
shocks accelerate CR protons and electrons. These CR electrons generate
\emph{primary} radio synchrotron emission. Hadronically interacting CR protons
with ambient gas generate charged pions that decay into secondary electrons and
positrons (hereafter referred to as secondary electrons), which radiate
\emph{secondary} synchrotron emission in the radio continuum. Hence, primary and
secondary radio emission are linked to star formation and therefore FIR
emission. Electrons accelerated by a magnetic field will radiate synchrotron
emission, so that we need to simultaneously understand the amplification and
saturation of magnetic fields to produce a predictive estimate for the
radio synchrotron luminosity.

Early work \citep{1989A&A...218...67V,1996A&A...306..677L} proposed that
galaxies act as primary electron ``calorimeters'', i.e., that all electrons lose
their energy to synchrotron and inverse Compton radiation before they escape
into the halo. Calorimeter theory has been questioned merely on the basis of
observed radio spectra, which are flatter than purely cooled electron spectra:
for an injection spectrum $f_\e(E)\propto{E}^{-\alpha_{\rmn{inj}}}$ (with
$\alpha_{\rmn{inj}}\approx2.2$) the steady-state synchrotron-cooled spectrum is
$f_\e(E)\propto{E}^{-\alpha_{\rmn{inj}}-1}$, yielding a synchrotron spectrum
$I_\nu\propto\nu^{-\alpha_\nu}$, where
$\alpha_\nu=\alpha_{\rmn{inj}}/2\approx1$--1.3. Instead, radio observations
yield flat spectra $\alpha_\nu\approx0.5$--$0.8$ in starburst galaxies, thus
presenting a challenge to the applicability of the calorimetric model.

Previous one-zone models \citep{2006ApJ...645..186T,2010ApJ...717....1L,
  2010ApJ...717..196L} suggest that the GHz synchrotron spectra are flatter than
expected from rapid cooling, even though they are calorimetric, because
relativistic bremsstrahlung and ionization losses flatten the electron/positron
spectrum. In order to maintain the linear FRC in this model, in which primary
synchrotron emission dominates the total luminosity at low SFRs, the
contribution of secondary radio emission has to significantly increase in
starburst galaxies, thus implying a change of the dominant radio emission
mechanism along the FRC \citep{2010ApJ...717....1L}. In these one-zone
models, the magnetic field strength, the CR electron and proton energy densities
are free parameters that are fit to reproduce observed radio and gamma-ray
emission spectra of individual galaxies
\citep{2004ApJ...617..966T,2005A&A...444..403D,2008A&A...486..143P,
  2009ApJ...698.1054D,2010ApJ...717....1L,2011ApJ...734..107L,2012ApJ...755..106P,
  2013ApJ...768...53Y,2015MNRAS.453..222Y,2016MNRAS.457L..29Y,2016ApJ...821...87E}.
More detailed one-dimensional flux-tube models of our Galaxy
\citep{2002A&A...385..216B} and two-dimensional (axisymmetric) models
\citep{2014A&A...564A..61M,2020MNRAS.494.2679B} make use of parametrized source
functions, and/or prescribed density and magnetic field distributions.

While these studies are well suited for studying the relative importance of
different emission mechanisms, they cannot provide non-parametric
three-dimensional emission models and self-consistent simulations of the
dynamical impact of CRs or magnetic fields on the hydrodynamics. In particular,
while these models reveal important links between non-thermal radio and
gamma-ray observables and theoretical scaling arguments
\citep{2006ApJ...645..186T,2010ApJ...717....1L}, they were unable to directly
probe common assumptions such as energy equipartition of magnetic fields, CRs
and turbulence and to which extent equipartition is a necessary condition for
the FRC because as soon as equipartition condition is invoked the dynamical
feedback on the hydrodynamics would have to be taken into account. Most
importantly, a large body of recent literature has made a convincing case that
CR driven winds could be (partially) responsible for feedback associated with
star formation, thereby regulating the amount of stars formed, modifying the
structure of galactic discs, and regulating the thermodynamic properties of the
circumgalactic medium. This was demonstrated in simulations of the CR-driven
Parker instability \citep{2003A&A...412..331H, 2016ApJ...816....2R}, in
vertically stratified boxes of the interstellar medium
\citep[ISM,][]{Simpson2016,2016ApJ...816L..19G,2018MNRAS.479.3042G,
  2018ApJ...856..112F,2019A&A...622A.143C,2020ApJ...903...77B}, in isolated
galaxy simulations
\citep{2012MNRAS.423.2374U,2013ApJ...777L..38H,2013ApJ...777L..16B,2014MNRAS.437.3312S,
  Pakmor2016b,2017ApJ...847L..13P,2017ApJ...834..208R,2017MNRAS.467..906W,
  2018MNRAS.475..570J,2018ApJ...868..108B,2019MNRAS.488.3716C,2020A&A...638A.123D,
  2021ApJ...910..126S,2022arXiv220312029T}, in galaxies that experience a
ram-pressure wind \citep{2020ApJ...893...29B}, and in cosmological simulations
of galaxy formation
\citep{2008A&A...481...33J,2014ApJ...797L..18S,2016MNRAS.456..582S,
  2020MNRAS.497.1712B,2020MNRAS.496.4221J,2020MNRAS.492.3465H}.

While an undeniable proof of the importance of CR-driven winds in galaxy
formation has still not been put forward, the radio and gamma-ray emission of
galaxies could provide decisive clues and may be the most direct way to confirm
these models. In particular, polarised radio haloes in edge-on galaxies
demonstrate the presence of poloidal magnetic field lines connecting the disc to
the halo and show that CR electrons escape into the circumgalactic medium via
diffusion and advection \citep{2000A&A...364L..36T,2009A&A...506.1123H,
  2019A&A...622A...9M,2020A&A...639A.111S,2020A&A...639A.112K}. This picture of
a dominant large-scale ordered poloidal field associated with the outflow is
confirmed by observations of the polarised thermal dust emission from
SOFIA/HAWC+ in combination with a potential field extrapolation
\citep{2021ApJ...914...24L}.  Radio synchrotron emission probes CR electrons,
which cannot directly provide dynamical feedback owing to their negligible
energy density. By contrast, CR protons and magnetic fields are observed to be
in pressure equilibrium with the turbulence in the mid-plane of the Milky Way
\citep{1990ApJ...365..544B}. Thus, they carry sufficient momentum and energy
density to deliver the required feedback on galaxy formation. This calls for a
unifying simulation approach that follows the CR proton energy density in galaxy
simulations while simultaneously linking the resulting CR distribution to
non-thermal observables.

To this end, here we perform three-dimensional MHD simulations in which we
follow the evolution of the CR energy density in space and time, taking into
account all relevant CR gain and loss processes. In post-processing we then
solve for the steady-state energy spectra of CR protons, primary
shock-accelerated CR electrons as well as secondary CR electrons and compute the
resulting multi-frequency emission from radio to gamma-rays. This yields
time-dependent, spatially resolved CR, radio and gamma-ray spectra in various
galaxies ranging in size from dwarfs to Milky Way-like galaxies. Comparing these
mock observations to multi-messenger data enables us to link fully dynamical
galaxy formation models to non-thermal observational data and to quantify how we
can use these non-thermal observables to calibrate CR and magnetic feedback in
galaxy formation.

The radio synchrotron emission in galaxies is tightly linked to CR transport and
the gamma-ray emission. For this reason, it cannot be considered in
isolation. This work builds upon three companion papers that use the same
simulations and modelling, and study (i) the spatial and spectral CR
distributions that are compared to Voyager and AMS-02 data \citep{2021WerhahnI},
(ii) gamma-ray emission maps, spectra and the FIR-gamma-ray correlation
\citep{2021WerhahnII} and (iii) the radio emission \citep{2021WerhahnIII}. In
particular the third paper is complementary to ours and focuses on (i)
quantifying the relative contribution of primary and secondary CR electrons to
the radio luminosity, (ii) how calorimeter theory can be reconciled with the
observed flat radio spectra in starburst galaxies by additionally considering
free-free absorption and emission at low and high radio frequencies,
respectively, and (iii) how the decreasing radio luminosities in starburst
galaxies at high gas densities due to the increasing relativistic bremsstrahlung
and Coulomb losses of CR electrons can be reconciled with the power-law FRC that
extends from quiescently star-forming to violently star-bursting galaxies. In
agreement with the findings by \citet{2010ApJ...717....1L},
\citet{2021WerhahnIII} confirm the ``conspiracy'' at high gas surface densities,
which implies that the decreasing {\em primary} synchrotron luminosity due to
the increasing bremsstrahlung and Coulomb losses in these dense starburst
galaxies is almost exactly counteracted by an increasing contribution of {\em
  secondary} radio emission with increasing SFR. In fact, \citet{2021WerhahnIII}
find in models with CR advection and anisotropic diffusion that primary CR
electrons generally dominate the radio synchrotron emission while the
contribution of secondary synchrotron emission increases from 5 to 30 percent
with increasing SFR.

In the present work, we elucidate the origin of the {\em global FRC} and show
how this relates to the saturated stage of the small-scale dynamo which is also
known as the fluctuation dynamo. We study whether the emerging magnetic pressure
balances the vertical disc gravity as a function of time, galactocentric radius
and galaxy mass and single out those disc radii that are primarily responsible
for the total synchrotron luminosity. We identify the main physical processes
responsible for the scatter in the FRC. Finally, in studying the morphology of
the magnetic field strength and synchrotron intensity, we assess whether our
simulations reproduce the {\em local FRC}.  The outline of this study is as
follows. In Section~\ref{sec:simulations}, we describe our simulations, the
methodology of computing steady-state spectra of CRs and the resulting
synchrotron emission. In Section~\ref{sec:MFs}, we analyse the evolution of CR
and magnetic energy densities as well as the kinematic and saturated regimes of
the small-scale dynamo.  In Section~\ref{sec:FRC}, we study the mean and scatter
of the global FIR--radio correlation and explain our results analytically. In
analysing the morphology of the radio emission, we then elucidate the physics of
the local FRC and conclude in Section~\ref{sec:conclusion}. In
Appendix~\ref{sec:app_dynamo}, we provide supporting material for our
discussions of the small-scale dynamo and assess the robustness of our results
for different initial magnetic field configurations in
Appendix~\ref{sec:B_init}. In Appendix~\ref{sec:resolution}, we perform a
resolution study of our CR and radio spectra.

\section{Simulations and cosmic ray modelling}
\label{sec:simulations}

\begin{table*}
\caption{Overview of the parameters of the different simulations. (1) We compare
  two models: one with only CR advection (`CR adv') and one with additionally
  anisotropic CR diffusion (`CR diff') that is characterised by (2) a parallel
  diffusion coefficient $D_\parallel$. (3) We vary the virial mass
  $M_{200}$ and (4) NFW concentration parameter $c_{200}$ of the haloes, (5) the
  initial number of Voronoi cells within the virial radius $N$, (6)
  the CR energy injection efficiency at SNRs $\zeta_{\mathrm{SN}}$, (7) the
  model used for the initial magnetic field configuration, and (8) the initial
  magnetic field strength $B_{\rmn{init}}$ or $B_{\rmn{IGM}}$, respectively. (9)
  Shown are the analyses in which the corresponding simulation model is being
  used (PS denotes power spectra).}
\begin{center}
\begin{tabular}{ccccccccl}
\hline
model & $D_\parallel$ & $M_{200}$
& $c_{200}$ & $N$ & $\zeta_{\mathrm{SN}}$ & $B_{\rmn{init}}$ model & $B_{\rmn{init}}$ or $B_{\rmn{IGM}}$ & analysis \\
& $\mathrm{[cm^2~s^{-1}]}$ & $[\mathrm{M_{\odot}}]$ & & & & & $[\mathrm{G}]$~~~ & \\
\hline
CR diff & $1\times10^{28}$ & $1\times10^{12}$ & 12 & $10^{7}$ & $0.05$ & $B=\rmn{const.}$ & $10^{-10}$ & FRC (Fig.~\ref{fig:FRC}), evolution (Figs.~\ref{fig:energies}, \ref{fig:energies2}, \ref{fig:B_rho_evol}, \ref{fig:B_growth_res}), PS (Fig.~\ref{fig:ps_halos}) \\
CR diff & $1\times10^{28}$ & $3\times10^{11}$ & 12 & $10^{7}$ & $0.05$ & $B=\rmn{const.}$ & $10^{-10}$ & FRC, FRC track (Fig.~\ref{fig:FRC}) \\
CR diff & $1\times10^{28}$ & $1\times10^{11}$ & 12 & $10^{7}$ & $0.05$ & $B=\rmn{const.}$ & $10^{-10}$ & FRC (Fig.~\ref{fig:FRC}), evolution (Figs.~\ref{fig:energies}, \ref{fig:energies2}, \ref{fig:B_growth_res}), PS (Fig.~\ref{fig:ps_halos}) \\
CR diff & $1\times10^{28}$ & $1\times10^{10}$ & 12 & $10^{7}$ & $0.05$ & $B=\rmn{const.}$ & $10^{-10}$ & FRC (Fig.~\ref{fig:FRC}), evolution (Figs.~\ref{fig:energies}, \ref{fig:energies2}, \ref{fig:B_growth_res}), PS (Fig.~\ref{fig:ps_halos}) \\
\hline                                                                                              
CR adv  & $0$              & $1\times10^{12}$ & 12 & $10^{7}$ & $0.05$ & $B=\rmn{const.}$ & $10^{-10}$ & FRC (Fig.~\ref{fig:FRC}), evolution (Figs.~\ref{fig:energies}, \ref{fig:energies2})  \\
CR adv  & $0$              & $3\times10^{11}$ & 12 & $10^{7}$ & $0.05$ & $B=\rmn{const.}$ & $10^{-10}$ & FRC (Fig.~\ref{fig:FRC}) \\
CR adv  & $0$              & $1\times10^{11}$ & 12 & $10^{7}$ & $0.05$ & $B=\rmn{const.}$ & $10^{-10}$ & FRC (Fig.~\ref{fig:FRC}), evolution (Figs.~\ref{fig:energies}, \ref{fig:energies2}) \\
CR adv  & $0$              & $1\times10^{10}$ & 12 & $10^{7}$ & $0.05$ & $B=\rmn{const.}$ & $10^{-10}$ & FRC (Fig.~\ref{fig:FRC}), evolution (Figs.~\ref{fig:energies}, \ref{fig:energies2}) \\
\hline                                                                                              
CR diff & $1\times10^{28}$ & $1\times10^{12}$ & 12 & $10^{7}$ & $0.05$ & $B=\rmn{const.}$ & $10^{-12}$ & evolution (Fig.~\ref{fig:energies}) \\
CR diff & $1\times10^{28}$ & $1\times10^{11}$ & 12 & $10^{7}$ & $0.05$ & $B=\rmn{const.}$ & $10^{-12}$ & evolution (Fig.~\ref{fig:energies}) \\
CR diff & $1\times10^{28}$ & $1\times10^{10}$ & 12 & $10^{7}$ & $0.05$ & $B=\rmn{const.}$ & $10^{-12}$ & evolution (Fig.~\ref{fig:energies}) \\
\hline
CR diff & $1\times10^{28}$ & $1\times10^{12}$ & 12 & $10^{6}$ & $0.05$ & $B=\rmn{const.}$ & $10^{-10}$ & evolution (Fig.~\ref{fig:B_growth_res}) \\
CR diff & $1\times10^{28}$ & $1\times10^{11}$ & 12 & $10^{6}$ & $0.05$ & $B=\rmn{const.}$ & $10^{-10}$ & evolution (Fig.~\ref{fig:B_growth_res}) \\
CR diff & $1\times10^{28}$ & $1\times10^{10}$ & 12 & $10^{6}$ & $0.05$ & $B=\rmn{const.}$ & $10^{-10}$ & evolution (Fig.~\ref{fig:B_growth_res}) \\
CR diff & $1\times10^{28}$ & $1\times10^{12}$ & 12 & $10^{5}$ & $0.05$ & $B=\rmn{const.}$ & $10^{-10}$ & evolution (Fig.~\ref{fig:B_growth_res}) \\
CR diff & $1\times10^{28}$ & $1\times10^{11}$ & 12 & $10^{5}$ & $0.05$ & $B=\rmn{const.}$ & $10^{-10}$ & evolution (Fig.~\ref{fig:B_growth_res}) \\
CR diff & $1\times10^{28}$ & $1\times10^{10}$ & 12 & $10^{5}$ & $0.05$ & $B=\rmn{const.}$ & $10^{-10}$ & evolution (Fig.~\ref{fig:B_growth_res}) \\
\hline                                                                                              
CR diff & $1\times10^{28}$ & $1\times10^{12}$ & ~~7& $10^{7}$ & $0.10$ & $B=\rmn{const.}$ & $10^{-10}$ & PS (Fig.~\ref{fig:ps}), curvature (Figs.~\ref{fig:B-K_evolution} -- \ref{fig:K_histograms}), maps (Figs.~\ref{fig:turbulence}, \ref{fig:maps}, \ref{fig:maps2}), \\
& & & & & & & & profiles (Figs.~\ref{fig:profiles} -- \ref{fig:B_HSE}), Appendix (Figs.~\ref{fig:d_cell}, \ref{fig:B_3D}) \\
CR diff & $1\times10^{28}$ & $3\times10^{11}$ & 12 & $10^{7}$ & $0.05$ & $B=\rmn{const.}$ & $10^{-12}$ & FRC track (Fig.~\ref{fig:FRC}) \\
CR diff & $1\times10^{28}$ & $1\times10^{11}$ & 12 & $10^{7}$ & $0.10$ & $B=\rmn{const.}$ & $10^{-12}$ & maps (Figs.~\ref{fig:maps}, \ref{fig:maps2}), profiles (Figs.~\ref{fig:profiles} -- \ref{fig:B_HSE})  \\
\hline                                                                     
CR diff & $1\times10^{28}$ & $1\times10^{12}$ & ~~7& $10^{6}$ & $0.10$ & Eq.~\eqref{eq:B} & $10^{-12}$ & Appendix (Figs.~\ref{fig:B_dipole}, \ref{fig:B_evol}, \ref{fig:B_rho}) \\
CR diff & $1\times10^{28}$ & $1\times10^{12}$ & ~~7& $10^{6}$ & $0.10$ & Eq.~\eqref{eq:B} & $10^{-14}$ & Appendix (Figs.~\ref{fig:B_dipole}, \ref{fig:B_evol}, \ref{fig:B_rho}) \\
CR diff & $1\times10^{28}$ & $1\times10^{12}$ & ~~7& $10^{6}$ & $0.10$ & Eq.~\eqref{eq:B} & $10^{-16}$ & Appendix (Figs.~\ref{fig:B_dipole}, \ref{fig:B_evol}) \\
CR diff & $1\times10^{28}$ & $1\times10^{12}$ & ~~7& $10^{6}$ & $0.10$ & $B=\rmn{const.}$ & $10^{-10}$ & Appendix (Figs.~\ref{fig:B_evol}, \ref{fig:B_rho}) \\
CR diff & $1\times10^{28}$ & $1\times10^{12}$ & ~~7& $10^{6}$ & $0.10$ & $B=\rmn{const.}$ & $10^{-12}$ & Appendix (Figs.~\ref{fig:B_evol}, \ref{fig:CR_spectra}, \ref{fig:syn_spectra}) \\
CR diff & $1\times10^{28}$ & $1\times10^{12}$ & ~~7& $10^{7}$ & $0.10$ & $B=\rmn{const.}$ & $10^{-12}$ & Appendix (Figs.~\ref{fig:CR_spectra}, \ref{fig:syn_spectra}) \\
\hline                                                                                              
\label{tab:simulations-overview}
\end{tabular}
\end{center}
\end{table*}

\subsection{Simulation code and setup}

We simulate the formation and evolution of isolated disc galaxies with the
unstructured moving-mesh code \textsc{Arepo} \citep{2010MNRAS.401..791S,
  2016MNRAS.455.1134P,2020ApJS..248...32W}, which follows the evolution of
magnetic fields with the ideal MHD approximation.  We use the implementation of
cell-centred magnetic fields in \textsc{Arepo} \citep{2011MNRAS.418.1392P},
which employs the HLLD Riemann solver \citep{2005JCoPh.208..315M} to compute
fluxes and the Powell 8-wave scheme \citep{1999JCoPh.154..284P} for divergence
cleaning \citep{2013MNRAS.432..176P}. This implementation has been shown to
reproduce several observed properties of magnetic fields in galaxies
\citep{2017MNRAS.469.3185P,2018MNRAS.481.4410P} and the circumgalactic medium
\citep{2020MNRAS.498.3125P}. Moreover, recent cosmological adaptive-mesh
refinement simulations of galaxy formation, which use constraint transport for
evolving the magnetic field equipped with a turbulent subgrid scheme to increase
the effective resolution \citep{2022MNRAS.513.6028L}, find consistent magnetic
field structures in disc galaxies in comparison to those obtained with
\textsc{Arepo} in the Auriga project \citep{2017MNRAS.469.3185P}.

The simulations in this study are similar to those in
\citet{2017ApJ...847L..13P}, model radiative cooling and star formation within a
pressurised ISM \citep{2003MNRAS.339..289S} and employ the one-moment CR
hydrodynamics algorithm \citep{Pakmor2016a,2017MNRAS.465.4500P}.  We model the
formation of disc galaxies with masses ranging from dwarf- to Milky Way-mass
galaxies (residing in dark matter haloes of masses $10^{10}$, $10^{11}$, and
$10^{12}~\msun$). Initially, the gas is in approximate hydrostatic equilibrium
with the dark matter potential and has a baryon mass fraction of
$\Omega_{\mathrm{b}}/\Omega_{\mathrm{m}}=0.155$. Dark matter and gas follow an
NFW mass density profile, $\rho_\rmn{NFW}$ \citep{1997ApJ...490..493N}, which we
slightly soften at the centre (below 0.1~kpc) to introduce a core into the
gas. The profile is parametrized by a concentration parameter
$c_{200}=r_{200}/r_{\mathrm{s}}$, where $r_{\mathrm{s}}$ is the characteristic
scale radius of the NFW profile and the radius $r_{200}$ encloses a mean density
equals 200 times the critical density necessary to close the universe. We assume
solid-body rotation of the dark matter halo that has an initial angular momentum
$J$, which is parametrized in terms of the dimensionless spin parameter
$\lambda=J|E|^{1/2}/(GM_{200}^{5/2})$, where $|E|$ is the total energy of the
halo, $G$ is Newton's constant, and we adopt a value $\lambda=0.3$. In our
standard simulations, the haloes initially contain $10^7$ gas cells within the
virial radius. Each cell has a target mass of $155~\msun\times{M}_{10}$, where
$M_{10}=M_{200}/(10^{10}~\msun)$. We ensure that the gas mass of all Voronoi
cells remains within a factor of two of the target mass by explicitly refining
and de-refining the mesh cells and also ensure that the volume of adjacent
Voronoi cells differs at most by a factor of ten.

We model CR protons as a second (relativistic) fluid with adiabatic index of
$4/3$ \citep{2017MNRAS.465.4500P}. Initially, CR protons are absent and CR
proton energy is instantaneously injected into the local environment of every
newly spawned stellar macro-particle with an efficiency $\zeta_{\rmn{SN}}=0.05$
and 0.1 of the kinetic supernova energy. While the high efficiency value has
been widely used, the smaller value derives from a combination of kinetic plasma
simulations at oblique shocks \citep{2014ApJ...783...91C} and three-dimensional
MHD simulations of CR proton acceleration at supernova remnant (SNR) shocks
\citep{Pais2018}, followed by a detailed comparison of simulated multi-frequency
emission maps and spectra from the radio to gamma rays to observational data
\citep{Pais2020a,Pais2020b,Winner2020}.  All simulations consider CR proton
losses as a result of Coulomb and hadronic CR interactions
\citep{2017MNRAS.465.4500P} and follow adiabatic changes of CR proton energy as
CR protons are advected with the gas (model `CR~adv'). Our model `CR~diff' {\it
  additionally} accounts for anisotropic diffusion of CR proton energy with a
coefficient $10^{28}~\mathrm{cm^{2}~s^{-1}}$ along the magnetic field and no
diffusion perpendicular to it \citep{Pakmor2016a}.\footnote{The hardening of the
logarithmic momentum slope of the CR proton spectrum at low Galactocentric radii
is interpreted as a signature of anisotropic diffusion in the Galactic magnetic
field \citep{Cerri2017,Evoli2017}. Secondary radioactive isotopes are produced
in CR spallation processes and have relatively short decay times. The observed
abundance of these unstable nuclei in AMS-02 data was used to determine the CR
residence time in the Galaxy and to constrain the quoted value of the CR
diffusion coefficient \citep{Evoli2019,Evoli2020}.}

In our standard simulations, the magnetic field is initialised as a uniform
homogeneous seed field along the $x$-axis with strength
$B_{\rmn{init}}=10^{-12}$ and $10^{-10}~\mathrm{G}$.  $B_{\rmn{init}}$
represents the pre-amplified magnetic field in a proto-galactic environment,
which has to be large enough to grow sufficiently during our collapse-driven
small-scale dynamo phase given our finite numerical resolution (see discussion
in Section~\ref{sec:growth_analysis}). Similarly, $B_{\rmn{init}}$ should be small enough
so that adiabatic compression does not boost the field strength to values that
modify the hydrodynamics. In Appendix~\ref{sec:B_init}, we explore the
robustness of our results to changes of the initial magnetic field
distribution. To this end we additionally simulate a configuration that is a
superposition of small magnetic dipoles aligned with the $z$ axis that have a
strength proportional to $\rho_\rmn{NFW}^{2/3}$, which may result from the
isotropic collapse of a proto-galaxy due to magnetic flux freezing. The emerging
model has a global large scale dipole-like magnetic topology. We find that the
magnetic dynamo grows more efficiently in this pre-compressed magnetic field
distribution. In particular, we can afford magnetic field strengths of the
intergalactic medium (IGM) that are $10^4$ times smaller than the initial
magnetic field in our homogeneous seed field model and still obtain the same
exponential dynamo growth rate. Most importantly, the resulting magnetic field
distributions can be mapped from one to the other model so that the results
presented in this work are insensitive to the specific choice of the initial
magnetic configuration.  For an overview of the individual models analysed in
this study, see Table~\ref{tab:simulations-overview}. In this work, we show
different radial profiles: in our terminology $R$ denotes a cylindrical disc
radius and $r$ is a three-dimensional radius.

\subsection{Steady-state spectra of cosmic rays}

In post-processing, we model the steady-state spectra $f_i(\x,E_i)$ as a
function of energy $E_i$ of (i) CR protons, (ii) secondary electrons and
positrons that result from hadronic CR-proton interactions with the ISM, and
(iii) primary shock-accelerated electrons at SNRs in every Voronoi cell of our
simulations at position $\x$. Following \citet{2021WerhahnI}, we solve the
diffusion-loss equation for CR protons, primary and secondary electrons,
respectively:
\begin{align}
\frac{\mathrm{}f_i(\x,E_i)}{\tau_{\mathrm{esc}}}-\frac{\mathrm{d}}{\mathrm{d}E_i}\left[f_i(\x,E_i)b(\x,E_i)\right]=q_i(\x,E_i),
\label{eq:diff-loss}
\end{align}
where $f_i(\x,E_i)=\dd N_i/(\dd V\dd E_i)$ is the differential number of CRs per
unit volume and energy, $q_i(\x,E_i)=\dd N_i/(\dd V\dd E_i\dd t)$ is the source
function of freshly injected CRs per unit volume, energy, and time, $E_i$ is the
CR energy and $i$ denotes the three CR populations. Motivated by diffusive shock
acceleration at SNRs, we assume a power-law momentum spectrum for the injection
of CR protons and primary electrons, $q_i(\x,E_i)= q_i[\x,p(E_i)]
\rmn{d}p_i/\rmn{d}E_i$. We adopt dimensionless momenta, $p_\e=P_\e/(m_\e c)$ and
$p_\p=P_\p/(m_\p c)$ for electrons and protons, respectively, where $m_\e$
($m_\p$) is the electron (proton) rest mass and $c$ denotes the speed of
light. The source functions are equipped with an exponential cutoff as follows:
\begin{align}
  q_i(\x,p_{i})\mathrm{d}p_{i} = C_{i}(\x) p_{i}^{-\alpha_{\rmn{inj}}}
  \exp[-(p_i/p_{\mathrm{cut},i})^{n}]\mathrm{d}p_{i},
\label{eq:source}
\end{align}
where $i=\{\mathrm{e,p}\}$ denotes the shock-accelerated CR species s (e denotes
primary and secondary electrons, p denotes protons), $\alpha_{\rmn{inj}}=2.2$ is
the injection spectral index of protons and electrons \citep{Lacki2013}, $n=1$
for protons and $n=2$ for primary electrons \citep{Zirakashvili2007,Blasi2010}
and we adopt cutoff momenta for protons
$p_{\mathrm{cut,p}}=1\,\mathrm{PeV}/(m_{\mathrm{p}}c^2)$ \citep{Gaisser1990} and
for electrons $p_{\mathrm{cut,e}}=20\,\mathrm{TeV}/(m_{\mathrm{e}}c^2)$
\citep{Vink2012}. Note that all our CR spectra extend from the non-relativistic
to the fully relativistic regime. In practice, we adopt
$p_{\e,\rmn{min}}=0.1/(m_\e c)$ and $p_{\p,\rmn{min}}=0.01/(m_\p c)$ and extend
the momentum range beyond the cutoff momenta.

We calculate the production spectra of secondary CR electrons
and positrons, $q_{\e^\pm}$ via equations~(B1) and (B6) in \citet{2021WerhahnI}
for two different energy regimes. At small kinetic proton energies,
$T_{\mathrm{p}}<10\,\mathrm{GeV}$, we combine the normalised pion energy
distribution \citep{Yang2018} with our own parametrization of the total cross
section for $\pi^{\pm}$ production \citep{2021WerhahnI}. At high energies,
$T_{\mathrm{p}}>100\mathrm{\,GeV}$, we use the model by
\citet{2006PhRvD..74c4018K} and perform a cubic spline interpolation in the
energy range in between.

In case of CR protons, we account for energy losses,
$b(\x,E_\p)=-(\mathrm{d}E_\p/\mathrm{d}t)(\x)$, owing to hadronic and Coulomb
interactions as well as CR escape due to advection and diffusion. Because
Eq.~\eqref{eq:diff-loss} is a linear equation in $f_i(\x,E_i)$ and
$q_i(\x,E_i)$, we re-normalise the steady-state spectra to match the simulated
CR energy density in each cell. The escape losses include CR advection and
diffusion, i.e.,
\begin{align}
\tau_{\mathrm{esc}}(\x)=\frac{1}{\tau_{\mathrm{adv}}^{-1}(\x) + \tau_{\mathrm{diff}}^{-1}(\x)}.
\label{eq:tau_esc}
\end{align}
The diffusion time-scale
$\tau_{\mathrm{diff}}(\x)=L_{\mathrm{CR}}^{2}(\x)/D$ is estimated using an
estimate for the diffusion length in each cell,
$L_{\mathrm{CR}}(\x)=(\varepsilon_{\mathrm{CR}}/\left|\nabla\varepsilon_{\mathrm{CR}}\right|)(\x)$.
We adopt an energy-dependent diffusion coefficient $D_i=D_0
(E_i/E_0)^{\delta}$, where $i=\mathrm{e,p}$,
$D_0=10^{28}\,\mathrm{cm^2~s}^{-1}$, $E_0=3\,\mathrm{GeV}$, and $\delta =
0.5$, which was inferred by fitting observed beryllium isotope ratios
\citep{Evoli2020}.\footnote{Note that the assumption of a weakly
energy-dependent diffusion coefficient in our steady-state modelling is
consistent with the constant diffusion coefficient of our CR-MHD
simulations. Those simulations evolve the full CR energy density, which is
dominated by GeV CRs and only attains a negligible contribution from high-energy
CRs that diffuse significantly faster.} We calculate the advection time-scale
$\tau_{\mathrm{adv}}(\x)=L_{\mathrm{CR}}(\x)/\varv_z(\x)$. Note that we only
account for the vertical velocity component with respect to the disc to estimate
the advection losses. This is justified because the radial and azimuthal
velocity differences of adjacent Voronoi cells in the disc are negligible in
comparison to the vertical velocities \citep[see figure~6
  of][]{2021WerhahnI}. CR transport via advection and anisotropic diffusion is
also strongly suppressed in the radial direction because the disc magnetic field
is mostly toroidal \citep{2013MNRAS.432..176P,Pakmor2016b} and because circular
rotation dominates the kinetic energy density of the gas (see below).

In addition to losses due to spatial advective and diffusive transport, CR
electrons (and positrons) also lose energy due to Coulomb interactions and the
emission of radiation.  We account for synchrotron, inverse Compton (IC) and
bremsstrahlung losses through the energy loss term $b(\x,E_\e)$. Bremsstrahlung
losses of electrons scale as $b_\rmn{brems}(\x)\propto n_\rmn{p}(\x) E_\e
\ln(2E_\e)$ (where $n_\rmn{p}$ is the proton number density, see
\citet{1979rpa..book.....R} for details).  Synchrotron and IC losses show the
same energy dependence: in the relativistic regime we obtain
$b_\rmn{sync}(\x)\propto B(\x)^2 E_\e^2$ and $b_\rmn{IC}(\x)\propto
B_{\rmn{ph}}(\x)^2 E_\e^2$ (where $B=\sqrt{\vec{B}}$ and
$B_{\rmn{ph}}=\sqrt{8{\uppi}\varepsilon_\rmn{ph}}$ are the strengths of the
magnetic field and equivalent magnetic field of a photon distribution with an
energy density $\varepsilon_\rmn{ph}$, respectively).  The photon energy density
$\varepsilon_\mathrm{ph}$ accounts for photons from the cosmic microwave
background (CMB) and stars. We assume that the UV light emitted by young stellar
populations is absorbed by warm dust with a temperature of $\sim20\,\mathrm{K}$
\citep{2000Calzetti} and re-emitted in the FIR with a Planckian black-body
distribution. Thus, we compute the energy loss rate in each cell $j$ by
integrating over the FIR flux arriving from all other $N$ cells $i$ with
$\dot{M}_{\star,i}>0$ at a distance $r_i$, and obtain
\begin{equation}
\varepsilon_{\star,j}=\sum_{i=1}^N\frac{L_{\mathrm{FIR},i}}{4\uppi r_{i}^{2}c},
\label{eq:photon}
\end{equation}
where ${L_{\mathrm{FIR},i}}\propto\dot{M}_{\star,i}$
\citep{1998ApJ...498..541K}, we use $r_i=[3V_i/(4\uppi)]^{1/3}$ as the distance if
the considered cell $j=i$ is actively star forming, and $V_i$ denotes the cell's
volume. In practice, we accelerate this computation with a tree code.

We link the primary electron to the proton population by means of a CR
electron-to-proton injection ratio, $K_{\mathrm{ep}}^{\mathrm{inj}}=0.02$, which
is defined to be the ratio of the corresponding injection spectra at the same
(physical) momentum $P_{0}$:
\begin{align}
  q_{\mathrm{e},\mathrm{prim}}[P_{0}/(m_\e c)]\mathrm{d}p_{\mathrm{e}}=
  K_{\mathrm{ep}}^{\mathrm{inj}}q_{\mathrm{p}}[P_{0}/(m_\p c)]\mathrm{d}p_{\mathrm{p}}.
\label{eq:K_ep_inj}
\end{align}
We choose $K_{\mathrm{ep}}^{\mathrm{inj}}$ so that it reproduces the {\em
  observed} value in the Milky Way at 10 GeV,
$K_{\mathrm{ep}}^{\mathrm{obs}}=0.01$, when averaging CR spectra around the
solar galactocentric radius in a simulation model that resembles the Milky Way
in terms of halo mass and SFR, and assume $K_{\mathrm{ep}}^{\mathrm{inj}}$ to be
a universal constant in all galaxies (see \citet{2021WerhahnI} for more
details).  Assuming the same injected spectral index of electrons and protons,
$2<\alpha_\rmn{inj}<3$, and a lower momentum cutoff that is much smaller than
$m_\rmn{p}c$ ($m_\rmn{e}c$) for protons (electrons), we obtain an injected
energy ratio of CR electrons and protons,
\begin{align}
  \zeta_\rmn{prim} = \frac{\eps_{\e,\rmn{inj}}}{\eps_{\p,\rmn{inj}}} = K_{\mathrm{ep}}^{\mathrm{inj}}
  \left(\frac{m_{\mathrm{p}}}{m_{\mathrm{e}}}\right)^{\alpha_{\mathrm{inj}}-2}
  \approx0.09,
\label{eq:K_ep-zeta}
\end{align}
where we adopted our value of $K_{\mathrm{ep}}^{\mathrm{inj}}=0.02$ and
$\alpha_{\mathrm{inj}}=2.2$. This result is consistent with the parameters used
in one-zone steady-state models in the literature
\citep[e.g.,][]{2010ApJ...717....1L}.

Ab initio, it is unclear that solving the diffusion-loss equation in each
individual computational Voronoi cell produces reliable results. In fact, this
procedure is only justified if and only if the characteristic time-scale of the
change in total energy density of CRs in our simulations,
$\tau_{\mathrm{CR}}=\varepsilon_{\mathrm{CR}}/ \dot{\varepsilon}_{\mathrm{CR}}$,
is longer than the time-scale associated with all cooling or escape processes
that maintain a steady state, $\tau_{\mathrm{all}} \lesssim \tau_{\mathrm{CR}}$.
Here, $\tau_{\mathrm{all}}$ is the combined time-scale of all relevant cooling
and diffusion processes at a given energy, $\tau_{\mathrm{all}}^{-1}=
\tau_{\mathrm{cool}}^{-1}+\tau_{\mathrm{diff}}^{-1}$.\footnote{Except for
galactic wind regions, the advection time-scale is everywhere larger than the
diffusion time-scale as is shown in figure~7 of \citet{2021WerhahnI}, justifying
our neglect of the advection process in $\tau_{\mathrm{all}}$.}  In figure~9 of
\citet{2021WerhahnI}, we find that the steady-state approximation applied to
each computational cell in our simulations is well justified in the ISM at and
above average densities while it breaks down in regions of low gas density, at
SNRs that freshly inject CRs, and in CR-driven galactic winds: these
environments imply fast changes in the CR energy density, which disturb the
steady-state configuration. While the CR proton energy is conservatively
transported in our simulations, reliably computing CR spectra in these regions
would require to dynamically evolve the spectral CR proton
\citep{2020MNRAS.491..993G,2022MNRAS.510.3917G} and electron distributions
\citep{2019MNRAS.488.2235W,Winner2020,2021ApJS..253...18O}, which likely deviate
from the steady state distribution at very small and large CR energies. Most
importantly, if we weight each Voronoi cell by the non-thermal radio synchrotron
or hadronic gamma-ray emission, we find that the majority of non-thermally
emitting cells obey the steady-state condition: $\tau_{\mathrm{all}} \lesssim
\tau_{\mathrm{CR}}$. This demonstrates that the steady-state assumption is well
justified in regions that dominate the radio synchrotron and gamma-ray
emission. Moreover, in Appendix~\ref{sec:resolution} we show that our CR and
radio spectra are numerically well converged. Thus, total radio luminosities and
intensity maps analysed in this work are reliable and robust.

\subsection{Radio synchrotron emission}

The synchrotron intensity depends on the strength of the transverse (with
respect to the line-of-sight) component of the magnetic field, $B_\perp$, and
the spatial and spectral CR electron distribution $f_\e(\x,p_\e)$. The intensity
of the omnidirectional emissivity per unit time, frequency, and volume,
$j_\nu(\x)\equiv{E}_\gamma\dd N_\gamma/(\dd t\,\dd\nu\,\dd^3x)$, is given by
\begin{align}
  j_{\nu}(\x)
  =\frac{\sqrt{3}e^{3}B_\perp(\x)}{m_{e}c^{2}}
  \int_{0}^{\infty}f_\e(\x,p_\e)F(\xi)\dd p_\e,
\label{eq:jnu}
\end{align}
where ${E}_\gamma$ is the photon energy, $e$ denotes the elementary charge and
$F(\xi)$ is the dimensionless synchrotron kernel that is given in terms of an
integral over a modified Bessel function \citep{1979rpa..book.....R} and
$\xi=\xi(p_\e)=\nu/\nu_\rmn{c}(p_\e)$ is the synchrotron frequency in units of a
critical frequency, $\nu_\rmn{c}$. In practice, we adopt an analytical
approximation for $F(\xi)$ (\citealt{2010PhRvD..82d3002A}, see
\citealt{2021WerhahnIII} for more details). The specific intensity $I_\nu$ is
obtained by integrating $j_{\nu}(\x)$ along the line-of-sight $s$ and is a
function of observational frequency $\nu$ and position on the sky,
$\rvec_\perp$, and reads (in units of
erg~s$^{-1}$~Hz$^{-1}$~cm$^{-2}$~sterad$^{-1}$)
\begin{equation}
  \label{eq:Inu}
  I_\nu(\rvec_\perp)=\frac{1}{4\uppi}\,\int_0^{\infty} j_\nu(\rvec_\perp,s)\dd s.
\end{equation}
The specific radio luminosity follows as a result of volume integration of the
emissivity, $L_{\nu}=\int_{V}j_{\nu}(\x)\,\dd^3x$. This
formalism enables us to self-consistently predict the radio emission from
simulated galaxies. In this work, we restrict ourselves to the total synchrotron
emissivity, $j_\nu(\x)=j_{\nu,\rmn{prim}}(\x) + j_{\nu,\rmn{sec}}(\x)$, which is
the sum of primary and secondary emission while we will scrutinise the relative
contributions of primaries and secondaries to the total emission in our
companion paper \citep{2021WerhahnIII}.

To understand the involved electron energies to order of magnitude, we relate
the synchrotron emission frequency to the electron Lorentz factor
$\gamma_\e$,\footnote{This formula relates the magnetic field strength and
electron energy more accurately to the characteristic emission frequency
$\nu_{\rmn{syn}}$ in comparison to the critical synchrotron frequency
$\nu_{\rmn{c}}$ (which was widely used in the radio astronomy community
before). Approximating the synchrotron kernel with a Dirac delta distribution
that is centred on $\nu_{\rmn{syn}}$ exactly returns the synchrotron emissivity
for an electron spectral index $\alpha_\rmn{e}=3$ and only shows relative
deviations up to 30 per cent for $\Delta \alpha_\rmn{e}=0.5$.}
\begin{equation}
  \label{eq:synch_photon}
  \nu_{\rmn{syn}}=2\nu_{\rmn{c}}
  =\frac{3eB_\perp}{2\uppi\,m_{\e}c}\,\gamma_\e^2\simeq1~\mbox{GHz}\,\frac{B_\perp}{1\,\upmu\mbox{G}}\,
  \left(\frac{\gamma_\e}{10^4}\right)^2.
\end{equation}
In the case of hadronic CR proton interactions with the ISM, the parent proton
and secondary electron energies are related by $E_\e\approx E_\p/16$. Hence, GHz
radio synchrotron emission in $\upmu$G magnetic field strengths probes 5~GeV
electrons, which are either directly accelerated at SNR shocks or hadronically
produced by 80~GeV protons.

\begin{figure*}
\centering
\includegraphics[width=\textwidth]{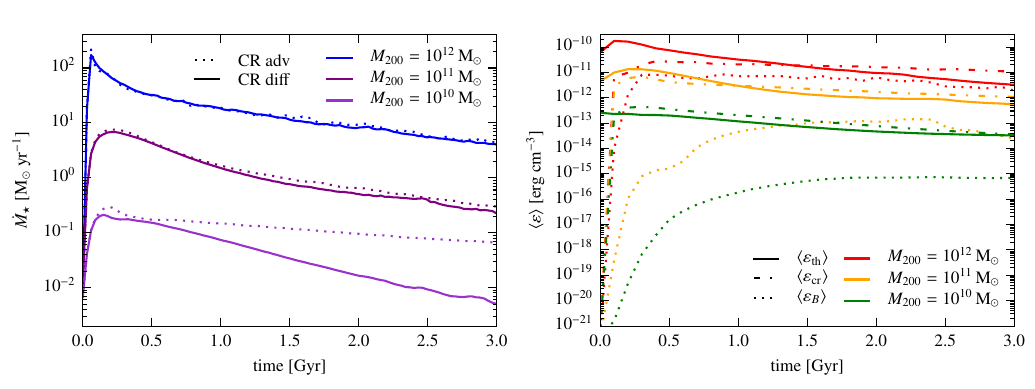}
\includegraphics[width=\textwidth]{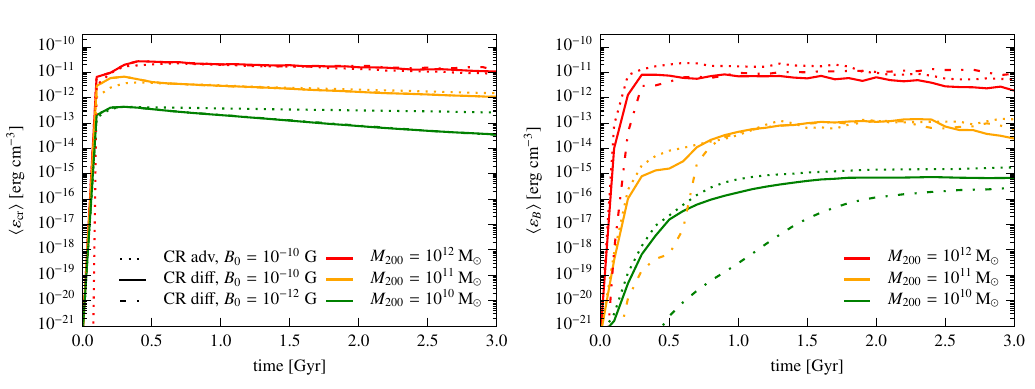}
\caption{Time evolution of the SFR (top left-hand panel) and the volume-averaged
  energy densities in a disc of radius 10~kpc and total height 1~kpc (other
  panels). Different halo masses ($10^{10},~10^{11}$, and $10^{12}~\msun$, all
  with concentration $c_{200}=12$) are colour coded.  Simulations with advective
  and anisotropic diffusive CR transport (solid) suppress star formation more
  strongly in smaller galaxies in comparison to simulations that only account
  for CR advection (dotted). The top right-hand panel shows the time evolution
  of the thermal, CR and magnetic energy densities in our anisotropic CR
  diffusion model with an initial magnetic field
  $B_{\rmn{init}}=10^{-10}\,\mathrm{G}$ and demonstrates that the magnetic
  saturation time scale increases for smaller galaxies and does not saturate at
  equipartition with the thermal energy density. In the bottom panels, we
  compare the CR (left) and magnetic energy density (right) for different
  simulations denoted in the bottom-left panel. While CR diffusion causes a net
  loss of CR energy from the disc, the initial magnetic field strength has no
  impact on the CR energy density.  The magnetic energy density grows first
  exponentially (via a small-scale dynamo) and saturates in the
  following. Because of the finite turbulent driving time of the initial
  starburst and the larger magnetic growth time in smaller galaxies, the
  magnetic field in the simulations with $B_{\rmn{init}}=10^{-12}\,\mathrm{G}$
  increases at a much slower rate (dash-dotted).}
\label{fig:energies}
\end{figure*}

\begin{figure*}
\centering
\includegraphics[width=\textwidth]{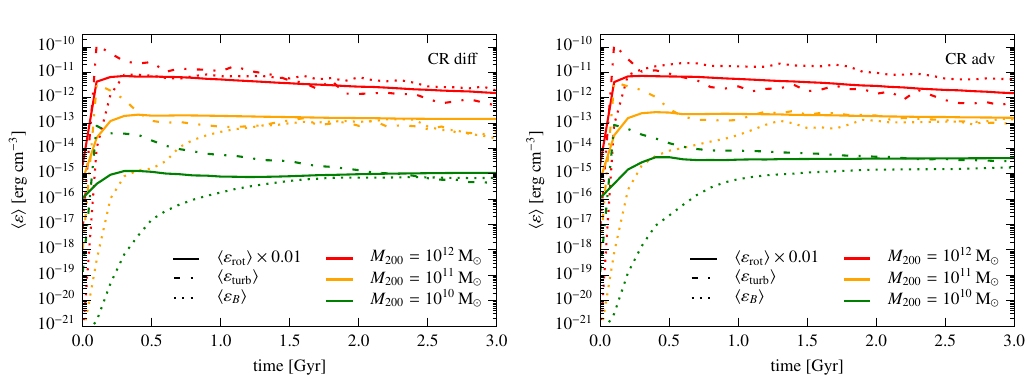}
\caption{Halo mass dependence of the magnetic saturation level. We show the time
  evolution of the volume-averaged magnetic energy density (dotted), the kinetic
  rotational and `turbulent' energy densities ($\eps_{\rmn{rot}} =\rho
  \varv_\varphi^2/2$, solid, and $\eps_{\rmn{turb}}=
  \rho\delta\varv^2/2\approx3/4\times\rho (\varv_z^2+ \varv_R^2)$,
  dash-dotted) in a disc of radius 10~kpc and total height 1~kpc for our models
  `CR diff' (left) and `CR adv' (right).  Simulations of different halo masses
  ($10^{10},~10^{11}$, and $10^{12}~\msun$, all with concentration $c_{200}=12$,
  $B_{\rmn{init}}=10^{-10}\,\mathrm{G}$ and $\zeta_\rmn{SN}=0.05$) are colour
  coded. All discs are rotationally supported so that $\eps_{\rmn{rot}}\approx
  100 \eps_{\rmn{turb}}$ at late times. A small-scale dynamo amplifies the
  magnetic field so that it comes into equipartition with the `turbulent' energy
  density after the initial growth phase. We see evidence for an additional
  magnetic amplification mechanisms in the $10^{12}~\msun$ halo at late times.}
\label{fig:energies2}
\end{figure*}

\section{Energy equipartition and the small-scale dynamo}
\label{sec:MFs}

First, we are studying the growth and saturation of CR and magnetic energy
densities in comparison to the thermal and kinetic energy density across
different halo masses. We will specifically work out the saturation level of the
magnetic field strength with halo mass. We carefully explore the characteristics
of the amplification mechanisms for galactic magnetic fields and identify
adiabatic compression in the initial stage, followed by a superposition of
small-scale dynamo processes to be responsible for magnetic field growth. This
is supported by a discussion of the numerical Reynolds number in our moving mesh
simulations.  Power spectrum analyses and magnetic curvature statistics provide
further insights into the kinematic and saturated regimes of the small-scale
magnetic dynamo.

\subsection{Growth and saturation of CR and magnetic energy densities}
\label{sec:mag_sat}

To mimic density inhomogeneities of cosmologically growing haloes in our
idealised setup, we sample the softened NFW mass density profile randomly while
ensuring equal mass per Voronoi cell. As a result, there is a distribution of
mass densities at any given radius, which breaks the axisymmetry of our
setup. At the beginning of our simulations, we switch on cooling of the slowly
rotating gas in approximate hydrostatic equilibrium.  As a result, the densest
gas at the halo centre cools fastest, collapses and experiences adiabatic
compression.  A shock forms once later collapsing gas encounters the compressed
dense gas at the centre. The (peanut-shaped) accretion shock propagates into the
slightly inhomogeneous circumgalactic medium \citep[see figure~5
  of][]{2003MNRAS.339..289S} and thermalises the kinetic energy from
gravitational infall, thereby reducing the gas velocities behind the shock.  The
accretion shock itself becomes corrugated as it interacts with the cooler and
denser, filamentary infalling structures that have initially been sourced by
small-scale overdensities. This leaves behind a hot and turbulent circumgalactic
atmosphere because the curved shock converts a fraction of the angular momentum
of the accreting gas into vorticity at the scale of the shock curvature
according to Crocco's theorem \citeyearpar{Crocco}. This vorticity cascades down
in scale and feeds a turbulent kinetic power spectrum that will amplify any
existing magnetic field as we will show in the following.

Conservation of specific angular momentum of the cooling gas causes a fraction
of high angular momentum gas to be constantly accreted along the equatorial
plane so that a centrifugally supported cool disc forms at around 150~Myr in the
Milky Way-like halo and somewhat later in the dwarf galaxies. The cool galactic
disc has a temperature of a few times $10^5$~K in our effective ISM model and
typical sound speeds of several tens of $\rmn{km~s}^{-1}$. This causes a
supersonic velocity shear between the rotationally supported cool disc and the
slower rotating and partially pressure supported hot circumgalactic medium at
temperatures $\gtrsim2\times10^6$~K, which excites and grows Kelvin-Helmholtz
surface and body modes
\citep{2016MNRAS.463.3921M,2019MNRAS.485..908B,2019MNRAS.489.3368B}. As a
result, turbulence is continuously injected at later times ($t\gtrsim200$~Myr)
through non-linearly interacting body modes on a range of scales at and below
the scale of the vertical extend of the velocity shear. As we will show below,
this causes a turbulent cascade, amplifies the magnetic fields further via a
small-scale dynamo \citep[which can be supplemented by a large-scale
  dynamo,][]{2005PhR...417....1B} and is eventually dissipated at the grid
scale.

The rotating cool disc provides favourable conditions for star formation so that
all our simulations exhibit strong initial starbursts that are followed by
exponentially declining SFRs in our large galaxies, independent of the CR
transport scheme as well as in the $10^{10}\,\msun$ halo in our model
`CR~diff'. By contrast, star formation starts to level off in the model `CR~adv'
in this dwarf galaxy (top left-hand panel of Fig.~\ref{fig:energies}). The
behaviour is mirrored in the evolution of the CR energy density in the panel
below. The reason of the suppressed SFR in small haloes are CR-driven winds that
efficiently remove gas from the disc in the `CR~diff' model: after CRs have
accumulated in the disc, their buoyancy bends and opens up the toroidal disc
magnetic field. CRs diffuse into the halo and accelerate the gas, thereby
driving an outflow solely through the CR pressure gradient force with increasing
strength towards smaller galaxies \citep[see also][ for a study of the halo-mass
  dependence of CR feedback]{2018MNRAS.475..570J}. We find only weak fountain
flows in model `CR~adv' \citep[see also][]{Pakmor2016b}.

The top right-hand panel of Fig.~\ref{fig:energies} shows the time evolution of
the average thermal, CR and magnetic energy densities in model `CR~diff'. After
the starburst, the injection of CR energy at SNRs quickly causes the CRs
to reach approximate equipartition with the thermal energy (top right-hand panel
of Fig.~\ref{fig:energies}). While CR and thermal energies balance each other in
the $10^{10}\,\msun$ halo, the CR component quickly dominates the overall energy
budget in larger galaxies to the point where it triples the thermal energy in
the Milky Way-mass galaxy. Note that these energy densities represent averages
in a disc of radius 10~kpc and total height 1~kpc and that the individual
pressures vary with galactocentric radius and height from the disc. While our
simulations predict that CR and thermal pressures reach equipartition at the solar
radius in our Milky Way-mass galaxy, they dominate over the thermal pressure at
larger radius and disc heights while they fall short of the thermal pressure at
smaller radii (see figure~1 of \citealt{2017ApJ...847L..13P}).

The approximate equipartition of CR and thermal energy density (within a factor
of three) is a direct consequence of CR physics and our pressurised ISM
\citep{2003MNRAS.339..289S}, which models the multi-phase ISM with an effective
equation of state so that it balances the vertical disc gravity. In fact, the
approximate equipartition suggests an attractor solution of a self-regulated
feedback loop: if the CR pressure accumulates until CRs dominate the energy
budget, they will buoyantly and diffusively escape into the halo and push on the
gas by means of their gradient pressure force. Conversely, if thermal pressure
dominates, it will loose its energy radiatively at a much faster rate in
comparison to magnetic fields and CRs, which have negligible radiative losses
and experience losses due to inelastic collisions with the ambient gas
\citep{2008A&A...481...33J}. This self-regulation picture implies a robust
physical attractor solution that our simulations seem to settle into, but to
which extend this attractor is realised and understanding the conditions for its
violations needs to be studied with CR hydrodynamics in the self-confinement
picture \citep{2019MNRAS.485.2977T,2022arXiv220312029T}.
 
In the first stage of our simulations, the magnetic energy density $\eps_B$
grows exponentially across more than ten orders of magnitude, which is followed
by slower growth until $\eps_B$ eventually saturates. The top right-hand panel
of Fig.~\ref{fig:energies} shows a smaller growth rate in smaller galaxies which
leads to a saturation level below equipartition in these dwarf galaxies. We find
saturation times of 0.3~Gyr and 1.5~Gyr for our haloes with masses $10^{12}$ and
$10^{11}~\msun$ in model `CR~diff'. In addition, the strong outflow in the
$10^{10}~\msun$ halo in model `CR~diff' quenches the dynamo, which reaches a
saturated mean field
${\langle}B\rangle=\sqrt{8\uppi{\langle}\eps_B\rangle}\approx0.1~\upmu$G
(averaged over the disc and after 3~Gyr), which is three times lower than in
model `CR~adv' \citep{2017MNRAS.465.4500P}. It is interesting to note that
already in the $10^{11}~\msun$ halo, the mean magnetic energy density saturates
below that of the CMB,
\begin{align}
  \label{eq:CMB}
  \eps_\rmn{CMB} = \frac{\uppi^2}{15}\,\frac{(k_\rmn{B} T_\rmn{CMB})^4}{(\hbar c)^3} (1+z)^4
  \approx 4.17\times10^{-13}(1+z)^4~\rmn{erg~cm}^{-3}
\end{align}
where $k_\rmn{B}$ is Boltzmann's constant, $T_\rmn{CMB}=2.725$~K, $\hbar$ is the
reduced Planck's constant, and $z$ denotes the cosmic redshift. This implies that
CR electrons in low-mass galaxies cool primarily via IC interactions on CMB
photons rather than on FIR photons or via synchrotron emission.

Most importantly, we observe that the magnetic field strength in galaxies
smaller than the Milky Way saturates at values significantly below equipartition
with the thermal pressure. Figure~\ref{fig:energies2} shows that the evolution
of the magnetic energy density saturates approximately in equipartition with the
poloidal kinetic energy density that is a proxy for the `turbulent' energy
density driven by gravitational collapse and non-linearly interacting
Kelvin-Helmholtz body modes in the disc,
$\eps_{\rmn{turb}}=\rho\delta\varv^2/2\approx3/4\times\rho(\varv_z^2+\varv_R^2)$,
where $\rho$ is the gas mass density. Initially, the accretion shock converts
the radial infall velocities into thermal energy and turbulence, which is
visible as a peak in $\eps_{\rmn{turb}}$ at 0.1~Gyr, after which time the
accretion shock leaves the vertical extent of the cylindrical averaging
region. In consequence, the poloidal kinetic energy density decreases because
gas accretion and the associated turbulent driving become weaker. However, the
velocity shear associated with the fast rotating galactic disc in the hot
circumgalactic medium maintains continuous turbulent driving and explains the
approximately constant $\eps_{\rmn{turb}}$, as will be explicitly shown in
Section~\ref{sec:growth_analysis} \citep[see also figure~6
  of][]{2019MNRAS.489.3368B}.

As we will show in Section~\ref{sec:dynamo}, a small-scale dynamo exponentially
amplifies the magnetic field so that it comes into approximate equipartition
with the gravo-turbulent energy density. A fraction of the initial potential
energy of the halo gas feeds the gravo-turbulence and hence drives the
small-scale dynamo either through the corrugated accretion shock or via the
velocity shear between hot circumgalactic medium and cool disc. In this picture,
we would expect that the saturated level of magnetic energy scales as
\begin{align}
  E_B \sim E_{\rmn{turb}} \sim  \frac{\eta\,M_{200}\, \varv_{200}^2}{2} \propto
  M_{200}^{5/3},
\end{align}
where $\varv_{200}=\sqrt{GM_{200}/r_{200}}$ is the virial velocity and $\eta$ is
an energy conversion efficiency. In fact, this halo mass scaling is consistent
with our findings in Fig.~\ref{fig:energies2}, in which we find magnetic energy
densities in our `CR diff' model (at 3 Gyr) of
$\eps_B\sim\{9\times10^{-16},4\times10^{-14},2\times10^{-12}\}~\rmn{erg~cm}^{-3}$
for the $10^{10},~10^{11}$, and $10^{12}~\msun$ haloes, implying that $\eta$ is
independent of halo mass in our simulations. Moreover, we see evidence for an
additional magnetic amplification mechanism over the small-scale dynamo in the
$10^{12}~\msun$ halo after 1.5~Gyr (0.3~Gyr) in the `CR diff' (`CR adv') model,
that is stronger in the `CR~adv' model and consistent with a large-scale dynamo
\citep{Pakmor2016b}.  This is expected because CR (an-)isotropic diffusion is
known to suppress any magnetic dynamo action because active CR transport (via
diffusion or streaming) causes magnetised plasma to move off of the disc via CR
pressure-gradient driven outflows, and carries magnetic flux alongside into the
circumgalactic medium so that this would have to be replenished by the dynamo
and decreases its overall efficiency in the disc \citep{Pakmor2016b}.

Note that our model underestimates supernova-driven turbulence because we only
directly inject CR energy in a way that can drive locally expanding bubbles
while we account for the remaining supernova energy in our effective equation of
state of the ISM. Accounting for this additional local energy injection may be
able to further amplify magnetic fields
\citep{2016MNRAS.457.1722R,2017MNRAS.471.2674R,2017ApJ...843..113B}.  We also
find that our discs are rotationally supported so that the kinetic rotational
energy density $\eps_{\rmn{rot}}=\rho \varv_\varphi^2/2\approx 100
\eps_{\rmn{turb}}$ at late times so that
$\eps_{\rmn{rot}}\approx3\eps_{\rmn{th}}$ in the $10^{10}~\msun$ halo and
$\eps_{\rmn{rot}}\approx50\eps_{\rmn{th}}$ in the $10^{12}~\msun$ halo in our
`CR diff' model. This explains the puffed-up appearance of dwarfs and the
flattened disc-like morphology of our Milky Way-mass galaxies.  In order to
separate the various processes of magnetic field growth, we now analyse the
correlation between gas density and magnetic field and quantify the scaling of
the magnetic growth rate with Reynolds number.

\subsection{Magnetic growth via adiabatic compression and the small-scale dynamo}
\label{sec:growth_analysis}

\begin{figure*}
\centering
\includegraphics[width=\textwidth]{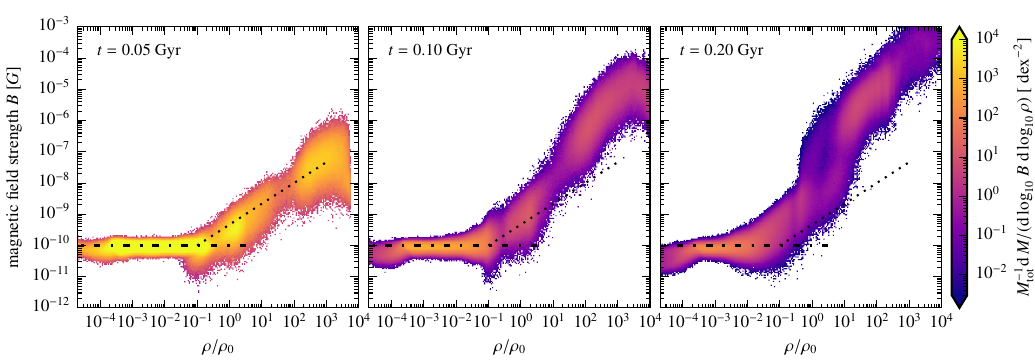}
\caption{Mass-weighted probability density of magnetic field strength, $B$, and
  mass density, $\rho$, for our high-resolution Milky Way-like halo of mass
  $10^{12}~\msun$ and concentration $c_{200}=12$. From left to right, the panels
  show three characteristic phases of magnetic field growth: (i) adiabatic
  growth phase, (ii) exponential growth phase in the kinematic regime (as will
  be shown in Fig.~\ref{fig:B_growth_res}), and (iii) growth of magnetic
  coherence scale (as will be demonstrated in Section~\ref{sec:dynamo}). All gas
  densities are scaled to the star formation threshold $\rho_0 = 4.05\times
  10^{-25}\,\rmn{g~cm}^{-3}$. The initial distributions are shown with a
  dash-dotted line while the adiabatically compressed magnetic field follows
  $B\propto\rho^{2/3}$ for isotropic collapse (dotted line). Note that the
  small-scale dynamo starts to grow the field above the adiabatically compressed
  value at around $t\approx0.07$~Gyr.}
\label{fig:B_rho_evol}
\end{figure*}

\begin{figure*}
\centering
\includegraphics[width=0.495\textwidth]{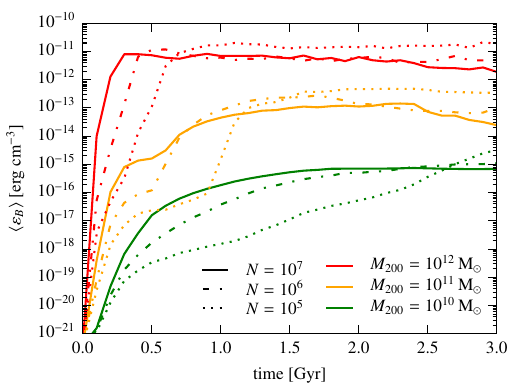}
\includegraphics[width=0.495\textwidth]{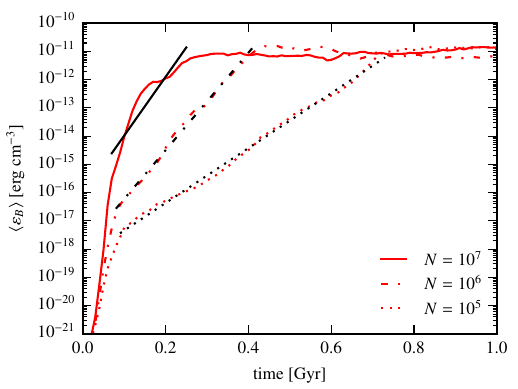}
\caption{Time evolution of the volume-averaged magnetic energy densities in a
  disc of radius 10~kpc and total height 1~kpc. Different halo masses
  ($10^{10},~10^{11}$, and $10^{12}~\msun$, all with concentration $c_{200}=12$)
  are colour coded and different numerical resolution levels are shown with
  different line styles. The panel on the left-hand side shows the overall
  evolution while the panel on the right-hand side shows a zoom on the
  $10^{12}~\msun$ halo at early times and additionally exponential fits to the
  exponential growth phase following the initial adiabatic compression
  phase. This demonstrates that the dynamo starts earlier and the exponential
  growth rate is larger for increasing resolution and halo masses.}
\label{fig:B_growth_res}
\end{figure*}

\begin{figure*}
\centering
\includegraphics[width=\textwidth]{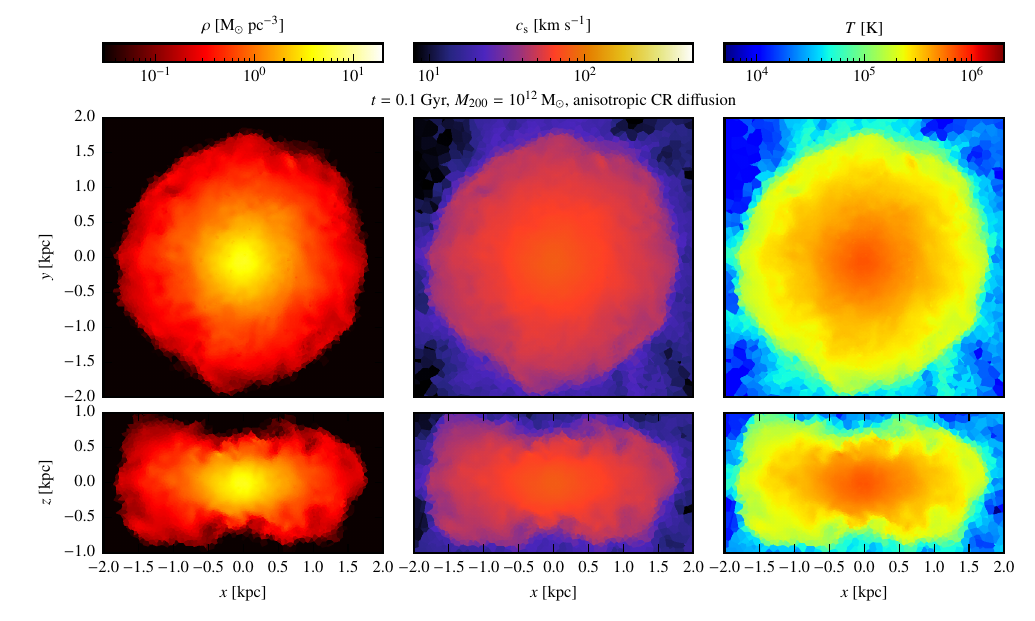}
\includegraphics[width=\textwidth]{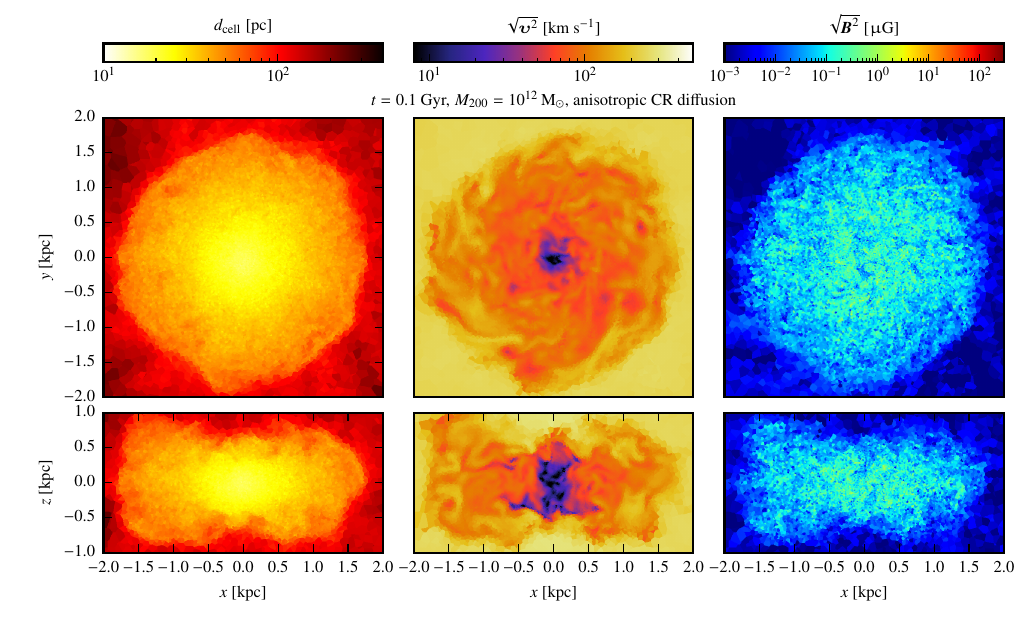}
\caption{Properties of the central region in our Milky~Way-mass galaxy
  ($M_{200}=10^{12}~\rmn{M}_\odot$, $c_{200}=7$) during the exponential growth
  phase in the kinematic regime at $t=0.1$~Gyr.  We show cross-sections in the
  mid-plane of the disc (face-on views) and vertical cut-planes through the
  centre (edges-on views) of the gas mass density, sound speed and temperature
  (top panels, from left to right) and the Voronoi cell diameters
  $d_\rmn{cell}$, gas velocity and magnetic field strength (bottom panels). The
  corrugated accretion shock dissipates kinetic energy from gravitational infall
  into heat, induces turbulence with an outer scale of
  $\mathscr{L}\approx1$~kpc, and drives a small-scale dynamo.}
\label{fig:turbulence}
\end{figure*}

\begin{figure*}
\centering
\includegraphics[width=\textwidth]{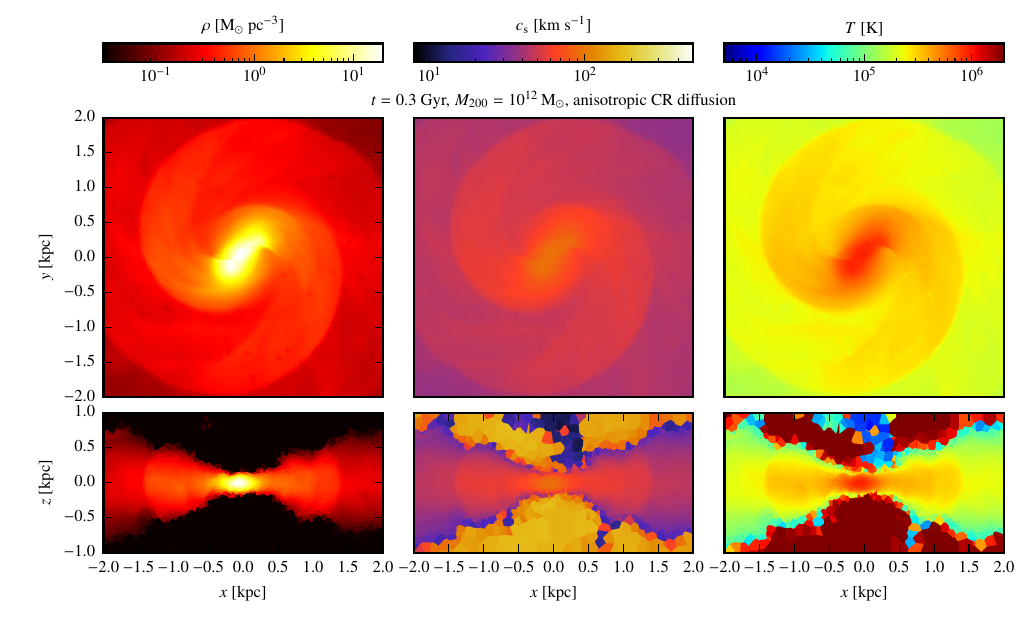}
\includegraphics[width=\textwidth]{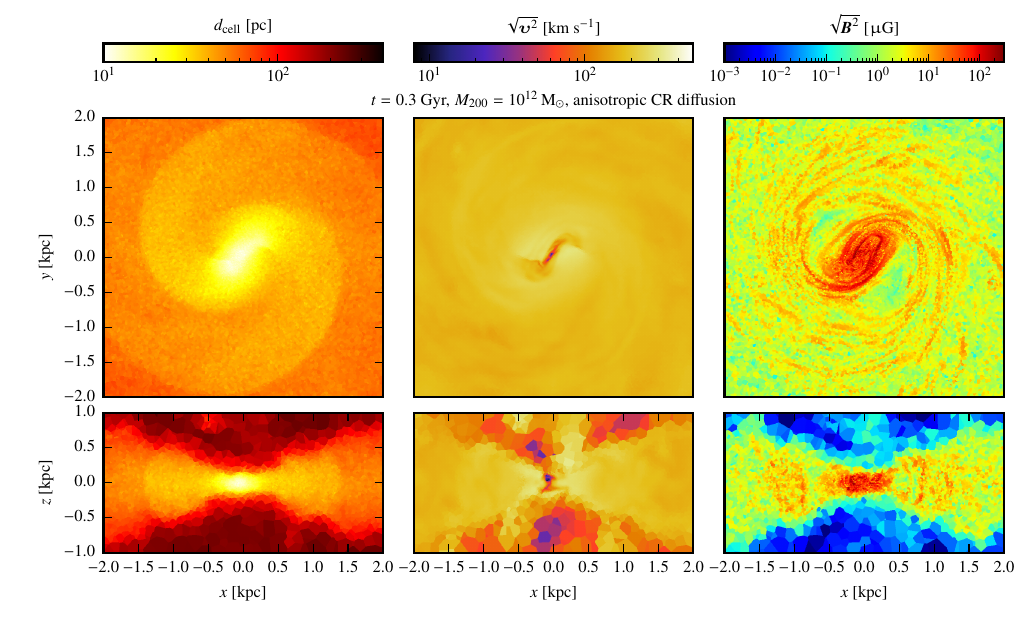}
\caption{Properties of the central region in our Milky~Way-mass galaxy
  ($M_{200}=10^{12}~\rmn{M}_\odot$, $c_{200}=7$) after saturation of the dynamo
  on small scales while the magnetic coherence scale increases (at
  $t=0.3$~Gyr). As in Fig.~\ref{fig:turbulence}, we show gas mass density, sound
  speed and temperature (top panels, from left to right) and the Voronoi cell
  diameters $d_\rmn{cell}$, gas velocity and magnetic field strength (bottom
  panels), all with identical colour scales as in Fig.~\ref{fig:turbulence}.
  There is a supersonic velocity shear between the rotationally supported cooler
  disc and the slower rotating and hotter circumgalactic medium that excites
  Kelvin-Helmholtz body modes (with a characteristic outer scale of turbulence
  of $\mathscr{L}\approx1$~kpc). Non-linear interactions of these modes also
  drive a small-scale dynamo in the disc.}
\label{fig:turbulence2}
\end{figure*}

First, we explore the correlation of magnetic field strength, $B$, and gas
density, $\rho$, during the initial stages of the simulation of our Milky
Way-like halo of mass $10^{12}~\msun$ and concentration $c_{200}=12$. After the
onset of cooling and the associated gravitational collapse, the initial magnetic
field is adiabatically compressed. The isotropic collapse toward the halo centre
causes the initial magnetic field to increase and to scale with the gas density
as $B\propto\rho^{2/3}$ (see left-hand panel of Fig.~\ref{fig:B_rho_evol}). At
around $t=0.07$~Gyr, the central region has settled to the ISM density so that
the gas is not any more adiabatically compressed. Instead, the small-scale
dynamo continues to grow the magnetic field. Because it acts fastest at the
smallest turbulent eddies, which are best resolved at the highest densities in
our quasi-Lagrangian simulation, at $t=0.1$~Gyr the magnetic field is clearly
elevated over the adiabatic scaling at densities $\rho\gtrsim100\rho_0$ (see
middle panel of Fig.~\ref{fig:B_rho_evol}).

Once the small-scale magnetic power saturates (at a large fraction of the
kinetic power), the coherence scale of the magnetic field grows to progressively
larger scales (as will be shown in Section~\ref{sec:dynamo}) and correspondingly
lower gas densities. This can be seen in the right-hand panel of
Fig.~\ref{fig:B_rho_evol} (at $t=0.2$~Gyr), which shows a departure of the
dynamo-grown field over the adiabatically compressed one already at densities
$\rho\gtrsim\rho_0$. Soon thereafter, the field saturates in equipartition with
the turbulent energy (cf.\ Fig.~\ref{fig:energies2}). The shape of the
$B$--$\rho$ correlation at saturation does not depend on our specific choice of
magnetic initial conditions as we explicitly show in Fig.~\ref{fig:B_rho}, where
we compare our constant initial field to a dipole configuration of an
adiabatically pre-compressed field. Figure~\ref{fig:B_rho} shows that deep in
the saturated regime of magnetic field growth (i.e., well after the starburst),
most gas at $\rho\gtrsim10\rho_0$ has been converted to stars, implying a much
reduced probability density in comparison to the starburst phase studied in
Fig.~\ref{fig:B_rho_evol}. Alongside this conversion, our subgrid model assumes
that the flux-frozen magnetic field is also locked up in those stars, thereby
reducing the disc magnetic field.

To study the initial magnetic growth phase, we compare the time evolution of the
magnetic energy density, $\eps_B$, averaged across the disc (of radius 10~kpc
and total height 1~kpc) in haloes of different masses and numerical resolution
(left-hand panel of Fig.~\ref{fig:B_growth_res}), but with identical
concentrations so that the halo profiles are exact gravitational replicas of
each other at different masses. Clearly, increased numerical resolution at each
halo mass enables us to resolve smaller eddies and hence faster magnetic growth
rates. While the saturation level of $\eps_B$ is converged for $N\geq10^6$
initial resolution elements, the simulations with initially $N=10^5$ cells
overproduces $\eps_B$ by a significant amount.

The right-hand panel of Fig.~\ref{fig:B_growth_res} shows a zoom on the
$10^{12}~\msun$ halo at early times. We see an initial exponential growth of the
magnetic field due to adiabatic compression of the seed magnetic field. Provided
the cooling time is shorter than the gravitational free-fall time-scale,
collapse and the associated adiabatic compression happens at the free-fall time,
which is identical for the simulations at different numerical resolutions of a
given halo and thus nearly independent of resolution. Subsequently, we see the
already discussed magnetic growth at high densities: the onset of this second,
exponential growth phase depends on resolution and starts at 50--80~Myr
(corresponding to $N=10^7$ and $N=10^5$, respectively). Most importantly, the
growth rate of this small-scale dynamo strongly depends on resolution and also
follows an exponential growth:
\begin{align}
  \label{eq:growth}
  \eps_B(t)=\eps_{B,0}\,\rmn{e}^{\Gamma_B t}.
\end{align}
For our simulations with an initial number of Voronoi cells of
$N=\{10^5,\,10^6,\,10^7\}$, we measure the exponential growth rate $\Gamma_B =
\{22.4,\,39.8,\,48.0\}~\rmn{Gyr}^{-1}$, which corresponds to a growth time of
$\tau_B = \Gamma_B^{-1} =\{44.6,\,25.1,\,20.8\}~\rmn{Myr}$. This basic picture
of resolution dependent growth remains robust to changes of the averaging
region, from a cylindrical disc to a sphere of different radii ($r_{\rmn{s}}=
\{1, 0.5, 0.25\}$~kpc) while the resulting growth rates are subject to larger
Poisson fluctuations as a result of the lower number of cells for the smaller
integration volumes.

The magnetic growth curve for the $10^{12}~\msun$ halo of the high-resolution
simulation ($N=10^7$) shows a distinct feature at 0.2~Gyr, which corresponds to
the time-scale of the formation of the rotationally supported cool disc. This
comes about because at this time, the outward propagating accretion shock is
unable to maintain a powerful turbulent driving in the centre, which slows down
the growth rate of the small-scale dynamo. However, the formation of the cooling
disc generates a large velocity shear between the fast rotating disc and the
slower rotating halo gas, which continuously injects fresh and powerful
turbulence that is able to re-ignite the small-scale dynamo (as we will show
below). The later onset of the small-scale dynamo in the lower resolution
simulations (with $N\leq10^6$) causes a blending of the two effects, which
therefore cannot any more be separated in the time evolution of the average
magnetic energy.

This interpretation is supported by Fig.~\ref{fig:turbulence}, which shows the
properties of the proto-galaxy during the exponential growth phase. The
corrugated accretion shock dissipates kinetic energy from gravitational infall
and heats the assembling galaxy to temperatures in excess of $2\times10^5$~K
corresponding to sound speeds of $50~\rmn{km~s}^{-1}$. Most importantly, the
velocity panels of Fig.~\ref{fig:turbulence} show that the curved shock
decelerates the supersonic infall and injects vorticity.  Interacting eddies
generate subsonic turbulence, which cascades kinetic energy down in scale to the
mesh size $d_{\rmn{cell}}$, which is rather homogeneous behind the accretion
shock and assumes the smallest values at the densest centre. As we will further
argue below, this kinetic turbulence drives the small-scale dynamo and grows the
magnetic field as can be seen in the bottom right-hand panels of
Fig.~\ref{fig:turbulence}. With increasing distance from the shock, turbulence
decays as can be inferred from the velocity maps in the galactic centre.

Once the accretion shock propagates outwards and more gas accretes in the
equatorial plane, it cools and forms a cooling and rotationally supported disk
several tens of Myrs later (Fig.~\ref{fig:turbulence2}). This implies a large
velocity shear between the fast rotating, radiatively cooling disc and the
slower rotating and hotter circumgalactic medium that has been thermalised by
the outwards propagating accretion shock. Locally, this situation can be
identified with a cold stream moving supersonically through a hot, dilute
circumgalactic medium, which is known to excite the Kelvin-Helmholtz instability
in the hydrodynamic case
\citep{2016MNRAS.463.3921M,2020MNRAS.494.2641M,2018MNRAS.477.3293P} as well as
in the MHD case \citep{2019MNRAS.485..908B,2019MNRAS.489.3368B}.  In the
supersonic regime, the Kelvin-Helmholtz instability not only manifests itself by
exciting the well-known {\em surface modes} but also by exciting {\em reflective
  or body modes} as (magneto-)acoustic waves reflect at the interface of the
dense stream to the dilute background, thus trapping the acoustic wave energy
within the stream. As a result, the waves grow in amplitude inside the stream
(or in our case the rotationally supported cool disc). This excites a broad
spectrum of unstable wave modes that grow into the non-linear regime and
interact with each other to inject subsonic turbulence, which further amplifies
the magnetic field through a second small-scale dynamo mode (bottom panels of
Fig.~\ref{fig:turbulence2}). Note that gravitational collapse of dense gas
clouds, star formation, energetic feedback and the centrifugal force in the
rotating frame modify the late-time behaviour of the non-linearly saturating
dynamo in comparison to the idealised simulations of supersonically moving cold
streams.

Using typical shear velocities of
$\mathscr{V}\sim\mbox{(50$-$100)}~\rmn{km~s}^{-1}$ across a scale of
$\mathscr{L}\sim1~\rmn{kpc}$ (bottom panels of Fig.~\ref{fig:turbulence2}),
where the lower (upper) value characterises an average (maximum) shear, we can
estimate the turbulent energy dissipation rate at the Kolmogorov scale,
$\varv_\ell^3/\ell\sim\mathscr{V}^3/\mathscr{L}
\sim\mbox{(0.7$-$5)}\times10^{-25}~\rmn{erg~s}^{-1}\,m_\rmn{p}^{-1}$. To order
of magnitude, this energy rate is generated on the vertical sound crossing time,
characterised by the thermal velocity $\varv_\rmn{th}\sim50~\rmn{km~s}^{-1}$ (in
our ISM subgrid model, see Fig.~\ref{fig:turbulence2}) and implying a dissipated
energy density $\eps_\rmn{diss}\sim\rho\mathscr{V}^3/\varv_\rmn{th}
\sim(2\times10^{-9}-2\times10^{-7})~\rmn{erg~cm}^{-3}$. This corresponds to
equivalent magnetic field strengths of
$B_\rmn{diss}\sim\mbox{(200$-$2000)}~\upmu$G available for tapping in by the
small-scale dynamo. Here, we use characteristic density values of
$80~\rmn{cm}^{-3}$ ($800~\rmn{cm}^{-3}$) for the outer (inner) disc in our
simulations, respectively, which amounts to mass densities of
$1~\rmn{M}_\odot~\rmn{pc}^{-3}$ ($10~\rmn{M}_\odot~\rmn{pc}^{-3}$), see
Fig.~\ref{fig:turbulence2}. These equivalent magnetic field strengths are about
a factor of 20 larger than the realised magnetic field strengths
$B\sim\mbox{(10$-$100)}~\upmu$G in our simulations (Fig.~\ref{fig:turbulence2})
and substantially larger than the volume-averaged magnetic energy density
$\eps_B\sim10^{-11}~\rmn{erg~cm}^{-3}$ in the disc (Fig.~\ref{fig:energies2}).

Our simulations solve the equations of ideal MHD, i.e., we do not explicitly
model physical viscosity and resistivity. Moreover, our simulations can only
resolve dynamo growth on scales larger than our numerical Voronoi grid, which is
substantially coarser in comparison to the astrophysical magnetic resistive
scale and therefore, our simulated magnetic fields grow much slower in
comparison to the astrophysical case.  Thus, numerical viscosity and resistivity
determine the growth rate of the small-scale dynamo. Employing the scaling
properties of Kolmogorov turbulence (see Appendix~\ref{sec:app_Gamma_scaling}),
we expect a growth rate in the kinematic regime of the small-scale dynamo (which
is equal to the eddy turnover rate at the dissipation scale for a magnetic
Prandtl number of unity) of
\begin{equation}
  \label{eq:Gamma1}
  \Gamma=\frac{\mathscr{V}}{\mathscr{L}}\,\rmn{Re}^{1/2},
\end{equation}
where $\mathscr{L}$ and $\mathscr{V}$ are the length and velocity scale at the
turbulent injection scales and Re is the physical Reynolds number, which is
given by
\begin{align}
  \label{eq:Reynolds}
  \rmn{Re} = \frac{\mathscr{L} \mathscr{V}}{\nu_\rmn{vis}}
  \sim \frac{3\mathscr{L} \mathscr{V}}{\lambda_{\rmn{mfp}} \varv_{\rmn{th}}},
\end{align}
where $\nu_\rmn{vis}\sim\lambda_{\rmn{mfp}}\varv_{\rmn{th}}/3$ is the kinetic
viscosity, $\lambda_{\rmn{mfp}}$ is the particle mean free path, and
$\varv_{\rmn{th}}$ is the thermal velocity. Thermal particles moving a mean free
path collide and randomise their velocities, which implies that
$\lambda_{\rmn{mfp}}$ is the typical length over which the fluid can communicate
changes in its shear stress. A fluid with a longer mean free path therefore more
easily opposes changes to its local shear velocity, i.e., is more viscous.  By
analogy with this property, we define the numerical Reynolds number via
\begin{align}
  \label{eq:Re_num}
  \rmn{Re}_\rmn{num} \equiv \frac{\mathscr{L} \mathscr{V}}{\nu_{\rmn{num}}}
  \sim \frac{3\mathscr{L} \mathscr{V}}{d_{\rmn{cell}} \varv_{\rmn{sig}}},
\end{align}
where $\nu_{\rmn{num}}$ is the numerical viscosity and $d_\rmn{cell}=
(6V/\uppi)^{1/3}$ is the diameter of the smallest characteristic Voronoi cell
(assuming a spherical cell volume $V$), i.e., it is abundantly present in a
singly connected region so that it governs therein the numerical dissipation
properties. The signal speed of the gas relative to the numerical mesh,
$\varv_{\rmn{sig}}$, differs among the various computational techniques. For a
spatially fixed mesh (homogeneous or adaptive Eulerian), the signal velocity is
the sum of bulk velocity relative to the mesh and thermal velocity,
$\varv_{\rmn{sig}}=\mathscr{V}+\varv_{\rmn{th}}$, so that the numerical Reynolds
number for Eulerian techniques is given by
\begin{align}
  \label{eq:Re_num_Eulerian}
  \rmn{Re}_\rmn{num,\,Euler} 
  \sim \frac{3\mathscr{L}}{d_{\rmn{cell}}}\,\frac{\mathscr{V}}{\mathscr{V}+\varv_{\rmn{th}}}
  \sim \frac{3\mathscr{L}}{2d_{\rmn{cell}}},
\end{align}
where we adopted the transsonic case $\mathscr{V}\approx\varv_{\rmn{th}}$ in the
last step, which is relevant for a small-scale dynamo in galactic discs that are
excited through Kelvin-Helmholtz surface and body modes evolving into the
non-linear regime \citep{2019MNRAS.485..908B,2019MNRAS.489.3368B}. However, for
a quasi-Lagrangian code where the mesh moves close to the speed of the gas, we
have $\varv_{\rmn{sig}}\approx \varv_{\rmn{th}}$ so that we obtain
\begin{align}
  \label{eq:Re_num_Lagrange}
  \rmn{Re}_\rmn{num,\,Lagrange} 
  \sim \frac{3\mathscr{L}}{d_{\rmn{cell}}}\,\frac{\mathscr{V}}{\varv_{\rmn{th}}}
  \sim \frac{3\mathscr{L}}{d_{\rmn{cell}}},
\end{align}
where we also adopted the transsonic case $\mathscr{V}\approx\varv_{\rmn{th}}$
in the last step and which is larger by a factor of two in comparison to the
Eulerian case. To avoid mesh twisting and large gradients in mesh resolution,
\textsc{Arepo} moves the generating points relative to the gas with a velocity
that is typically small in comparison to $\varv_{\rmn{th}}$ for a galaxy
simulation that does not resolve the cold molecular phase.

To estimate $\rmn{Re}_\rmn{num}$, we adopt a cell diameter
$d_{\rmn{cell}}\approx 10$~pc appropriate for the fastest growing magnetic field
in the centre (Fig.~\ref{fig:d_cell}) and a thermal velocity of our star-forming
subgrid ISM at $2\times 10^5$~K of
$\varv_{\rmn{th}}\approx50~\rmn{km~s}^{-1}$. We approximate the outer scale of
turbulence by $\mathscr{L}\approx1$~kpc, which is the characteristic length
scale of the initial accretion shock (Fig.~\ref{fig:turbulence}) and also
corresponds to the scale of the thick disc (Fig.~\ref{fig:turbulence2}). To
estimate the injection velocity at the outer scale, we note that in both
scenarios discussed (an exponentially growing small-scale dynamo in the
post-shock regime of the initial accretion shock as well as the non-linearly
growing Kelvin-Helmholtz instability), we expect
$\mathscr{V}\approx\varv_{\rmn{th}}$. As a result, we find a typical value of
$\rmn{Re}_\rmn{num}\sim3\times10^2$. Thus, we obtain turbulent velocities of
$\varv_\ell=\mathscr{V}\,\rmn{Re}_\rmn{num}^{-1/4}\approx12~\rmn{km~s}^{-1}$
(Eq.~\ref{eq:v_scaling}) at the resolution scale $\ell$ in the galaxy
centre. The eddy turnover rate at this (numerical) Kolmogorov scale is:
\begin{equation}
  \label{eq:Gamma2}
  \Gamma \sim \frac{\varv_\ell}{2 \uppi\, r_\rmn{eddy}}
  \sim 65 ~\rmn{Gyr}^{-1} \approx 1.35\, \Gamma_B,
\end{equation}
where we assume that the smallest turbulent eddies are starting to be resolved
for a radius that extends across three cells, $r_\rmn{eddy}=3 d_{\rmn{cell}}$
and $\Gamma_B = 48~\rmn{Gyr}^{-1}$ is the measured exponential growth rate of
the magnetic field of our high-resolution simulation (Eq.~\ref{eq:growth}). This
theoretically expected growth rate is 35 per cent larger than the measured value
because of numerical dissipation and because stars represent sinks of magnetic
energy in our model, thereby reducing the dynamo efficiency in comparison to the
theoretical maximum.\footnote{The magnetic growth rate of the initial,
accretion-shock driven dynamo is somewhat larger than the average growth rate
until saturation, $\Gamma_B$ (see Fig.~\ref{fig:B_growth_res}). This is related
to the somewhat larger turbulent velocities during this phase
(Fig.~\ref{fig:turbulence}), which reflect the cooling, post-shock state rather
than the cooled, equilibrium ISM state and hence, allows for larger turbulent
eddy velocities and dynamo growth rates.}  Clearly, astrophysical Reynolds
numbers for an accretion-driven dynamo in proto-galaxies are of order
$\rmn{Re}\sim10^{11}$, implying growth times that are a factor $2\times10^4$
shorter \citep{2013A&A...560A..87S}.

If we re-normalise the numerical growth rates to the simulation with the lowest
resolution, $\Gamma_{B,5} = 22.4~\rmn{Gyr}^{-1}$ (where $\Gamma_{B,5}$ denotes
$\Gamma_{B}$ of our $N=10^5$ simulation), we would expect to obtain growth
rates of $\Gamma_{B,6} = \Gamma_{B,5} 10^{1/6} = 32.9~\rmn{Gyr}^{-1}$ (a factor
0.83 smaller than the measured growth rate) and $\Gamma_{B,7} = \Gamma_{B,5}
10^{1/3} = 48.2~\rmn{Gyr}^{-1}$ (consistent with the measured growth rate). This
excellent agreement of simulations and small-scale dynamo theory is reassuring
that our MHD solver produces reliable results and provides support of the
picture that the small-scale dynamo grows the magnetic field after an initial
phase of adiabatic compression. To understand the scale dependence of the
magnetic dynamo, we now turn to a power spectrum analysis.

\subsection{Power spectrum analysis of the small-scale dynamo}
\label{sec:dynamo}

\begin{figure*}
\centering
\includegraphics[width=\textwidth]{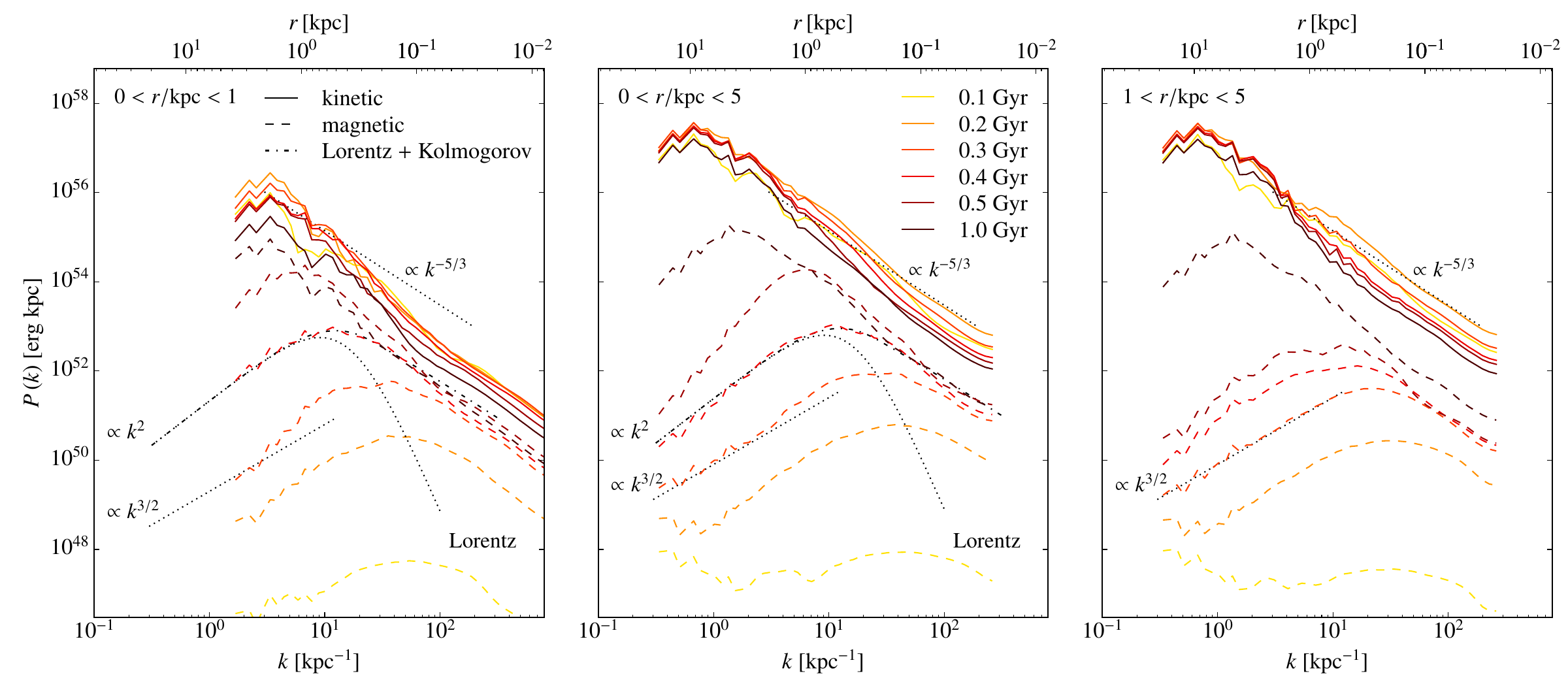}
\caption{Comparison of kinetic (solid) and magnetic (dashed) power spectra for
  different analysis regions (indicated in the top left) for the `CR~diff'
  simulation of the $10^{12}~\msun$ halo with $c_{200}=7$ and
  $B_{\rmn{init}}=10^{-10}$G.  The kinetic energy power spectrum decays with
  time because the thermal and CR pressure that surround the forming disc slows
  gravitational infall onto the disc and the Kelvin-Helmholtz driven turbulence
  weakens with time. On the contrary, the magnetic power spectrum grows
  exponentially until it saturates on small scales while it grows further on
  larger scales, implying a shift of the coherence scale to larger scales.  The
  black dotted lines show the slopes of a \citet{Kazantsev1968} spectrum
  ($\propto k^{3/2}$) in the kinematic regime at large scales and a
  \citet{Kolmogorov1941} spectrum ($\propto k^{-5/3}$), which is theoretically
  expected at non-linear scales in a small-scale dynamo and realised in our
  simulation if we omit all gas cells in the central 1~kpc sphere (right-hand
  panel).  When we include the central region (left-hand and middle panels), the
  magnetic power spectrum follows a Lorentzian function on large scales, which
  is the Fourier transform of the exponential magnetic profile with scale length
  $r_0=2\uppi/k_0=0.4$~kpc. The power of this profile is then cascaded to
  smaller scales with a rate given by MHD turbulence theory (dash-dotted, see
  Eq.~\ref{eq:power_spec2}). }
\label{fig:ps}
\end{figure*}

Initially weak seed magnetic fields can be amplified due to stretching, twisting
and folding of field lines driven by turbulent eddies
\citep{1983flma....3.....Z,1995stf..book.....C}, where the exponential
amplification time is given by the turbulent eddy turnover time. For weak seed
magnetic fields, the kinematic limit applies and we expect a scaling of the
magnetic energy spectrum $P(k)\dd k \propto k^{3/2}\dd k$ according to the
dynamo theory by \citet{Kazantsev1968}. Provided the initial magnetic energy is
smaller than the kinetic energy of the smallest turbulent eddies and neglecting
compressible effects, we can apply the Kazantsev theory to the entire inertial
range of turbulence between the injection scale and viscous cut-off scale.

In order to study whether our galaxy simulations grow the magnetic field through
this mechanism, we plot the evolution of the magnetic and kinetic power spectra
in Fig.~\ref{fig:ps}. Following \citet{2017MNRAS.469.3185P,2020MNRAS.498.3125P},
we compute these power spectra by taking the absolute square of the Fourier
components of $\sqrt{\rho}\,\bs{\varv}$ and $\bs{B}/\sqrt{8\uppi}$,
respectively, for gas within a sphere of radius $r_\rmn{s} = 1$ and 5 kpc
(indicated in the legends of Fig.~\ref{fig:ps}). This is done within a
zero-padded box of size $\pm 2r_\rmn{s}$ across so that the fundamental mode has
a wavelength of 4 and 20 kpc, respectively. Accordingly, drops in the power
spectra on scales greater than 2 and 10 kpc are an artefact of this
zero-padding.  By considering only gas within 1 and 5 kpc of the galactic
centre, we isolate the regions in which the greatest amplification takes place.

After the kinematic magnetic power spectrum has been exponentially amplified to
the point where the magnetic and turbulent energy of the smallest eddies have
achieved equipartition, the back-reaction through the magnetic tension force is
strong enough to suppress the stretching process of these eddies at the
equipartition scale. On larger scales, the magnetic power spectrum continues to
be amplified to the equipartition scale of the corresponding modes with a slower
growth rate. Because in Kolmogorov turbulence, there is more turbulent kinetic
energy available on larger scales, the magnetic energy grows to a larger
amplitude. Hence, the magnetic coherence scale grows as a function of time,
which appears like inverse cascade but in fact, it does not represent an inverse
cascade in its strict definition because there are no integrals of motion in a
fluctuating dynamo, which would necessarily create an inverse cascade.

On scales smaller than the current equipartition scale, we are entering the
non-linear stage of the small-scale dynamo, that is characterised by an
equipartition of magnetic and turbulent kinetic energy so that the initially
hydrodynamic turbulence is modified to become MHD turbulence
\citep{1995ApJ...438..763G}. Until the kinetic injection scale has reached the
equipartition condition, we observe in Fig.~\ref{fig:ps} a co-existence of the
kinematic dynamo spectrum on large scales and a spectrum $P(k)\dd k \propto
k^{-5/3}\dd k$ on scales smaller than the equipartition scale
\citep{2005PhR...417....1B}.

In our case, the dark matter gravitational potential causes a density
stratification on which the kinetic and magnetic turbulence is imprinted. The
turbulence is driven by rotational gravitational infall, cooling and star
formation. The power spectrum probes thus the combination of the large-scale
density and magnetic profiles and turbulent fluctuations on smaller scales. The
three-dimensional distribution of the root-mean square magnetic field strength
obeys a steep exponential profile in the centre at early times, $B(r)=B_0
\e^{-k_0 r}$, where $r=\sqrt{\x^2}$ is the spherical radius and $k_0=2\uppi/r_0$
is the wavenumber corresponding to the scale length $r_0$ of the exponential
magnetic profile (see Fig.~\ref{fig:B_3D}). The Fourier transformation of the
exponential `form factor' is given by the Lorentzian profile in wave number,
$k=|\kvec|$,
\begin{align}
\label{eq:power_spec1}
\mathcal{F}\left(\e^{-k_0 r}\right)(k)
&=\int_{-\infty}^\infty\rmn{d}^3x \,\e^{-k_0 \sqrt{x^2+y^2+z^2}}\e^{-\rmn{i} \kvec\bcdot\x}\nonumber\\
&=4\uppi\int_0^\infty\rmn{d}r r^2 \e^{-k_0 r} \frac{\sin(kr)}{kr}
=\frac{8\uppi k_0}{(k^2+k_0^2)_{\phantom{0}}^2}.
\end{align}

\begin{figure*}
\centering
\includegraphics[width=\textwidth]{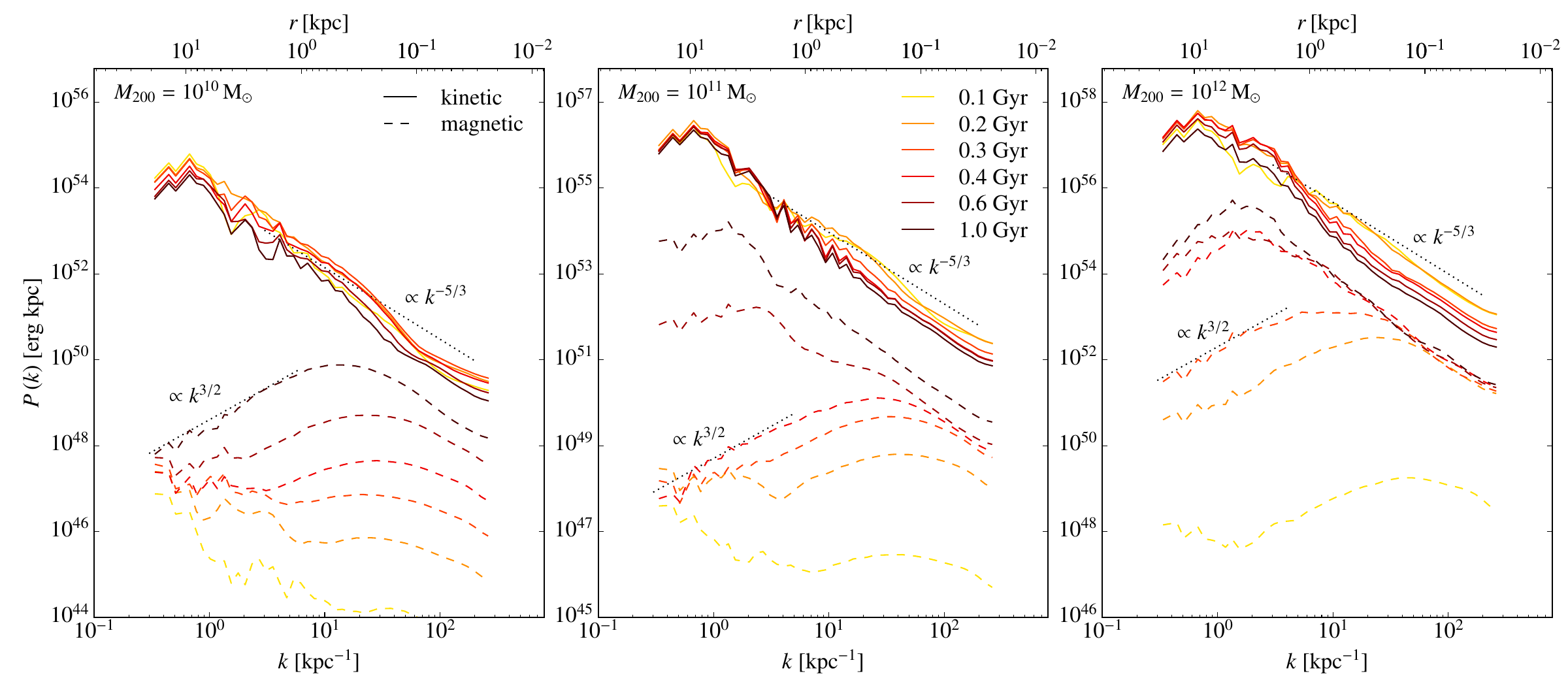}
\caption{Comparison of kinetic and magnetic power spectra for different halo
  masses ($10^{10},~10^{11}$, and $10^{12}~\msun$, all with a concentration
  parameter $c_{200}=12$, $B_{\rmn{init}}=10^{-10}$G, and the model with
  advective and anisotropic diffusive CR transport). We compute the power
  spectra in a spherical shell bounded by $1<r/\rmn{kpc}<5$. The larger
  concentration parameter in comparison to $c_{200}=7$ used in the analysis of
  Fig.~\ref{fig:ps} causes a larger adiabatic compression of $B_{\rmn{init}}$
  and turbulent power upon gravitational collapse to the central region and
  hence a faster growth of the magnetic power spectra at early times. The
  magnetic power spectra match the expectation for a subsonic small-scale dynamo
  and show the slopes of a \citet{Kazantsev1968} spectrum ($\propto k^{3/2}$) in
  the kinematic regime at large scales and a \citet{Kolmogorov1941} spectrum
  ($\propto k^{-5/3}$, black dotted lines) at small, non-linear scales. The
  lower amplitude of the turbulent kinetic power spectrum results in a slower
  magnetic growth rate so that the increase of the magnetic coherence scale to
  larger scales is stalled in smaller galaxies.}
\label{fig:ps_halos}
\end{figure*}

The power spectrum is the mean absolute square of the Fourier transform,
multiplied by $k^2$ (to account for the volume element in Fourier space), which
defines the asymptotic slope on large scales. On these scales, the Lorentzian
rises more steeply than the \citet{Kazantsev1968} power spectrum, $P(k) \propto
k^{3/2}$, and should thus dominate the power on large scales while it drops
towards small scales as $P(k) \propto k^{-6}$. Due to non-linear interactions of
wave modes, however, this power should cascade down in wave number space at a
rate that is given by the theory of MHD turbulence, $P(k) \propto k^{-5/3}$.  We
supplement the Lorentzian by the \citet{Kolmogorov1941} spectrum and conjecture
the following functional form of the power spectrum for intermediate times after
small scales have reached the stage of non-linear evolution and until the
magnetic and kinetic power spectra have reached equipartition at the kinetic
injection scale,
\begin{align}
\label{eq:power_spec2}
P(k) = P_0\,\left\{\frac{k^2k_0^6}{(k^2 + k_0^2)_{\phantom{0}}^4}
+ \frac{1}{3}\,\tanh\left[\left(\frac{k}{k_0}\right)^2\right]
    \frac{1}{1 + (k/k_0)^{5/3}}\right\},
\end{align}
where $P_0$ is a normalisation. We adopt a $\rmn{tanh}[(k/k_0)^2]$ profile to
cut off the turbulent power on large scales ($k<k_0$). The pre-factor $1/3$
ensures that both terms in Eq.~\eqref{eq:power_spec2} contribute equally to the
roll-over wave number $k_0$. The $k^2$ scaling in the argument of the hyperbolic
tangent matches the $k^2$ slope of the scaled Lorentzian on large scales, thus
ensuring a smooth transition from the Lorentzian form factor to the turbulent
spectrum on small scales ($k>k_0$). The left-hand and central panels of
Fig.~\ref{fig:ps} show a comparison of kinetic and magnetic power spectra for a
sphere of radius 1 and 5~kpc, respectively. The model of
Eq.~\eqref{eq:power_spec2} provides an excellent fit to the magnetic power
spectrum at times $\lesssim0.5$~Gyr (with $r_0=0.4$~kpc at 0.4~Gyr, see
Fig.~\ref{fig:B_3D} and Table~\ref{tab:fit2}) while the power spectrum is
clearly inconsistent with the \citet{Kazantsev1968} model. However, if we
exclude a sphere of radius 1~kpc that hosts the central steep magnetic profile,
the resulting magnetic power spectrum provides an excellent fit to the combined
model with a \citet{Kazantsev1968} spectrum in the kinematic regime at large
scales and a \citet{Kolmogorov1941} spectrum at small, non-linear scales
(right-hand panel of Fig.~\ref{fig:ps}).

In Fig.~\ref{fig:ps_halos}, we thus adopt this choice of a spherical shell
bounded by $1<r/\rmn{kpc}<5$ and calculate kinetic and magnetic power spectra
for different halo masses ($10^{10},~10^{11}$, and $10^{12}~\msun$ for our model
with advective and anisotropic diffusive CR transport). The simulation of the
$10^{12}~\msun$ halo is identical to that analysed in Fig.~\ref{fig:ps} except
for the concentration parameter, which we increase from $c_{200}=7$ to 12 in
order to study its effect on the magnetic power spectrum.  The larger
concentration parameter implies a deeper gravitational potential, which causes a
larger adiabatic compression of $B_{\rmn{init}}$ and an increased level of
turbulent power driven by gravitational collapse toward the central region. As a
result, we observe faster growth of the magnetic power spectrum at early times
for the larger halo concentration $c_{200}= 12$ (cf.~right-hand panels of
Fig.~\ref{fig:ps} and Fig.~\ref{fig:ps_halos}).

In agreement with our findings in Fig.~\ref{fig:energies2}, the saturated
small-scale dynamo state is reached later in smaller galaxies, which are
characterised by an equipartition between magnetic and turbulent
energy. Similarly to the magnetic power spectrum, the kinetic power spectrum on
large scales is dominated by disc rotation because $\eps_{\rmn{rot}}\approx 100
\eps_{\rmn{turb}}$ (Fig.~\ref{fig:energies2}) so that only scales
$\lambda\lesssim1$~kpc probe kinetic turbulence.

Provided galaxies gain mass at a rate comparable to the rate at which they form
stars, gravitational infall seeds turbulence as demonstrated through a
combination of numerical simulations and analytical arguments
\citep{2010A&A...520A..17K}. This process is particularly relevant in the
extended outer discs beyond the star-forming radius of large (Milky Way-like)
galaxies and during the epoch of galaxy assembly, where cold flow accretion
likely seeds turbulence \citep{2008ApJ...687...59G}. Our model of a collapsing
sphere of gas in an NFW potential is a toy model that is simple enough so that
we can isolate individual effects and study the impact of gas accretion on
magnetic field growth and how the radio emission evolves from starburst systems
to quiescently star-forming disc galaxies. Previously, a similar accretion
driven small-scale dynamo has been identified in isolated disc galaxies
\citep{2019MNRAS.483.1008S} and in cosmological zoom-in simulations
\citep{2017MNRAS.469.3185P,2017MNRAS.472.4368R}.

We note that our models only exhibit one phase of gas accretion and star
formation and we adopt a pressurised ISM without fully modelling
supernova-driven turbulence. In reality, cosmologically growing dwarf galaxies
may have several star-forming phases, which could perhaps further amplify
sub-equipartition magnetic fields so that they may reach a larger fractional
saturation. Moreover, our effective ISM description that only accounts for
direct CR energy injection underestimates supernovae-driven turbulence. Energy
injection as a result of supernovae and radiation feedback can also drive strong
gas turbulence and result in a small-scale dynamo as identified in isolated disc
galaxies \citep{2016MNRAS.457.1722R,2017MNRAS.471.2674R,2017ApJ...843..113B}.
Future galaxy simulations with radiation and CR hydrodynamics that explicitly
model the energy and momentum injection by supernovae and account for radiative
transfer of stellar UV emission will enable us to separate the contribution of
gravitationally driven gas accretion and star formation feedback to the
small-scale dynamo.

\subsection{Magnetic curvature: insights into the small-scale dynamo}
\label{sec:curvature}

\begin{figure*}
\centering
\includegraphics[width=\textwidth]{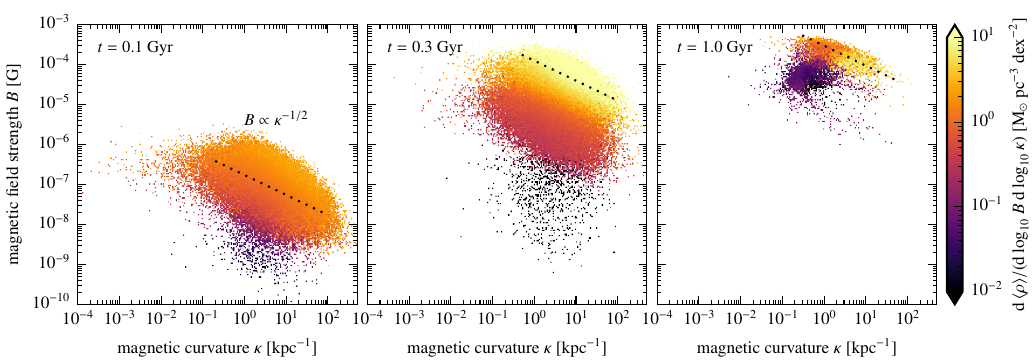}
\caption{Emergence of the magnetic field and curvature in the galaxy centre
  ($r<1$~kpc). We show the mean density in the plane spanned by magnetic field
  strength $B$ and curvature $\kappa$ in our Milky Way-like halo (with mass
  $10^{12}~\msun$ and concentration $c_{200}=7$). From left to right, the panels
  show the three characteristic phases of the magnetic dynamo: (i) exponential
  growth phase in the kinematic regime, (ii) growth of the magnetic coherence
  scale, and (iii) saturation phase of the magnetic dynamo. Starting with the
  kinematic phase through the growth phase of the magnetic coherence scale,
  there is a clear anticorrelation of $B^2$ and $\kappa$ visible at the largest
  density, which tightens up at the saturation phase, in particular at high
  curvature and gas density (which has a smaller dynamical timescale). We
  observe the expected correlation $B\propto \kappa^{-1/2}$
  (Eq.~\ref{eq:curvature}) at various densities, which signals a superposition
  of different dynamo processes on different length and time scales
  characterised by different densities.}
\label{fig:B-K_evolution}
\end{figure*}

\begin{figure*}
\centering
\includegraphics[width=\textwidth]{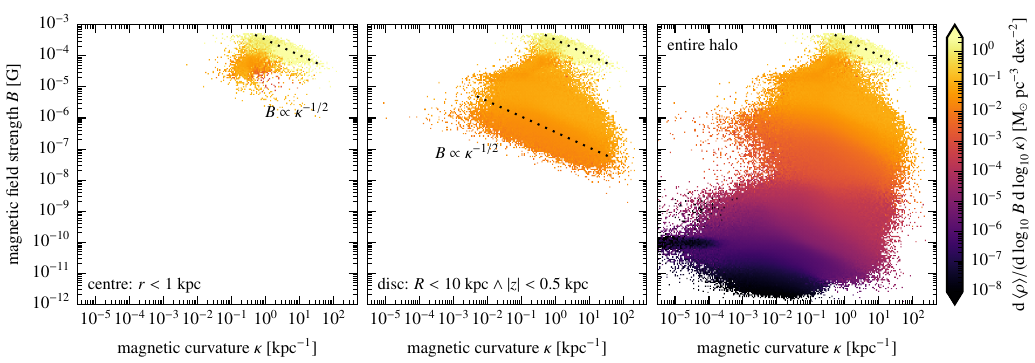}
\includegraphics[width=\textwidth]{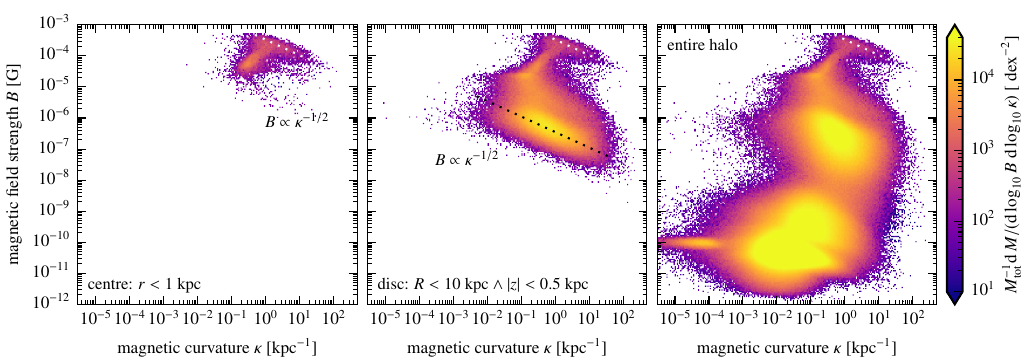}
\caption{Separating different dynamo processes by spatial cuts for the saturated
  dynamo regime at 1~Gyr in our Milky Way-like halo (of mass $10^{12}~\msun$ and
  concentration $c_{200}=7$). The top panels show the mean density in the plane
  spanned by magnetic field strength $B$ and curvature $\kappa$ while the bottom
  panels show the mass-weighted probability density.  From left to right, we
  show the central sphere of radius 1 kpc (left), the disc region (middle) and
  the entire halo (right). This clearly demonstrates that while there is a
  complex superposition of different dynamo processes in the entire halo, these
  can be deconstructed by applying spatial cuts and restricting to a narrow
  range in mean density, which mirrors the behaviour of a small-scale dynamo in
  incompressible, homogeneous turbulent boxes \citep{2004ApJ...612..276S}.}
\label{fig:B-K_regions}
\end{figure*}

While the power spectrum is well suited for analysing the physical scales of
magnetic field growth, it is not optimised for exploring the associated forces
and to fully characterise the small-scale dynamo during galaxy formation. Thus,
to complement our power-spectrum analysis, we turn to the Lorentz force density,
which reads in terms of $\vec{B}$ in the MHD approximation:
\begin{align}
  \vec{f}_\rmn{L} = \frac{1}{c}\,\vec{j}\btimes\vec{B}
  = \frac{1}{4\uppi}\left(\bnabla\btimes\vec{B}\right)\btimes\vec{B}
  = \frac{1}{4\uppi}\left(\vec{B}\bcdot\bnabla\right)\vec{B}-\frac{1}{8\uppi}\bnabla B^2 ,
  \label{eq:Lorentz1}
\end{align}
where the two terms on the right-hand side are often (erroneously) attributed to
the magnetic curvature and pressure forces, respectively. In order to fully
separate the effects of magnetic curvature and pressure -- which are mixed up in
this representation -- we write $\vec{B}=B\vec{b}$, where $\vec{b}$ is the unit
vector in the direction of $\vec{B}$ and obtain \citep{2013arXiv1301.5572S}
\begin{align}
  \vec{f}_\rmn{L} 
  &= \frac{B^2}{4\uppi}\left(\vec{b}\bcdot\bnabla\right)\vec{b}
  +\frac{1}{8\uppi} \vec{b}(\vec{b}\bcdot\bnabla)B^2-\frac{1}{8\uppi}\bnabla B^2\nonumber\\ 
  &= \frac{B^2}{4\uppi}\left(\vec{b}\bcdot\bnabla\right)\vec{b}
  -\frac{1}{8\uppi}\bnabla_\perp B^2
  \equiv \vec{f}_\rmn{c} + \vec{f}_\rmn{p},
  \label{eq:Lorentz2}
\end{align}
where we define the gradient perpendicular to the magnetic field lines,
$\bnabla_\perp=(\bm{\mathsf{1}}-\vec{b}\vec{b})\bcdot\bnabla$. The second term,
$\vec{f}_\rmn{p}$, acts like a pressure force perpendicular to the magnetic
field lines and the first term, $\vec{f}_\rmn{c}$, is the magnetic curvature
force that also acts in a plane orthogonal to the field line. To see this, we
locally identify a curved field line with its curvature circle so that we can
locally define an azimuthally directed field $\vec{B}=B\vec{e}_\varphi$ in
cylindrical coordinates $(R,\varphi,z)$. Hence, in this case we obtain
$(\vec{b}\bcdot\bnabla)\vec{b}
=(\vec{e}_\varphi\bcdot\bnabla)\vec{e}_\varphi=-\vec{e}_R/R$ so that the
curvature force always points towards the centre of the curvature circle and
aims to reduce the curvature by pulling the field line straight with a force
that is the greater the smaller the curvature radius is.

Hence, it is advisable to define a magnetic curvature via
\begin{align}
  \label{eq:curvature}
  \bm{\kappa} \equiv (\vec{b}\bcdot\bnabla)\vec{b}
  = \frac{\left(\bm{\mathsf{1}}-\vec{b}\vec{b}\right)\bcdot(\vec{B}\bcdot\bnabla)\,\vec{B}}{B^{2}}
  = \frac{4\uppi\,\vec{f}_\rmn{c}}{B^{2}},
\end{align}
which immediately defines the curvature radius via
\begin{align}
  \label{eq:curvature_radius}
  R_\rmn{c}\equiv\frac{1}{\kappa}=\frac{1}{|\bm{\kappa}|}
  = \frac{1}{|(\vec{b}\bcdot\bnabla)\vec{b}|}.
\end{align}
Equation~\eqref{eq:curvature} may seem to suggest that large curvature forces
$f_\rmn{c} =|\vec{f}_\rmn{c}|$ and small magnetic field strengths imply a large
magnetic curvature. However, idealised simulations of incompressible, driven MHD
turbulence show that in the regime of high curvature, $\kappa$ is not strongly
correlated with a large curvature force but instead with a small value of the
magnetic field strength while small magnetic curvature is correlated with a low
level of curvature force \citep[see figure 8 of][]{2019PhPl...26g2306Y}.  Hence,
the anticorrelation between $B^2$ and $\kappa$ is a consequence of the curvature
force normal to the magnetic field line. A large curvature force rapidly
straightens out any curved field line, which apparently precludes the
possibility of a joint presence of a high curvature and a large magnetic field.
Heuristically, this means that a strong magnetic field resists bending.

Most importantly, a magnetised plasma that does not experience any driving will
evolve into a state that minimises magnetic tension and curvature (of course
subject to the magnetic helicity constraint). Hence, the continued presence of
magnetic curvature requires an active process such as a small-scale dynamo to
build up and maintain a high level of magnetic curvature. In the following, we
will study the emergence of magnetic curvature and its correlation properties
with field strength and curvature force during galaxy assembly and growth,
aiming at characterising the processes growing magnetic fields and whether this
is consistent with a single or even multiple small-scale dynamos.

Figure~\ref{fig:B-K_evolution} shows the correlation of $B$ and $\kappa$ in the
galaxy centre ($r<1$~kpc), colour coded by the mean density in the pixels. At
constant density, there is the expected correlation $B\propto \kappa^{-1/2}$
(Eq.~\ref{eq:curvature}) at high curvature, which is consistent with a
small-scale dynamo \citep{2004ApJ...612..276S}. The correlation weakens toward
low curvature where the curvature force $f_\rmn{c}$ starts to correlate with
$\kappa$. We verified that there is no strong correlation of $f_\rmn{c}$ and
$\kappa$ at high curvature, which confirms the finding of
\citet{2019PhPl...26g2306Y} who performed simulations of incompressible, driven
MHD turbulence.

In contrast to those simplified setups, we demonstrate the emergence of this
anticorrelation of $B^2$ and $\kappa$ at various densities in
Fig.~\ref{fig:B-K_evolution}, suggesting a superposition of multiple small-scale
dynamo modes. At the beginning, there is one small-scale dynamo mode growing the
magnetic field in the centre (possibly associated with turbulence driven by the
accretion shock). As this dynamo saturates the field in the centre at around
150~Myr (Fig.~\ref{fig:B_growth_res}), other (small-scale) dynamo modes are
excited at lower densities (in the central region but also in the disc and halo)
that grow the magnetic field simultaneously to the saturated inner dynamo,
however at a reduced rate. Those dynamos are likely driven by turbulence that
has been injected by Kelvin-Helmholtz surface and body modes that are
excited by the supersonically rotating cool galactic disc with respect to the
hot halo gas.

\begin{figure*}
\centering
\includegraphics[width=\textwidth]{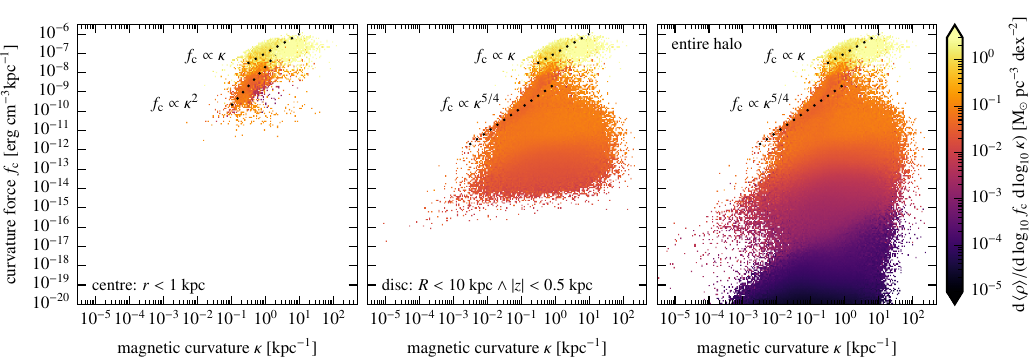}
\includegraphics[width=\textwidth]{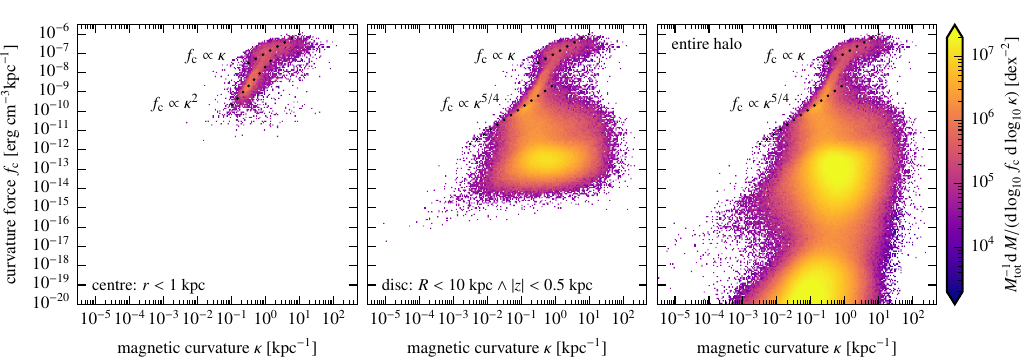}
\caption{Correlation of the curvature force density $f_\rmn{c}$ with curvature
  strength $\kappa$. The top panels show the mean density in the
  $f_\rmn{c}$--$\kappa$ plane while the bottom panels show the mass-weighted
  probability density.  From left to right, we show the central sphere of radius
  1 kpc (left), the disc region (middle) and the entire halo (right). While
  there is little correlation at high curvature, the correlation at constant
  density tightens at intermediate curvature $\kappa\sim1~\rmn{kpc}^{-1}$ and
  shows the expected scaling $f_\rmn{c}\propto\kappa$
  (Eq.~\ref{eq:curvature}). The steepening of this correlation to
  $f_\rmn{c}\propto\kappa^2$ at lower curvature in the central region suggests a
  non-trivial $B(\kappa)$ scaling.}
\label{fig:fc-K_regions}
\end{figure*}

\begin{figure*}
\centering
\includegraphics[width=\textwidth]{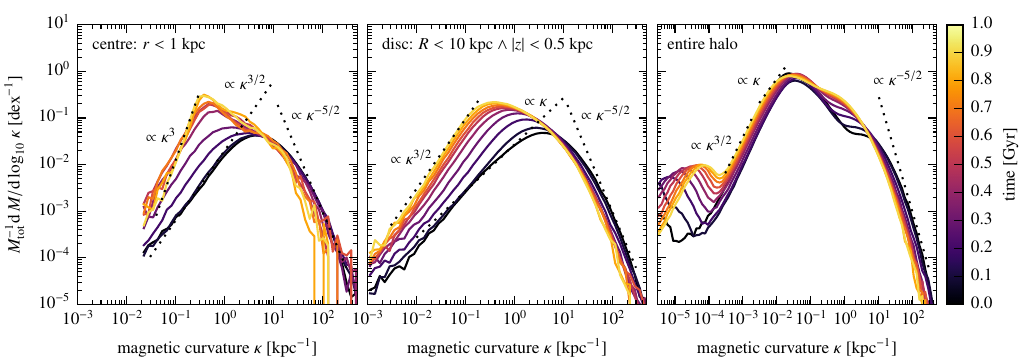}
\caption{Mass-weighted probability density of magnetic curvature for the central
  1 kpc (left), the disc region (middle) and the entire halo (right). In all
  cases (and excluding very small curvatures $\kappa<10^{-3}~\rmn{kpc}^{-1}$),
  we see that after the initial exponential growth phase (i.e., for
  $t\gtrsim150$~Myr), there is a growth phase of the magnetic coherence scale,
  which thus reduces the magnetic curvature scale. The asymptotic behaviour is
  consistent with turbulent MHD simulations by \citet{2019PhPl...26g2306Y}. Note
  the extended abscissa in the rightmost panel for the entire halo.}
\label{fig:K_histograms}
\end{figure*}

This picture becomes even richer if we increase the size of the analysing region
from the central region to the full disc (with a disc radius $R<10$~kpc and
vertical height $|z|<0.5$~kpc) and the entire halo in
Fig.~\ref{fig:B-K_regions}. There, we show the mean density (top panels) and the
mass-weighted probability density (bottom panels) in the $B$--$\kappa$ plane at
the saturation phase of the magnetic dynamo in the centre ($t=1$~Gyr). The disc
and halo regions show a larger range in densities and hence, dynamical
time-scales, which shows a complex superposition of different dynamo
processes. Focusing on the disc region (top middle panel), there are at least
three different dynamo processes operating, each separated by the characteristic
density: the upper sequence corresponds to the small-scale dynamo in the central
region while the lower branches are related to dynamo processes in the outer
disc.

Restricting to a narrow range in mean density, we see the expected correlation
$B\propto \kappa^{-1/2}$ at high curvatures, respectively, which is consistent
with MHD simulations of the small-scale dynamo in incompressible and homogeneous
turbulence of high magnetic Prandtl number and low Reynolds number
\citep{2004ApJ...612..276S}. At lower curvature, the correlations substantially
weaken, which is also seen in complementary MHD simulations of high Reynolds
number and a magnetic Prandtl number of unity \citep[see figure 8
  of][]{2019PhPl...26g2306Y}.

Figure~\ref{fig:fc-K_regions} shows the correlation of the magnitude of the
curvature force $f_\rmn{c}$ with the curvature $\kappa$: at intermediate
magnetic curvature $\kappa\sim1~\rmn{kpc}^{-1}$, we observe a correlation of $\kappa$
and $f_\rmn{c}$.  This linear correlation steepens toward lower curvature to
$f_\rmn{c}\propto\kappa^2$ in the centre at the lowest curvature, which is a
consequence of the linear correlation $B^2\propto\kappa$ in
Fig.~\ref{fig:B-K_regions} and likely related to a compressible MHD effect.  At
lower values of the curvature force density
$f_\rmn{c}\lesssim10^{-10}~\rmn{erg~cm}^{-3}~\rmn{kpc}^{-1}$ (in the disc region
and the entire halo), there is little correlation so that the main diagnostic of
the dynamo is indeed the anticorrelation of $\kappa$ and $B^2$ as derived in
Eq.~\eqref{eq:curvature} and realised in our simulations at constant density
(Figs.~\ref{fig:B-K_evolution} and \ref{fig:B-K_regions}).

There have been analytical derivations of the probability density of $\kappa$
\citep{2001PhRvE..65a6305S, 2004ApJ...612..276S,2019PhPl...26g2306Y}, which we
would like to confront to our simulations. Figure~\ref{fig:K_histograms} shows
the time evolution of the mass-weighted probability density of $\kappa$ for
different spatial regions: the central sphere of radius 1 kpc, the disc region
and the entire halo (from left to right). After the initial exponential growth
phase for $t\gtrsim150$~Myr and excluding very small curvatures
$\kappa<10^{-3}~\rmn{kpc}^{-1}$, the main effect is a shift of the peak of the
distribution to smaller values of $\kappa$ or equivalently to larger curvature
radii. This is the signature of a small-scale dynamo that is characterised by
continuous growth of magnetic power on large scales until saturation at the
corresponding larger kinetic turbulent energy, which increases the magnetic
coherence scale with time. We also observe a change in the asymptotic power-law
slopes and interesting structures in the probability density of the entire halo.

In order to quantify this behaviour, we note that the three spatial components
of the magnetic field fluctuations are independent (subject to the
$\bnabla\bcdot\vec{B}=0$ constraint) and thus obey quasi-Gaussian distributions.
Hence, the random variable $x=B^2=B_x^2+B_y^2+B_z^2$ follows a $\chi^2$
distribution with $k=3$ degrees of freedom,
\begin{align}
  f_k(x) = \frac{x^{k/2-1}\e^{-x/2}}{2^{k/2}\Gamma(k/2)},
  \quad x>0,
  \label{eq:chi_squared}
\end{align}
and $f_k(x)=0$ otherwise. $\Gamma(k)$ denotes the gamma function. We can thus
derive the asymptotic limit $\kappa\to\infty$ of the probability density of
$\kappa$ by using a Taylor expansion in the limit of $B\to0$
\citep{2019PhPl...26g2306Y}:
\begin{align}
  f(\kappa)\rmn{d}\kappa = f_3(B^2)\left|\frac{\rmn{d} B^2}{\rmn{d}\kappa}\right|\rmn{d}\kappa \to
  \frac{(4\uppi f_\rmn{c})^{3/2}}{\sqrt{2\uppi}}\,\kappa^{-5/2}\rmn{d}\kappa.
  \label{eq:PDF_high_kappa}
\end{align}
Using the fact that in the regime of large curvature, $\kappa$ is not strongly
correlated with the curvature force (see Fig.~\ref{fig:fc-K_regions}), we obtain
the scaling of $f(\kappa\to\infty)\propto \kappa^{-5/2}$ for all analysed
regions in Fig.~\ref{fig:K_histograms}, which agrees very well with the
theoretical expectation of Eq.~\eqref{eq:PDF_high_kappa} in the limit
$\kappa\to\infty$.

In order to derive the asymptotic limit $\kappa\to0$ of the probability density
of $\kappa$, we note that the curvature force is confined to a plane orthogonal
to $\vec{B}$. We can thus assume that $f_\rmn{c}^2$ is $\chi^2$ distributed,
with two degrees of freedom and use a Taylor expansion in the limit
$f_\rmn{c}\to0$:
\begin{align}
  f(\kappa)\rmn{d}\kappa =
  f_2(f_\rmn{c}^2)\left|\frac{\rmn{d} f_\rmn{c}^2}{\rmn{d}\kappa}\right|\rmn{d}\kappa \to
  \left(\frac{B^2}{4\uppi}\right)^2\,\kappa\rmn{d}\kappa.
  \label{eq:PDF_low_kappa}
\end{align}
Idealised simulations of incompressible, driven MHD turbulence show that in the
regime of small magnetic curvature ($\kappa\to0$), the curvature and magnetic
field strength are uncorrelated, implying $f(\kappa\to0)\propto \kappa$
\citep{2019PhPl...26g2306Y}.

By contrast, in Figs.~\ref{fig:B-K_regions} and \ref{fig:fc-K_regions} we
observe a superposition of various magnetic dynamos as well as correlations of
curvature and magnetic field strength in the regime of lower curvature. Those
result from adiabatic compression and/or pressure forces and gravity that can
modify the behaviour of $f(\kappa)$ in interesting ways. Hence, projecting the
two-dimensional probability density $f(\kappa,B)$ along $B$ (see
Fig.~\ref{fig:B-K_regions}) results in several bumps in $f(\kappa)$ as shown in
Fig.~\ref{fig:K_histograms}. Those result from different dynamo processes (with
the exception of the left-most bump of $f(\kappa)$ for the entire halo, which
arises from the slightly modified initial conditions). During the exponential
growth phase at the centre ($t\lesssim150$~Myr), we observe a correlation of
$f_\rmn{c}\propto\kappa^{5/4}$ or equivalently $B^2\propto\kappa^{1/2}$ at small
values of $\kappa$, which results in a steeper slope of $f(\kappa\to0)\propto
\kappa^{3/2}$. This limiting behaviour is also observed at 1 Gyr in our disc
region and our entire halo.  During the saturation phase of the magnetic dynamo
at the centre ($t=1$~Gyr), we observe a stronger correlation of
$f_\rmn{c}\propto\kappa^{2}$ or equivalently $B^2\propto\kappa$, which results
in a steeper slope of $f(\kappa\to0)\propto \kappa^{3}$.

We conclude that the high-curvature limit of our simulations agrees with
theoretical predictions of the small-scale dynamo in incompressible simulations
of turbulence of high Reynolds number \citep{2019PhPl...26g2306Y} while we
obtain different scalings in comparison to simulations in the subviscous range
of high magnetic Prandtl numbers \citep{2004ApJ...612..276S} who base their
mathematical descriptions on studies of the time evolution of curvature in line
and surface elements for a simple model turbulence
\citep{1991JFM...225..529D}. We show that gravitational collapse and the
inside-out formation of a galactic disc excite a superposition of different
small-scale dynamo modes. In consequence, in the limiting regime at low
curvature the probability density shows a complex scaling behaviour, which is
subject to a compressibly modified magnetic field. This is a direct consequence
of the more complex velocity field in our simulation that is shaped by the
gravitational collapse, CRs and ISM physics, which imprint more structure in the
magnetic field. Contrarily, the velocity field in turbulent homogeneous boxes is
randomly driven on large scales so that the velocity field is initially
uncorrelated. Any structure in the velocity and magnetic field in those
idealised boxes is thus generated by a small-scale dynamo while it is
additionally shaped by compressible motions, gravity and ISM physics in our
galaxy simulations.

\section{FIR--radio correlation}
\label{sec:FRC}

We now analyse the radio synchrotron emission and start the discussion with the
global FRC and sources of scatter. To understand this correlation we will show
an analytical calculation of the FRC and analyse individual maps and profiles of
the radio emission and other quantities of our simulated galaxies, which help in
scrutinising whether our simulations reproduce the local FRC.

\begin{figure}
\centering
\includegraphics[width=0.49\textwidth]{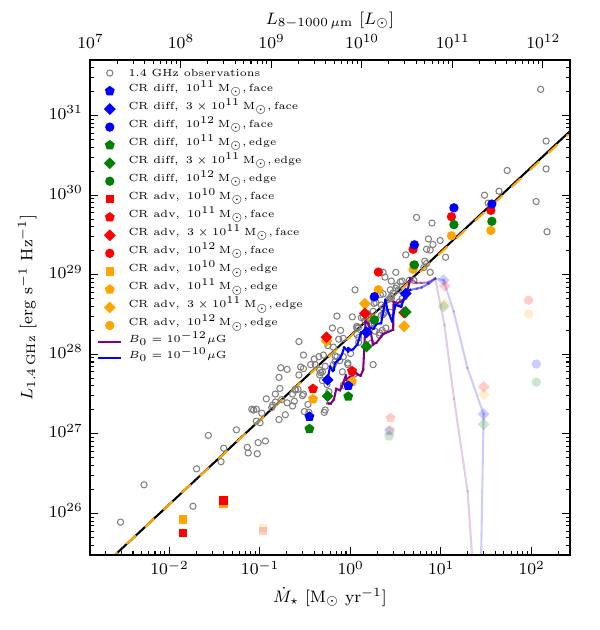}
\caption{Correlation between the radio luminosity $L_{1.4\,\rmn{GHz}}$ and FIR
  luminosity $L_{\rmn{FIR}}$ (top axis) or SFR (bottom axis) of star-forming
  galaxies.  The observed galaxies \citep[open circles,][]{2003ApJ...586..794B}
  exhibit a best-fit relation (orange dashed) that is slightly steeper than the
  linear calorimetric relation (Eq.~\ref{eq:FIR-radio}) and perfectly agrees
  with our analytically derived FRC (black solid,
  Eq.~\ref{eq:FRC_analytics4}). We overplot the emission of our simulated
  galaxies that only account for advective CR transport (`CR adv', red and
  orange) and simulations in which we additionally follow anisotropic CR
  diffusion (`CR diff', blue and green). Semi-transparent symbols correspond to
  a phase of a growing magnetic dynamo whereas fully-coloured symbols
  characterise the stages of a saturated dynamo and fall on the observed
  relation. The solid blue (purple) lines delineate time evolution tracks of
  simulated galaxies ($3\times10^{11}~\rmn{M}_\odot$, face-on projection) with
  initial magnetic field strength $10^{-10}~\upmu$G ($10^{-12}~\upmu$G). We
  contrast edge-on projections (orange and green) to face-on views (red and
  blue), which mostly exhibit a higher luminosity because the entire toroidal
  disc field contributes to the synchrotron intensity.  }
\label{fig:FRC}
\end{figure}

\subsection{Global FIR--radio correlation}

The FIR luminosity ($L_{\rmn{FIR}}$, $8-1000~\upmu$m) is related to the
SFR, $\dot{M}_\star$, of spiral galaxies
\citep{1959ApJ...129..243S,1998ApJ...498..541K,1998ARA+A..36..189K},
\begin{equation}
  \label{eq:FIR-SFR}
  \frac{\dot{M}_\star}{\msun~\rmn{yr}^{-1}}=\epsilon\,1.7\times10^{-10}\,\frac{L_{\rmn{FIR}}}{\rmn{L}_\odot}.
\end{equation}
This conversion assumes that thermal dust emission calorimetrically traces the
emission of young stars, and the factor $\epsilon=0.79$ derives from the
\citet{2003PASP..115..763C} initial mass function.

In Fig.~\ref{fig:FRC} we correlate the total specific radio synchrotron
luminosity $L_{1.4\,\rmn{GHz}}$ with the SFR of our differently-sized galaxies
at various times, which also correspond to different FIR luminosities as is
shown at the top horizontal axis. We select simulation times such that we sample
the SFR in logarithmically-equidistant steps of one e-folding, starting with the
maximum during the initial starburst. Our simulations in the saturated
small-scale dynamo regime (shown with the fully-coloured data points) match the
observed FRC extremely well.

Observational data \citep[as compiled by][]{2003ApJ...586..794B} exhibit a
best-fitting relation that is slightly steeper than the linear calorimetric
relation,
\begin{equation}
  \label{eq:FIR-radio}
  L_{1.4\,\rmn{GHz}}=\tilde{L}_0\left(\frac{L_{\rmn{FIR}}}{10^{10}\rmn{L}_\odot}\right)^{\alpha_\rmn{FRC}},
\end{equation}
where $\tilde{L}_0=2.26\times{10}^{28}\,\rmn{erg~s}^{-1}~\rmn{Hz}^{-1}$ and
$\alpha_\rmn{FRC}=1.055$. Combining Eqs.~\eqref{eq:FIR-SFR} and \eqref{eq:FIR-radio}
yields
\begin{equation}
  \label{eq:FIR-radio2}
  \frac{L_{1.4\,\rmn{GHz}}}{\dot{M}_\star}=\frac{L_0}{\msun~\rmn{yr}^{-1}}
  \left(\frac{\dot{M}_\star}{\msun~\rmn{yr}^{-1}}\right)^{\alpha_\rmn{FRC}-1},
\end{equation}
where $L_0=1.68\times{10}^{28}\,\rmn{erg~s}^{-1}~\rmn{Hz}^{-1}$.

The purple and blue lines in Fig.~\ref{fig:FRC} trace the time evolution of the
SFR and synchrotron luminosity $L_{1.4\,\rmn{GHz}}$ of a simulated galaxy
($3\times10^{11}~\rmn{M}_\odot$) with initial magnetic field strengths
$B_{\rmn{init}}=10^{-12}~\upmu$G and $10^{-10}~\upmu$G,
respectively.\footnote{The line segments are spaced by ${\Delta}t=0.1$~Gyr (the
first symbols shown are at $t=0.3$ and 0.2~Gyr for the purple and blue lines,
respectively).} During the kinematic phase of the dynamo, the model with the
larger value of $B_{\rmn{init}}$ grows faster and produces a larger radio
luminosity. At the time of saturation, the tracks of simulations with different
$B_{\rmn{init}}$ start to converge, but retain memory of their different
evolution and show radio luminosities that differ by a factor of up to two as a
result of the stochasticity of the star formation process triggered by the
different magnetic realisations of both models.

\subsection{Analytics}
\label{sec:analytics}

To understand the reason for this success of matching the FRC, we will show that
it is a necessary consequence of electron calorimetry, thus generalising the
findings of \citet{2007ApJ...654..219T} to also include primaries in addition to
secondary electrons. First, we relate the CR proton luminosity, $L_\p $, to the
SFR via
\begin{align}
  \label{eq:Lp}
  L_\p = \zeta_\rmn{SN}\dot{M}_\star\frac{E_\rmn{SN}}{M_\star}
  =  5.5\times10^{-4}\,\zeta_{\rmn{SN},0.05} L_\rmn{FIR},
\end{align}
where we used Eq.~\eqref{eq:FIR-SFR} to eliminate the SFR and
$\zeta_{\rmn{SN},0.05}=\zeta_{\rmn{SN}}/0.05$ is the kinetic energy fraction of
SNRs injected into CR protons \citep{Pais2018}. Assuming a
\citet{2003PASP..115..763C} initial mass function and that stars with a mass of
$8-40~\msun$ explode as supernovae \citep[while they directly collapse to a
  black hole above $40~\msun$,][]{1999ApJ...522..413F}, there is about one
core-collapse supernova per $100~\msun$ of newly formed stars.\footnote{We
assume an initial mass function between 0.08 and 140~$\msun$
\citep{2016ApJ...824...82C}.}  Thus, we obtain a canonical supernova energy
release per unit stellar mass of
$E_\rmn{SN}/M_\star=10^{51}~\rmn{erg}/(100~\msun)=1\times10^{49}~\rmn{erg}~\msun^{-1}$. The
total luminosity of primary and secondary electrons is given by
\begin{align}
  \label{eq:Le}
  L_\e = \left(\zeta_\rmn{prim} + \zeta_\rmn{sec}\eta_\rmn{cal,p}\right)L_\p
  = \zeta_\rmn{e}L_\p\approx 0.27 L_\p,
\end{align}
where $\zeta_\rmn{prim}\approx0.09$ is the energy ratio of primary CR electrons
and protons given by Eq.~\eqref{eq:K_ep-zeta},
$\zeta_\rmn{sec}=\eps_\rmn{sec}/\eps_\p\approx0.25$ is the energy ratio of
secondary CR electrons and protons (as we will show below), and
$\eta_{\mathrm{cal,p}}=\Lambda_\rmn{hadr}/(\Lambda_\rmn{hadr}+\Lambda_\rmn{Coul})\approx0.73$
is the calorimetric fraction of CR proton energy that cools via hadronic
interactions as supposed to Coulomb interactions \citep[where
  $\Lambda_\rmn{hadr}$ and $\Lambda_\rmn{Coul}$ are the hadronic and Coulomb
  cooling rates,][]{2017MNRAS.465.4500P}. In hadronic interactions,
approximately 2/3 of the CR proton energy is channeled into charged pions, which
subsequently decay into high-energy neutrinos and electron-positron pairs so
that the pairs receive 1/4 of the pion energy \citep{2002cra..book.....S}. To
arrive at the result for $\zeta_\rmn{sec}$, we also account for the nuclear
enhancement factor of 1.4 to 1.6, that accounts for heavier nuclei in the
composition of CRs and the ISM \citep{Biallas1976,StephensBadhwar1981}.  In the
following, we adopt $\zeta_\rmn{e}=0.27$ and $\zeta_{\rmn{SN}}=0.05$ as fiducial
values and drop the explicit dependence on these variable in our calculation. In
order to connect this bolometric electron luminosity to the spectral radio
luminosity, we have to calculate the bolometric energy fraction emitted into the
radio band. To this end, we write down the volume-integrated injected electron
distribution (which we assume to be a simple power-law momentum distribution):
\begin{align}
  \label{eq:fe}
  Q_\e (p_\e)= \frac{\dd N_\e}{\dd p_\e\dd t}
  = \int q_\e (p_\e)\dd V
  = \C_\e p_\e^{-\alpha}\theta(p_\e-p_\rmn{min}),
\end{align}
where $p_\rmn{min}$ is the dimensionless low-momentum cutoff, $\C_\e$ is the
normalization (in units of s$^{-1}$), and $\theta (p)$ denotes the Heaviside
step function. The CR electron luminosity is given by
\begin{align}
\label{eq:eps1}
L_\e&=\int_0^{\infty}Q_\e(p_\e)\,T_\e(p_\e)dp_\e=\frac{\C_\e\,m_{\e}c^2}{\alpha-1}\\
&\quad\times\left[\frac{1}{2}\,\B_\frac{1}{1+p_\rmn{min}^2}\left(\frac{\alpha-2}{2},\frac{3-\alpha}{2}\right)+
  p_\rmn{min}^{1-\alpha}\left(\sqrt{1+p_\rmn{min}^2}-1\right)\right],\nonumber\\
&\equiv \C_\e\,m_{\e}c^2 A_\rmn{bol} (p_\rmn{min},\alpha)
\label{eq:eps2}
\end{align}
where $T_\e(p_\e)=\left(\sqrt{1+p_\e^2}-1\right)\,m_{\e}c^2$ is the kinetic electron
energy, $\B_y(a,b)$ denotes the incomplete beta function
\citep{1965hmfw.book.....A}, and we assume $\alpha>2$. The bolometric energy
fraction $A_\rmn{bol} (p_\rmn{min},\alpha)^{-1}$ depends only weakly on
$p_\rmn{min}$ (provided that $p_\rmn{min}\lesssim1$) and enables us to
rewrite the electron source distribution via
\begin{align}
  \label{eq:A_bol}
  E_\e^2Q(E_\e)\equiv\frac{E_\e\dd N_\e}{\dd\ln\gamma_\e\,\dd t}
  \approx m_\e c^2 \C_\e \gamma_\e^{2-\alpha}
  = \frac{L_\e}{A_\rmn{bol}}\,\gamma_\e^{2-\alpha},
\end{align}
where $\gamma_\e$ is the electron Lorentz factor. In the second step, we adopted
the electron distribution in the relativistic limit (Eq.~\ref{eq:fe}), which is
valid for synchrotron-emitting electrons.\footnote{Note that
$A_\rmn{bol}=\ln(p_\rmn{max}/p_\rmn{min})$ for a power-law electron distribution
with $\alpha=2$ and $p_\rmn{min}\gg 1$ so that the factor
$\gamma_\e^{2-\alpha}=1$. If the spectral index were instead $\alpha=2.2$, the
simplified assumption $\alpha=2$ would have overestimated the synchrotron
luminosity by a factor of 6 for $\gamma_\e=8\times10^3$.}  Assuming that the
synchrotron cooling time of CR electrons is shorter than their escape times
\citep{1989A&A...218...67V}, we find an FRC of the form
\begin{align}
  \label{eq:FRC_analytics2}
  \nu L_\nu (\rmn{GHz}) &= \frac{E_\gamma\dd N_\gamma}{\dd\ln\nu\,\dd t} 
  =\eta_{\rmn{syn}}\,\frac{E_\e\dd
    N_\e}{2\,\dd\ln\gamma_\e\,\dd t}
  \approx\frac{\eta_{\rmn{syn}}}{2A_\rmn{bol}}\,\gamma_\e^{2-\alpha}L_\e\\
  &\approx 8.36\times10^{-7}\,\left(\frac{\eta_{\rmn{syn}}}{0.30}\right)
  \,\left(\frac{\gamma_\e}{8.4\times10^3}\right)^{2-\alpha} L_{\rmn{FIR}},
  \label{eq:FRC_analytics3}
\end{align}
where we have used Eqs.~\eqref{eq:Lp} and \eqref{eq:Le} in the last step.  Here,
$\dd\ln\nu=2\,\dd\ln\gamma_\e$ according to Eq.~\eqref{eq:synch_photon},
$\eta_{\rmn{syn}}$ denotes the calorimetric energy fraction of electrons that
cool and radiate synchrotron emission, and we adopted in the last step
$\alpha=2.2$, $p_\rmn{min}=1$, and electrons with a Lorentz factor
$\gamma_\e=8.4\times10^3$ that emit 1.4~GHz synchrotron radiation in a magnetic
field of $2.4~\upmu$G (see Eq.~\ref{eq:synch_photon})\footnote{For simplicity,
we adapt $B$ instead of $B_\perp$ to relate the magnetic field strength to the
characteristic electron energy. Accounting for the orientation of the magnetic
field would slightly modify the normalisation of the FRC, which is however
degenerate with the other input parameters and hence justifies our choice.}. In
our models, $\eta_{\rmn{syn}}$ ranges from 0.1 to 0.7 at SFRs
$\dot{M}_\star\gtrsim1~\msun~\rmn{yr}^{-1}$, with slightly smaller (larger)
values realised in our `CR diff' (`CR adv') models \citep{2021WerhahnIII}. We
note that the lowest values of $\eta_{\rmn{syn}}$ in the `CR diff' model are
realised in our $10^{10}~\msun$ halo and could become larger if we use an
improved ISM model in cosmological simulations of dwarf galaxies with (i) star
formation- and cosmic accretion-driven turbulence that helps to further grow the
disc magnetic field as well as (ii) an improved CR transport model which
delivers a realistic (spatially and temporally varying) CR diffusion coefficient
in the self-confinement picture \citep{2019MNRAS.485.2977T}.

Interestingly, the factor $\gamma_\e^{2-\alpha}$ predicts a slightly
super-linear FRC because of the different saturation values of the magnetic
field strength with halo mass and SFR (see Fig.~\ref{fig:energies}). In our `CR
diff' model, we find saturated magnetic field strengths of
$[0.14,1.8,7,14]~\upmu$G for SFRs of $[0.01,0.4,5,30]~\msun~\rmn{yr}^{-1}$.
This implies that the Lorentz factor of the electrons that emit 1.4~GHz
synchrotron radiation of $\gamma_\e\approx[35,9.6,4.8,3.5]\times10^3$ according
to Eq.~\eqref{eq:synch_photon} which we fit in order to obtain a functional
dependence of the 1.4~GHz synchrotron emitting Lorentz factor on the SFR,
\begin{align}
  \label{eq:gamma-SFR}
  \gamma_\e = 8.4\times10^3\,\left(\frac{\dot{M}_\star}{\msun~\rmn{yr}^{-1}}\right)^{-0.3}.
\end{align}
Using the \citet{1998ApJ...498..541K} relation of Eq.~\eqref{eq:FIR-SFR} as well
as Eq.~\eqref{eq:gamma-SFR}, we can rewrite Eq.~\eqref{eq:FRC_analytics3} and
obtain the specific radio luminosity at 1.4~GHz,
\begin{align}
  \label{eq:FRC_analytics4}
  L_{1.4\,\rmn{GHz}} 
  \approx~ &1.7\times 10^{28}\,\left(\frac{\eta_{\rmn{syn}}}{0.30}\right)\,
  \left(\frac{\dot{M}_\star}{\msun~\rmn{yr}^{-1}}\right)^{1.06}
  \rmn{erg~s}^{-1}~\rmn{Hz}^{-1},
\end{align}
where we adopted $\alpha=2.2$. In general, the power-law index in this
correlation depends on the electron spectral index and reads $\alpha_\rmn{FRC}=1
+ 0.3(\alpha-2)$. This correlation coincides with the observed super-linear FRC
by \citet{2003ApJ...586..794B} which obeys the fitting relation of
Eq.~\eqref{eq:FIR-radio2}.  Figure~\ref{fig:FRC} shows a comparison of observed
(orange dashed) and theoretical (black solid) mean FRC, which are nearly
identical. \citet{2021WerhahnIII} find that $\eta_{\rmn{syn}}$ increases with
time along the evolutionary path of a galaxy (for the $M_{200}\gtrsim
3\times10^{11}~\msun$ haloes), because of the decreasing gas densities (see
Section~\ref{sec:maps}) and hence the increasing cooling times of bremsstrahlung
and Coulomb interactions. This implies that the galaxies evolve from the lower
to the upper envelope of the FRC as their SFR decreases with time. There are
several other processes that cause scatter in the FRC which we will analyse now.

\subsection{Scatter of the global FIR--radio correlation}

We identify three physical effects that contribute to the scatter in the FRC:
(i) initial magnetic field strength in combination with stochasticity of the
star formation process and intermittent galactic winds, which modulate the
magnetic and CR energy as well as the resulting synchrotron luminosity in the
inner regions that are affected by the wind, (ii) different halo masses and thus
different specific SFRs at a given absolute SFR and (iii) the galactic
inclination in combination with an anisotropic magnetic field distribution cause
variations of the radio intensity.

\begin{enumerate}
\item At early times in the kinematic dynamo regime, when the magnetic field
  strength is still exponentially growing (transparently coloured points in
  Fig.~\ref{fig:FRC}), the galaxies fall short by orders of magnitude of the
  observed FRC. However, in the saturated dynamo regime the tracks of
  simulations with different $B_{\rmn{init}}$ start to converge, but
  nevertheless retain memory of their different evolution.\footnote{Note that
  this statement is specific to the idealised setup used here and may not be
  generalisable to cosmological simulations.} While the observed FRC of nearby
  galaxies \citep[within 255~Mpc,][]{2003ApJ...586..794B} probes this saturated
  dynamo regime, our simulations imply a residual scatter if galaxies had
  different seed magnetic fields at the formation time or a different formation
  history and magnetic dynamo efficiencies. This could potentially be caused by
  the stochasticity of the star formation process. On top of this average trend,
  we see an upwards deviation of the purple track in Fig.~\ref{fig:FRC} at
  around 1.6~$\msun~\rmn{yr}^{-1}$. This is due to a drop in the wind speed,
  which retains more CRs and a larger magnetic energy in the central region, and
  as a result, temporarily boosts the radio synchrotron emissivity by a factor
  of two.
 
\item As the small-scale dynamo has saturated, the galaxies move largely along
  the observed relation toward smaller SFRs and radio luminosities with a
  tendency to move from the lower to the upper envelope of the observational
  scatter (see Fig.~\ref{fig:FRC}). There are two reasons for this behaviour
  after the starburst phase. First, as explained in Section~\ref{sec:analytics},
  smaller SFRs imply smaller gas densities and smaller collisional loss rates so
  that the calorimetric energy fraction emitted via radio synchrotron increases
  \citep{2021WerhahnIII}. Second, the decreasing gas accretion onto the disc and
  thus the decreasing SFRs also lowers the CR proton injection rate. By
  contrast, there appears to be a magnetic dynamo that counteracts this
  decreasing trend of the CR and thermal energy densities so that the magnetic
  energy density remains constant with time. Thus, the total synchrotron
  emissivity decreases less strongly than the SFR and a galaxy evolves from the
  lower towards the upper envelope of the FRC. Thus, these two effects combined
  predict that at a given SFR, less massive systems with high specific SFRs to
  populate the lower envelope of the FRC in comparison to more massive systems
  that are in a late evolutionary phase with small specific SFRs.

\item There is a third effect that contributes to the FRC scatter by a factor of
  up to two. In the saturated dynamo regime, the toroidal disc field dominates
  over the poloidal components by an order of magnitude
  \citep{Pakmor2016b}. Observing such a galaxy face-on maximises the available
  synchrotron intensity (since the toroidal field is transverse to the
  line-of-sight in this configuration), whereas the intensity drops by a factor
  of two for an edge-on geometry (since half the toroidal field is now aligned
  with the line-of-sight). Conversely, a strong outflow generates a field
  morphology with substantial poloidal components and reduces the factor in
  between face-on and edge-on geometries as can be seen by the reverse ordering
  of the red and orange data points in the $3\times10^{11}~\msun$ halo at an SFR
  of around $1.5~\msun~\rmn{yr}^{-1}$. Note that supernova-driven turbulence may
  change the global field topology in the spiral-arm regions in the disc to a
  preferred turbulent magnetic configuration instead of a dominant toroidal
  morphology, which would significantly weaken the described effect. However,
  observations of the polarised thermal dust emission in M82 find a dominant
  large-scale ordered potential field associated with the outflow
  \citep{2021ApJ...914...24L} so that the described effect may be most important
  for ordered poloidal fields in galactic winds.
\end{enumerate}

Finally, we find that galaxies in the model `CR~diff' (`CR~adv') populate the
lower (upper) envelope of the observed relation after they have entered the
saturated dynamo regime (fully-coloured symbols). There are two effects that
cause this behaviour. First, the CR energy density is larger in the `CR~adv'
model at small galaxy masses (see Fig.~\ref{fig:energies}). This also causes a
separation of the FIR--gamma-ray correlation at low SFRs
\citep{2017ApJ...847L..13P,2021WerhahnII}. Second, we find that the outflows
driven by anisotropically diffusing CRs quench the magnetic dynamo and increase
its growth time in comparison to model `CR~adv' \citep[similarly to the case of
  isotropic diffusion, which quenches the dynamo in comparison to the
  anisotropic diffusion case,][]{Pakmor2016b}. Hence, galaxies in model `CR~adv'
reach the observational relation earlier and saturate at higher field
strengths. The $10^{10}~\msun$ halo in model `CR~diff' never reaches the FRC and
peaks around $L_{1.4~\rmn{GHz}}\approx10^{24}~\rmn{erg~s}^{-1}\rmn{Hz}^{-1}$.
Apparently, modelling CR diffusion with a constant diffusion coefficient
$10^{28}~\rmn{cm}^2~\rmn{s}^{-1}$ removes the gas too fast from this low-mass
disc so that the dynamo lags behind, suggesting a more subtle CR transport
process in dwarfs. In summary, an increasing (constant) CR diffusion coefficient
moves galaxies to the lower envelope of the {\it simulated} FRC.

A more realistic modelling of CR transport in the CR self-confinement picture
leads to a superposition of advection, diffusion and streaming with a spatially
and temporally varying effective diffusion coefficient
\citep{2018ApJ...854....5J,2019MNRAS.485.2977T,2022MNRAS.509.4803T,
  2020ApJ...890L..18T,2021MNRAS.503.2242T}.  Future work is needed to study
whether there is a range of effective CR diffusion coefficients at a fixed SFR,
which would imply a third source of scatter in the {\it observed} FRC by
indirectly affecting the FRC through its influence on the magnetic dynamo.
Interestingly, observational studies of the local FRC demonstrate that the radio
emissivity depends on CR transport properties in different environments
\citep{2012ApJ...756..141B,2017MNRAS.471..337B,2014AJ....147..103H,
  2019A&A...622A...8H}, although the level of agreement with simulations would
have to be carefully studied in future work.

\subsection{Radio morphology of discs and CR-driven outflows}
\label{sec:maps}

\begin{figure*}
\centering
\includegraphics[width=0.985\textwidth]{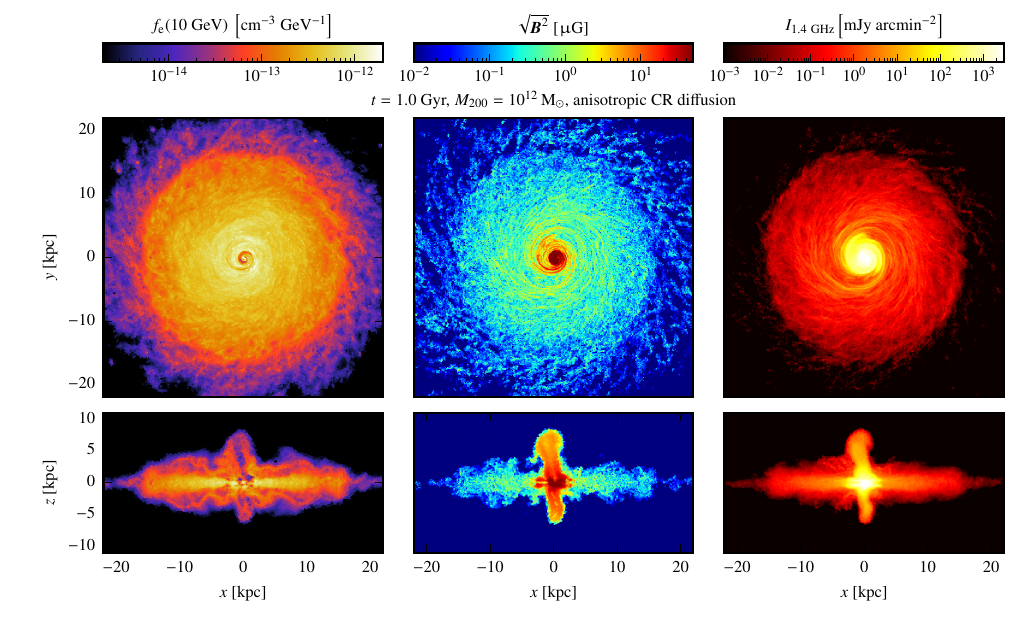}
\includegraphics[width=0.985\textwidth]{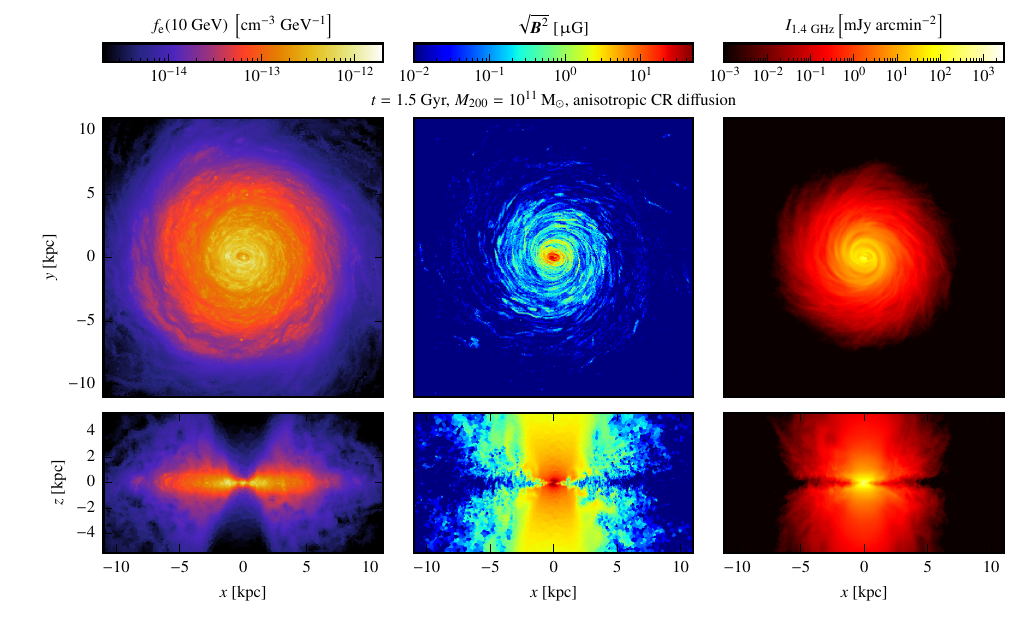}
\caption{Properties of the gas disc in our Milky~Way-mass galaxy at 1~Gyr
  ($M_{200}=10^{12}~\msun$, $c_{200}=7$, top panels) and in our smaller galaxy
  at 1.5~Gyr ($M_{200}=10^{11}~\msun$, $c_{200}=12$, bottom panels). The MHD
  simulations account for CR injection at SNRs, follow CR advection with the gas
  and anisotropic diffusion along magnetic fields relative to the gas.  We show
  cross-sections in the mid-plane of the disc (face-on views) and vertical
  cut-planes through the centre (edge-on views) of the total (primary and
  secondary) CR electron spectral density at 10~GeV (left), total magnetic field
  strength (middle) and the total radio synchrotron brightness,
  $I_\rmn{1.4\,\rmn{GHz}}$ (right). The magnetically-loaded outflows are visible
  as bubbles in the Milky~Way-mass galaxy and as an X-shaped radio morphology
  for the smaller galaxy (edge-on views on the right-hand side).}
\label{fig:maps}
\end{figure*}

\begin{figure*}
\centering
\includegraphics[width=0.985\textwidth]{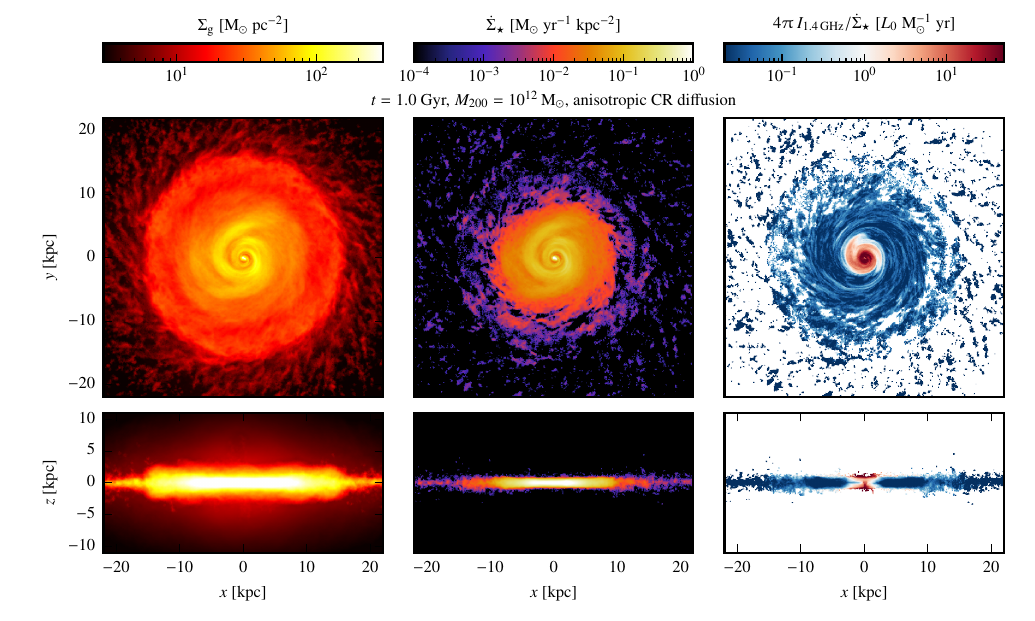}
\includegraphics[width=0.985\textwidth]{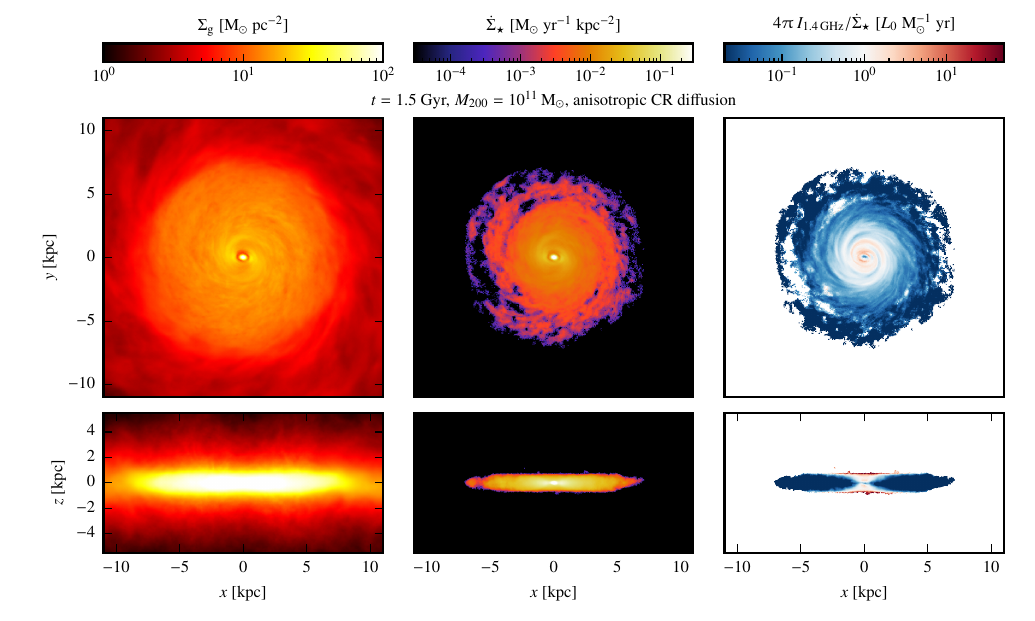}
\caption{Shown are face-on and edge-on views of the surface mass density of gas
  (left), SFR surface density (middle) and the ratio of the radio synchrotron
  brightness-to-SFR surface density,
  $4\uppi\,I_{\rmn{1.4\,GHz}}/\dot{\Sigma}_\star$ (right). Note that we scale
  this ratio to the mean radio flux $L_0$ of the observed FRC at an SFR of
  $1\,\msun\,\rmn{yr}^{-1}$ (Eq.~\ref{eq:FIR-radio2}) so that red (blue) colours
  indicate an excess (deficiency) in comparison to the observed FRC. We use the
  same simulation snapshots as in Fig.~\ref{fig:maps}.  }
\label{fig:maps2}
\end{figure*}

In Fig.~\ref{fig:maps}, we show face-on and edge-on views of the total CR
electron spectral density at 10~GeV and the magnetic field strength (left-hand
and middle panels) for our galaxies with halo masses $10^{12}$ and
$10^{11}~\msun$. While the Milky~Way-mass galaxy shows bubble-like CR-driven
outflows, there are fast outflows in the smaller galaxy, which evacuate a
bi-conical, X-shaped region in the circumgalactic medium (visible in yellow
colours in the bottom-right panel of Fig.~\ref{fig:maps}). In both galaxies, the
outflows are strongly magnetised and and rich in CR electrons and protons
\citep[cf.~Fig.1 in][ for the Milky~Way-mass galaxy with halo mass
  $M_{200}=10^{12}~\msun$ and concentration
  $c_{200}=7$]{2017ApJ...847L..13P}. After the formation of the dense gas disc
the CR pressure gradient is mostly vertically aligned so that it accelerates the
ISM and drives outflows perpendicular to the stellar and gaseous discs. Thus,
these CR-driven winds are an emergent phenomenon and the result of the outflows
taking the path of least resistance away from the galaxies. Both galaxies show
magnetic spiral arms which have counterparts in the surface gas density
(cf.~Figs.~\ref{fig:maps} and \ref{fig:maps2}).

As we will show below, the magnetic dynamo grows slower at larger radii and we
likely underestimate the dynamo growth at large galactocentric radii with our
pressurised ISM \citep{2003MNRAS.339..289S}, that neglects most of
supernova-driven turbulence with the exception of our explicit injection of CR
energy with an efficiency $\zeta_\rmn{SN}=0.05$ and 0.1 of the canonical
supernova energy. Correctly resolving the momentum and thermal energy input of
supernova explosions that are clustered \citep{2016MNRAS.456.3432G} and maintain
memory of their birthplaces at the junction of filaments \citep[with sub-parsec
  widths,][]{2011A&A...529L...6A,2012A&A...540L..11S,
  2021MNRAS.tmp.2569P,2021MNRAS.508.2736W} within molecular clouds is only
possible for the mass resolution of a few solar masses in isolated stratified
boxes at moderate (Milky Way-like) surface mass densities
\citep{2021MNRAS.504.1039R} or for dwarf galaxies
\citep{2021MNRAS.501.5597G}. Only at this mass resolution, it is possible to
form a multi-phase ISM with a volume-filling hot ($10^7$~K) phase and to
identify a mean-field dynamo \citep{2013MNRAS.430L..40G}.

The total radio emission is shown in the right-hand panels of
Fig.~\ref{fig:maps}. The emission of the Milky~Way-mass galaxy is dominated by
the disc and a bright bulge that sources magnetically-loaded outflows, which are
visible as faint bubbles.  On the contrary, the disc is almost invisible in the
smaller galaxy at $1.5$~Gyr, which shows instead an X-shaped radio morphology in
the outflow region that arises because of the overlap of the cylindrical outflow
geometry that is mirrored in the magnetic field distribution and the conical
cavities in the CR electron spectrum.

In Fig.~\ref{fig:maps2}, we compare face-on and edge-on views of the gas surface
mass density and the SFR surface density (left-hand and middle
panels). The additional CR pressure support inflates the gas discs and thus
enhances the gas density above and below the discs. By contrast, only gas in a
thin disc exceeds the critical gas density necessary to form stars
\citep{1998ApJ...498..541K}. The larger centrifugal support in the Milky
Way-mass galaxy in comparison to the smaller galaxy implies a thinner stellar
disc (see the discussion in Section~\ref{sec:mag_sat}). Note that the 
edge-on projection of the gas density yields a maximum gas surface mass density
of $\Sigma_\rmn{g}\approx3\times10^9\,\msun~\rmn{kpc}^{-2}$, a factor of 30
larger than the maximum value of the face-on projection.  This exemplifies the
strong dependence of $\Sigma_\rmn{g}$ on galactic inclination in all but the
very compact, ultra-luminous infrared galaxies and suggests the choice of a
parametrization that is independent of inclination (such as the SFR) for an
independent variable to be correlated with the gamma-ray or radio emission. We
will return to the right-hand panels of Fig.~\ref{fig:maps2} in
Section~\ref{sec:local_FRC}.

\begin{figure*}
\centering
\includegraphics[width=\textwidth]{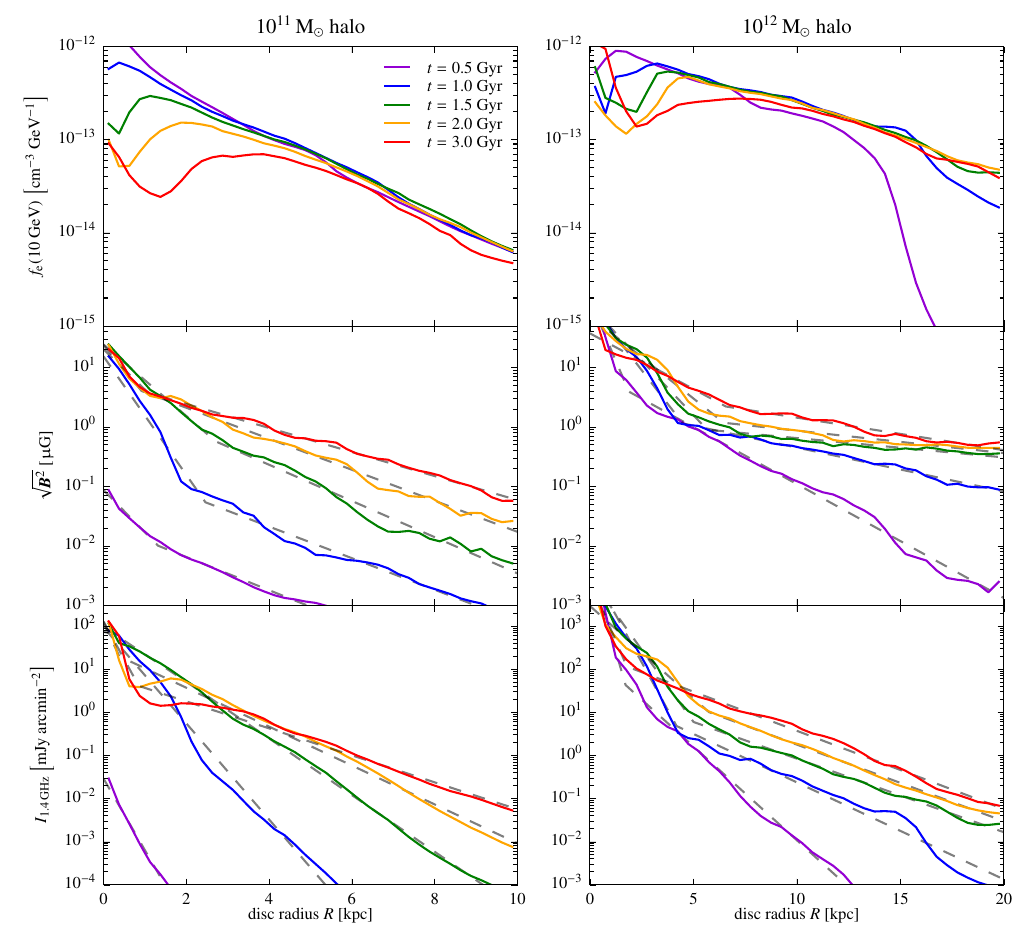}
\caption{Radial disc profiles of our small galaxy ($M_{200}=10^{11}~\msun$,
  $c_{200}=12$, left column) and our Milky~Way-mass galaxy
  ($M_{200}=10^{12}~\msun$, $c_{200}=7$, right column) at different times (as
  indicated in the legend). We show volume-averaged profiles of the total
  (primary and secondary) CR electron spectral density at 10~GeV (top), root
  mean square magnetic field strength (middle), both averaged in a disc of total
  height 1~kpc, and the total face-on radio synchrotron brightness at
  $\nu=1.4$~GHz (bottom). The profiles show a CR electron spectral density that
  decreases in radius and time, which is a consequence of the decreasing SFR, a
  growing magnetic field in the outer disc with time, which together imply
  increasing synchrotron surface brightness profiles with time. The radial
  profiles of magnetic field strength and synchrotron emission are well fit by
  joint exponentials (dashed lines, Eqs.~\ref{eq:Bfit} and \ref{eq:Ifit}).  }
\label{fig:profiles}
\end{figure*}

\begin{figure*}
\centering
\includegraphics[width=\textwidth]{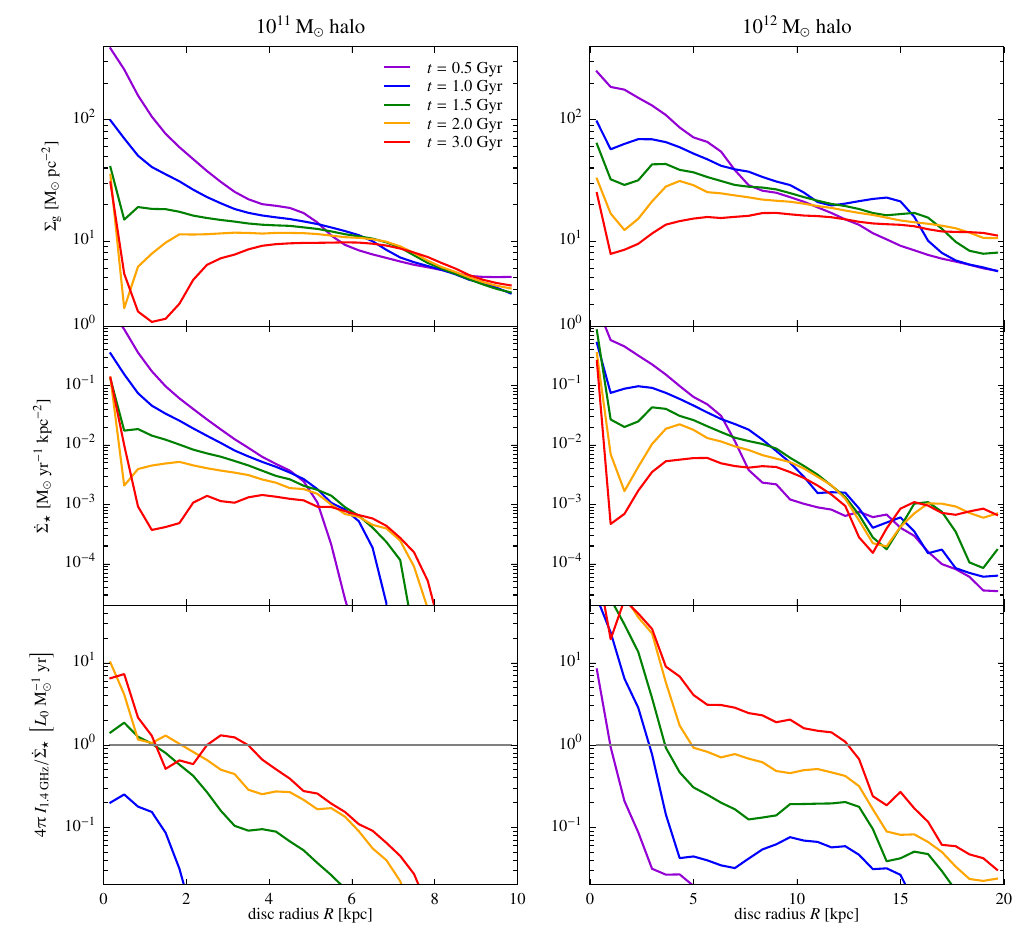}
\caption{Radial disc profiles of the surface mass density of gas (top), SFR
  surface density (middle) and the ratio of the radio synchrotron
  brightness-to-SFR surface density,
  $4\uppi\,I_{\rmn{1.4\,GHz}}/\dot{\Sigma}_\star$ (bottom), scaled to the mean
  radio flux $L_0$ of the observed FRC at an SFR of $1\,\msun\,\rmn{yr}^{-1}$
  (Eq.~\ref{eq:FIR-radio2}). We use the same simulations and snapshots as in
  Fig.~\ref{fig:profiles}.}
\label{fig:profiles2}
\end{figure*}

\begin{table*}
  \caption{Fitting parameters for the profiles of the disc magnetic field and
    the specific synchrotron intensity  as a function of disc radius
    $R$. The joint exponential profiles in Eqs.~\eqref{eq:Bfit} and
    \eqref{eq:Ifit} are characterised by a normalization ($B_0$ and $I_0$), a
    transition radius $r_{\rmn{tr}}$, and the scale radii of the inner and outer
    exponentials, $r_0$ and $r_1$.}
  \begin{center}
\begin{tabular}{cc | cccc | cccc}
\hline
\hline
 &  & \multicolumn{4}{c|}{magnetic profile $B (R)$} & \multicolumn{4}{c}{synchrotron intensity $I_{1.4\,\rmn{GHz}}(R)$} \\ 
 halo & simulation time & $B_0$ & $r_{\rmn{tr}}$ & $r_0$ & $r_1$ & $I_0$ & $r_{\rmn{tr}}$ & $r_0$ & $r_1$ \\
\phantom{\big|}%
[$\msun$] & [Gyr] & [$\upmu$G] & [kpc] & [kpc] & [kpc] & [mJy arcmin$^{-2}$] & [kpc] & [kpc] & [kpc] \\
\hline
$10^{11}$ & 0.5 &~~0.1 & 1.3 & 0.6 & 1.6    &~~$2.9 \times10^{-2}$ & 1.9 & 0.3 & 0.3 \\
$10^{11}$ & 1.0 & 15.5 & 2.5 & 0.4 & 1.6    & $7.4 \times10^{1}$ & 0.1 & 1.9 & 0.4 \\
$10^{11}$ & 1.5 & 24.5 & 2.7 & 0.7 & 1.4    & $9.6 \times10^{1}$ & 1.8 & 0.7 & 0.7 \\
$10^{11}$ & 2.0 & 24.0 & 0.9 & 0.5 & 1.7    & $1.2 \times10^{2}$ & 0.4 & 0.2 & 1.0 \\
$10^{11}$ & 3.0 & 20.4 & 1.1 & 0.6 & 2.1    & $1.3 \times10^{2}$ & 0.8 & 0.2 & 1.4 \\
\hline                                             
$10^{12}$ & 0.5 & 137.3 & 1.9 & 0.5 & 2.3   & $2.9 \times10^{4}$ & 1.7 & 0.3 & 1.0 \\
$10^{12}$ & 1.0 & 120.9 & 4.5 & 1.0 & 5.9   & $2.2 \times10^{4}$ & 4.2 & 0.5 & 1.9 \\
$10^{12}$ & 1.5 & 109.4 & 5.7 & 1.2 & 13.4~~& $1.3 \times10^{4}$ & 5.0 & 0.6 & 2.5 \\
$10^{12}$ & 2.0 &~~74.2 & 6.5 & 1.6 & 11.5~~& $2.9 \times10^{3}$ & 5.7 & 1.0 & 2.4 \\
$10^{12}$ & 3.0 &~~37.6 & 6.5 & 2.3 & 7.9   & $1.1 \times10^{4}$ & 1.8 & 0.4 & 2.4 \\
\hline
\end{tabular}
\end{center}
\label{tab:fit}
\end{table*} 

To study the emergence of magnetic fields and synchrotron emission, we show
average radial profiles of the CR electron spectral density $f_\e$ at 10~GeV, the
magnetic field strength $B$, and the synchrotron intensity at different times
in Fig.~\ref{fig:profiles} for our two haloes of $10^{11}$ and
$10^{12}~\msun$. The initial gravitationally driven collapse of gas implies
large central gas densities that ensues a vigorous starburst, as can be inferred
from the average radial profiles of the surface mass density of gas
$\Sigma_\rmn{g}$ and SFR surface density
$\dot{\Sigma}_\star$ at different times in Fig.~\ref{fig:profiles2}. The CR
electron spectrum mirrors this behaviour of the gas density and decreases at the
centre while it remain nearly constant at larger radii. By contrast, the
magnetic field strength grows with time in the outer disc. The growth rate is
faster in the $10^{12}~\msun$ halo in comparison to the smaller halo. The
synchrotron emissivity mirrors the behavior of $B$.

The radial profile of the disc magnetic field can be well fit with two
joint exponentials \citep{2017MNRAS.469.3185P} that describe $B$ in two regions
separated by a transition radius $r_{\rmn{tr}}$, i.e., a four-parameter fit defined by
\begin{equation}
  \label{eq:Bfit}
  B(R)=\left\{
  \begin{array}{ll}
    B_0\rmn{e}^{-R/r_0},&\rmn{if}~~R<r_{\rmn{tr}},\\
    B_0\rmn{e}^{-r_{\rmn{tr}}/r_0-(R-r_{\rmn{tr}})/r_1},&\rmn{if}~~R\geq{r_{\rmn{tr}}}.
  \end{array}\right.
\end{equation}
Here, $B_0$ is the central field strength, $r_0$ and $r_1$ are the scale radii
of the inner and outer exponentials. This functional form is well developed
after the dynamo has saturated (even before outflows are launched) and suggests
that two different amplification processes are responsible for growing magnetic
fields in the corresponding regions. In our Milky Way-sized galaxy, the
transition radius $r_{\rmn{tr}}$ grows with time, while we only observe this
behaviour at initial times (for $t\lesssim1.5$~Gyr) in the $10^{11}~\msun$ halo.
Not surprisingly, the radial profile of the specific synchrotron intensity can
also be well described by a joint exponential profile,
\begin{equation}
  \label{eq:Ifit}
  I_\nu(R)=\left\{
  \begin{array}{ll}
    I_0\rmn{e}^{-R/r_0},&\rmn{if}~~R<r_{\rmn{tr}},\\
    I_0\rmn{e}^{-r_{\rmn{tr}}/r_0-(R-r_{\rmn{tr}})/r_1},&\rmn{if}~~R\geq{r_{\rmn{tr}}}.
  \end{array}\right.
\end{equation}
where $I_0$ is the central synchrotron intensity and the other parameters have
the same meaning as in Eq.~\eqref{eq:Bfit}. In contrast to the magnetic profile,
the transition radius $r_{\rmn{tr}}$ of the synchrotron profile shows a
non-monotonic behaviour with time because the CR electron spectrum $f_\e$
obeys an opposite trend with time in comparison to $B$, i.e., the central
normalisation of $f_\e$ decreases with time. We summarise all fitting parameters
in Table~\ref{tab:fit}.

\subsection{Local FIR--radio correlation}
\label{sec:local_FRC}

In order to evaluate whether the FRC also holds on smaller scales within our
galaxies, we need to first look at the limits of our ISM subgrid-scale model. In
this model, star-forming gas is treated as a two phase medium of cold clouds
embedded in an ambient hot medium, which is treated with an effective equation
of state of a pressurised ISM and does not explicitly model molecular
clouds. Hence, we show maps and profiles of the ratio of the radio synchrotron
brightness-to-SFR surface density,
$4\uppi\,I_{\rmn{1.4\,GHz}}/\dot{\Sigma}_\star$ (right-hand panels of
Fig.~\ref{fig:maps2} and bottom panels of Fig.~\ref{fig:profiles2}). We scale
the radio emission to the mean radio flux $L_0$ of the observed FRC at a
fiducial SFR of $\dot{M}_{\star\,0}=1\,\msun\,\rmn{yr}^{-1}$
(Eq.~\ref{eq:FIR-radio2}). Hence, we expect a ratio of order unity at this
fiducial SFR and slightly larger (smaller) values at larger (smaller) SFRs due
to the non-linearity of the FRC with an FRC slope of $\alpha=1.055$
(Eq.~\ref{eq:FIR-radio}, \citealt{2003ApJ...586..794B}) or $\alpha=1.11$ in a
larger catalogue of 9,500 low-redshift galaxies
\citep{2021MNRAS.504..118M}. Thus, a 100 times larger (smaller) SFR implies a 65
per cent increase (decrease) relative to the fiducial FRC ratio
$L_0/\dot{M}_{\star\,0}$.

After saturation of the dynamo in our $10^{11}~\msun$ halo, we observe the
central few kpc to obey the local FRC. The centre of this galaxy shows an excess
of the local FRC ratio that mimics the observed complex behavior with local
extrema corresponding to various galactic structures, such as highly
star-forming spiral arms emphasizing the strong environmental dependence of the
thermal and non-thermal radio emission \citep[e.g., in M51,
  see][]{2011AJ....141...41D}.  These observations show a sub-linear FRC in the
low-density interarm and outer region and super-linear behaviour in the central
3.5~kpc which is also visible in the bottom right-hand panel of
Fig.~\ref{fig:maps2}. By contrast, the $10^{12}~\msun$ halo severely exceeds the
local FRC in the centre while it agrees within the scatter at late times and
radii $5 \lesssim R /\rmn{kpc}\lesssim 12$. This could either signal a
shortcoming in our star-formation and/or radio modelling or provide a hint that
the centres of starbursts exceed the FRC by a substantial amount.

\begin{figure*}
\centering
\includegraphics[width=\textwidth]{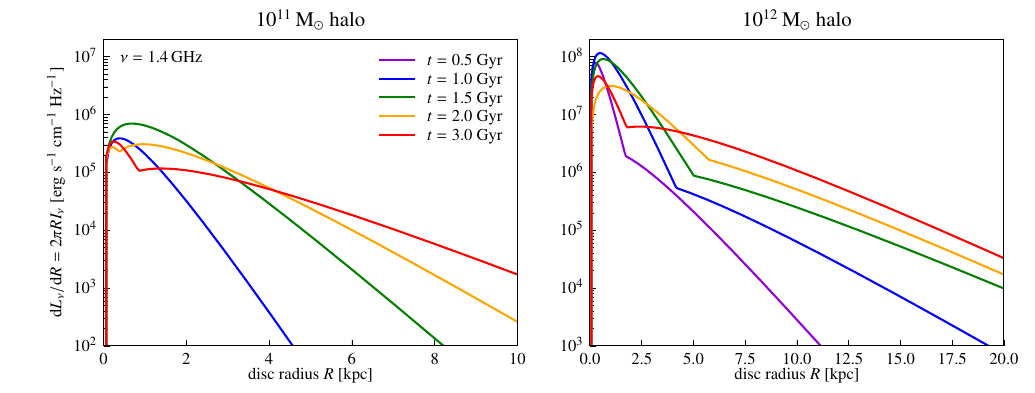}
\caption{Differential contribution to the total radio synchrotron luminosity per
  radius for the small galaxy ($M_{200}=10^{11}~\msun$, $c_{200}=12$, left
  column) and our Milky~Way-mass galaxy ($M_{200}=10^{12}~\msun$, $c_{200}=7$,
  right column) at different times (as indicated in the legend). We use the
  fitting functions of Eq.~\eqref{eq:Ifit}.}
\label{fig:dLnu_by_dr}
\end{figure*}

We observe that the local FRC decreases steeply towards large radii (see the
bottom panels of Fig.~\ref{fig:profiles2}). However, these radii show small
values of the surface brightness below $\upmu\rmn{Jy~arcmin}^{-2}$ (see the
bottom panels of Fig.~\ref{fig:profiles}) which are very challenging to
observe. Most importantly, these regions contribute negligibly to the total
radio synchrotron luminosity as can be inferred from their differential
contribution to the total radio synchrotron luminosity,
$\mathrm{d}L_\nu/\mathrm{d} R = 2\uppi R I_\nu~$ for our two haloes of $10^{11}$
and $10^{12}~\msun$ at different times in Fig.~\ref{fig:dLnu_by_dr}. One can
analytically show that $\mathrm{d}L_\nu/\mathrm{d} R$ for a single exponential
profile, $I_0\exp(-R/r_0)$, is maximsed at the scale radius. This changes for
our joint exponential profiles of Eq.~\eqref{eq:Ifit}: while the synchrotron
luminosity in the $10^{11}~\msun$ halo is dominated by the external exponential,
the inner exponential dominates in our Milky Way-mass galaxy at early times
$t\lesssim2$~Gyr because of the very steep central profile of the magnetic field
strength. At later times, the emission in outer parts of the disc increases
while it always has a subdominant contribution to the total synchrotron
luminosity.

\subsection{Hydrostatic pressure contribution by magnetic fields}
\label{sec:HSE}

\begin{figure*}
\centering
\includegraphics[width=\textwidth]{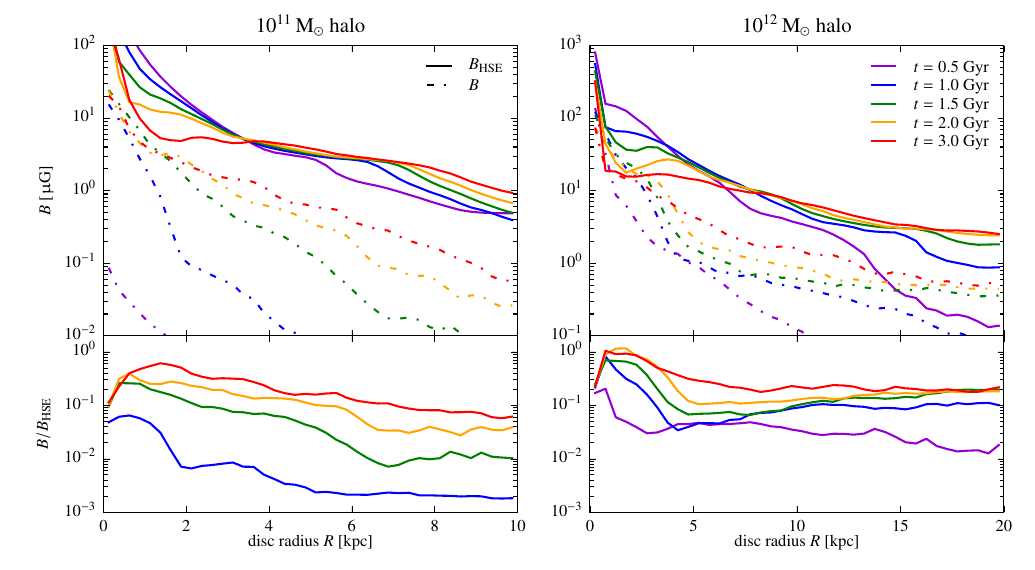}
\caption{The top panels show radial disc profiles of the volume-averaged
  magnetic field strength in a disc of total height 1~kpc (dash-dotted) and
  equivalent hydrostatic magnetic field (solid) that is required to balance the
  vertical disc gravity, $B_\rmn{HSE} = (8\uppi^2 G \Sigma_\rmn{g}
  \Sigma_\rmn{tot})^{1/2}$ for the small galaxy ($M_{200}=10^{11}~\msun$,
  $c_{200}=12$, left column) and our Milky~Way-mass galaxy
  ($M_{200}=10^{12}~\msun$, $c_{200}=7$, right column) at different times (as
  indicated in the legend). The bottom panels show the ratio $B/B_\rmn{HSE}$,
  which indicates that the magnetic field plays an important role in the overall
  pressure balance of the ISM in the central region but not at large disc
  radii. }
\label{fig:B_HSE}
\end{figure*}

This immediately poses the question whether the integrated magnetic pressure
provides a sizeable contribution to the ISM energy budget or whether it is even
able to aid in driving the outflows seen in our galaxies.  Assuming axisymmetry,
the gravitational field of the disc is given by $\Phi_{\rm grav}(R)\approx2\uppi
G\Sigma_{\rm tot}(R)$, where $\Sigma_\rmn{tot}$ is the total surface density of
stars and gas. This enables us to define a hydrostatic equivalent magnetic field
that balances the vertical gravitational force per unit area on the interstellar
gas via
\begin{align}
  \frac{B_\rmn{HSE}^2(R)}{8\uppi}
  &= \uppi G \Sigma_\rmn{g}(R) \Sigma_\rmn{tot}(R), \\
  B_\rmn{HSE} &\approx 15
  \left(\frac{\Sigma_\rmn{g}}{10\,\msun~\rmn{pc}^{-2}}\right)^{1/2}
  \left(\frac{\Sigma_\rmn{tot}}{100~\msun \rmn{pc}^{-2}}\right)^{1/2}\,\upmu\rmn{G}.
\end{align}
 Note that on circular orbits in the disc plane, the gravitational force due to
 the dark matter halo is exactly balanced by the centrifugal force. Hence, a
 field of strength $B_\rmn{HSE}$ would hydrostatically counteract the gravity of
 the disc while any smaller field strength would either signal a system that is
 out of equilibrium or require a thermal and/or turbulent pressure to counteract
 gravity.

In Fig.~\ref{fig:B_HSE}, we compare the profile of the magnetic field strength
to the equivalent hydrostatic magnetic field, $B_\rmn{HSE}$, that is required to
balance the vertical disc gravity for different times in our two fiducial
haloes. The $B_\rmn{HSE}$ profile shows that already at $t=0.5$~Gyr, the disc
mass has largely assembled up to disc radii of about 5~kpc (10~kpc) for our
$10^{11}$ ($10^{12}~\msun$) haloes so that the disc gravity barely changes at
these intermediate radii, but stars continue to form out of the available
gas. At later times, the disc gravity moderately grows at larger radii due to
gas accretion and decreases in the centre due to CR-driven galactic winds that
move gas from the disc into the circumgalactic medium. The bottom panels of
Fig.~\ref{fig:B_HSE} show the contribution of the magnetic field to counteract
disc gravity via the magnetic pressure force. Only at late times for radii
1--2~kpc does the magnetic field come close to $B_\rmn{HSE}$ but otherwise stays
subdominant, i.e., $B\lesssim B_{\mathrm{HE}}$. This implies that while magnetic
pressure can contribute significantly to the total pressure balance of the ISM,
it does not play an active role in driving disc winds. This can also be seen by
the comparison of pure MHD simulations of isolated galaxies that are contrasted
to simulations with purely advective CRs in figure~5 of
\citet{2017MNRAS.465.4500P}, which demonstrates that magnetic and CR pressures
(in the advective transport approximation) are not powerful enough in order to
drive galactic winds and hence, we require e.g., CR transport relative to the
gas for launching galactic winds. We caution that additional, supernova-driven
ISM turbulence could grow the magnetic field further and increase is
contribution to the overall energy balance.

\section{Discussion and conclusions}
\label{sec:conclusion}

In this work, we studied (i) the growth of the galactic seed magnetic field
through adiabatic compression and emergent small-scale dynamos driven by
turbulence resulting from gravitational collapse of the proto-galaxy and
velocity shear between the centrifugally supported galactic disc and hot
circumgalactic medium. Using this insight, we deconstructed (ii) the radio
synchrotron emission in three-dimensional MHD simulations of forming galaxies
that follow the CR energy density in the one-moment approximation with the goal
to understand the origin and physical processes underlying the FRC. Solving the
steady-state equation for CR protons, primary, and secondary electrons enables
us to reliably model the emergence of the radio emission from their formation to
the starburst appearance to the quiescently star-forming stage across a range of
galaxy masses from dwarfs to Milky Way sizes.

\subsection{Small-scale magnetic dynamo in galaxy formation}

To understand the emergence of galactic magnetic fields, we perform thorough
statistical studies of the time evolution of magnetic energy, $B$--$\rho$
correlations, power spectra, and statistics of magnetic curvature $\kappa$ and
curvature force density $f_\rmn{c}$, for various spatial regions (centre, disc
and entire halo), halo masses and at different characteristic times.

In particular, we carefully study the numerical Reynolds number of our
simulations, reaching values of $\rmn{Re}\sim3\times10^2$. We demonstrate that the
growth rate of the small-scale dynamo does not only agree with our theoretical
estimate and scales with the numerical grid size exactly as expected but also
saturates at the turbulent energy density. Thus, this provides quantitative
evidence that our employed numerical method of moving mesh magneto-hydrodynamics
with Powell cleaning  delivers accurate results.

We find that the magnetic energy density is exponentially amplified over ten or
fourteen orders of magnitude in the Milky Way-mass galaxy, depending on the
choice of our initial seed magnetic field. Initially, the magnetic field grows
via adiabatic compression, nearly independent of numerical resolution. In a
second phase, the small-scale dynamo emerges and grows the magnetic field
fastest at the highest resolution in the galaxy centre. The small-scale dynamo
starts earlier and the exponential growth rate is larger for increasing
resolution and halo masses. Initially, turbulence is injected at the corrugated
accretion shock, which propagates outwards. As the associated turbulence is
decaying, the large velocity shear between the supersonically rotating cool disc
with respect to the (partially) pressure-supported hot circumgalactic medium
excites Kelvin-Helmholtz surface and body modes, which non-linearly interact and
inject additional turbulence which continuously drives additional small-scale
dynamos.

While our model represents a simplified model for the formation of a disc
galaxy, we note that it exhibits all the essential ingredients, including a
rotationally supported star-forming disc that is embedded in a warm
circumgalactic medium, which has been thermalised by an accretion shock. In a
more realistic galaxy, the disc-halo interface may be more structured and
exhibit a smoother gradient. However, this only changes the spectrum and exact
growth rates of the Kelvin-Helmholtz unstable modes (and the associated
small-scale dynamo) and not the qualitative picture put forward here, as can be
seen by introducing a smooth transition layer between a cold, dense stream
moving at supersonic speed through the ambient hot, dilute circumgalactic medium
\citep{2019MNRAS.485..908B,2019MNRAS.489.3368B}.

Using curvature statistics, we can clearly demonstrate the superposition of
various small-scale dynamo modes at different densities and scales.  Once the
dynamo mode at the highest density comes into (approximate) equipartition with
the turbulent kinetic energy at that scale, magnetic power continues to grow on
large scales until saturation at the corresponding larger kinetic turbulent
energy, which increases the magnetic coherence scale with time to eventually
saturate its growth.  Magnetic saturation is independent of our adopted initial
magnetic field, which ranges from a homogeneous seed field to a magnetic dipole
field that follows a strength proportional to $\rho_\rmn{NFW}^{2/3}$. The latter
may result from the isotropic collapse of a proto-galaxy due to magnetic flux
freezing. We demonstrate that there exists a mapping between initial field
strength $B_\rmn{init}$ and the IGM strength $B_\rmn{IGM}$ of our pre-collapsed
model for the initial magnetic field so that the energy growth rates as well as
the $B$--$\rho$ correlations are nearly identical (although that mapping differs
for both set of comparisons).

In our model, gravitational collapse drives a starburst that quickly brings the
CR and thermal energy density into approximate equipartition.  We find that the
magnetic field saturates at a slower rate in smaller galaxies and does not reach
equipartition with the thermal and CR energy densities. Instead, its energy
density saturates with the gravo-turbulent energy density that we approximate
with the (subdominant) poloidal energy density and scales according to
$E_B\propto M_{200}^{5/3}$.  Interestingly, the rotational kinetic energy
density in these centrifugally-dominated discs is about 100 times larger than the
`turbulent' kinetic energy density. The emerging magnetic pressure balances the
vertical disc gravity in the centre but not at large radii. However, we
anticipate that cosmological simulations of dwarf galaxies with an improved ISM
model \citep[e.g.,][]{2021MNRAS.501.5597G} in combination with a two-moment CR
hydrodynamic scheme \citep[e.g.,][]{2019MNRAS.485.2977T} can further amplify the
magnetic field strength in these systems via supernova- and cosmic
accretion-driven turbulence as well as modulate the impact of CR transport on
the magnetic dynamo amplification with a more consistent CR diffusion coefficient
in the self-confinement picture.

We show analytically, that the largest scales in the magnetic power spectrum are
dominated by the exponential profile of the magnetic field in the galaxy centre
and cascades down in scale consistent with the prediction from MHD turbulence.
After excluding the central region where the magnetic field plays an important
role in the overall pressure balance, we show that the magnetic growth in the
kinematic phase follows a \citet{Kazantsev1968} spectrum. In agreement with
simulations of the fluctuating small-scale dynamo \citep{2005PhR...417....1B},
the Kazantsev spectrum turns over at the scale where the magnetic and turbulent
energy have achieved equipartition so that the back-reaction exerted by the
magnetic tension force suppresses the stretching process of eddies that are
smaller than the equipartition scale. MHD turbulence imposes a spectrum on
smaller scales, which follows a \citet{Kolmogorov1941} slope.

\subsection{The correlation of radio synchrotron and FIR emission}

We find that the total radio synchrotron emission of the Milky~Way-mass galaxy
is dominated by the disc and a bright bulge that results in a CR-driven,
magnetically-loaded outflow which is visible in form of faint radio bubbles.  On
the contrary, the edge-on view of the smaller galaxy with halo mass
$10^{11}~\msun$ shows a characteristic X-shaped radio morphology in the outflow
region, which results from combining the cylindrical magnetic outflow geometry
with the CR electron distribution that shows bi-conical, low-density cavities.

We show that quiescently star-forming and star-bursting galaxies with a
saturated small-scale dynamo are on the FRC while galaxies with quenched
dynamos---that in our simulations are realised by a strong CR-driven outflow in
combination with a single starburst event caused by a one-time accretion phase
and turbulent driving---fail to reach the magnetic energy that is
necessary to reach the FRC. The intrinsic scatter in the FRC arises due to the
following effects.
\begin{enumerate}
\item The radio luminosity varies due to somewhat different magnetic saturation
  levels that result from different seed magnetic fields at the formation time
  or a different formation history and magnetic dynamo efficiencies. In
  addition, intermittent galactic wind velocities modulate the magnetic and CR
  energy in the inner galactic regions that are affected by the wind and
  temporarily change the resulting synchrotron luminosity.

\item Different galaxy (and stellar) masses at a given SFR: the gas accretion
  rate decreases after the starburst phases and so does the SFR.  In
  consequence, there are two effects that decrease the radio synchrotron
  luminosity at a smaller rate in comparison to the SFR. First, smaller SFRs
  imply smaller gas densities and smaller collisional loss rates so that the
  calorimetric energy fraction emitted via radio synchrotron increases. Second,
  the amount of injected CR energy decreases as the SFR drops but the magnetic
  field strengths are maintained by a dynamo action at a constant level.  As a
  result, a galaxy is not exactly evolving along the relation towards smaller
  FIR flux but instead progresses towards the upper envelope of the
  scatter. Thus, at a given SFR, our model predicts less massive systems with
  high specific SFRs to populate the lower envelope of the FRC in comparison to
  more massive systems that are in a late evolutionary phase and show small
  specific SFRs.
    
\item A varying radio intensity with galactic inclination in combination with an
  anisotropic magnetic field distribution: while the face-on radio luminosity
  reaches about twice the value of the edge-on luminosity for toroidal field
  configurations (in quiescent star-forming discs), the edge-on luminosity
  becomes significant for substantial poloidal configurations (in strong
  outflows).
\end{enumerate}

Our simulations show that different CR transport models modulate the magnetic
dynamo efficiency and the CR electron density via CR-driven winds. Thus, an
increasing (constant) CR diffusion coefficient moves galaxies to the lower
envelope of the {\it simulated} FRC. It remains to be seen whether this effect
contributes to the scatter in the {\it observed} FRC and whether there is a
range of effective CR transport speeds at a fixed SFR.

Identifying the underlying physical processes responsible for the scatter in the
FRC opens up the possibility of constructing a {\em non-thermal fundamental
  plane} of star-forming galaxies. Adding a third parameter (or a combination of
parameters) in addition to the FIR and radio luminosities might enable to
construct a lower-dimensional manifold relative to which the scatter can be
reduced, thus enabling more precise SFR estimates with radio data. Our
simulations suggest that observables that are sensitive to the specific SFR as
well as the galactic inclination are prime candidates for such a third
parameter (combination).

In agreement with observations, regions of several 100 pc within star-forming
galaxies broadly obey the FRC out to disc radii, which contribute significantly
to the radio emission.  We find that galaxy centres show an excess of the local
FRC ratio that mimics the observed complex behaviour with local extrema
corresponding to highly star-forming galactic structures.  These observations
show a sub-linear FRC in the low-density interarm and outer region and
super-linear behaviour in galaxy centres. Interestingly, the central parts of
starbursts exceed the FRC by a substantial amount.

Future work that follows the CR proton and electron spectra in space and time is
needed to consolidate these findings that are based on a steady-state
assumption. In order to more realistically model the magnetic dynamo, future
work also needs to explicitly model the supernova-driven ISM turbulence
that yields a multi-phase medium. Nevertheless, we believe that we identified
the critical physical processes that cause the emergence of radio emission in
galaxies and the origin of the FRC. This will pave the road for using
observations of non-thermal radio emission to learn about the impact of CR
feedback in galaxy formation.

\section*{Acknowledgements}
We thank Joe Whittingham, Timon Thomas, Mohamad Shalaby and Anvar Shukurov for
useful discussions, as well as Rainer Beck and two anonymous referees for
comments on the original manuscript. CP, MW, and PG acknowledge support by the
European Research Council under ERC-CoG grant CRAGSMAN-646955 and ERC-AdG grant
PICOGAL-101019746. This research was in part supported by the Munich Institute
for Astro- and Particle Physics (MIAPP) which is funded by the Deutsche
Forschungsgemeinschaft (DFG, German Research Foundation) under Germany's
Excellence Strategy – EXC-2094 – 390783311.

\section*{Data Availability}
The data underlying this article will be shared on reasonable request to the corresponding author.

\bibliographystyle{mnras}
\bibliography{bibtex/paper}

\appendix

\section{Supporting material for discussions of the small-scale dynamo}
\label{sec:app_dynamo}

In this appendix, we provide additional supporting material for the discussion
of the small-scale dynamo in Sections~\ref{sec:growth_analysis} and
\ref{sec:dynamo} by (i) deriving the exponential growth rate of the small-scale
dynamo in ideal MHD, (ii) scrutinising the time evolution of the numerical
resolution in our high-resolution simulations, which is required to estimate the
effective Reynolds numbers in our simulations, and (iii) by analysing the radial
profile of the magnetic field strength $B$ during the initial stages of
exponential growth.

\subsection{Magnetic growth rate in Kolmogorov turbulence}
\label{sec:app_Gamma_scaling}

Assuming incompressible turbulence that is driven at the outer scale
$\mathscr{L}$ with a velocity $\mathscr{V}$, we define the Kolmogorov length
$\ell$ (where kinetic energy is dissipated with a kinetic viscosity
$\nu_\rmn{vis}$):
\begin{equation}
\ell\equiv\left(\frac{\nu_{\rmn{vis}}^3}{\dot\epsilon}\right)^{1/4},
\quad{\rm where}\quad
\dot\epsilon=\frac{\varv_\lambda^3}{\lambda}=\frac{\mathscr{V}^3}{\mathscr{L}}
\end{equation}
is the energy flow rate per unit mass that is valid on all scales $\lambda$ and
constant (because energy does not accumulate at any intermediate scale). Hence,
we can also identify $\dot\epsilon=\mathscr{V}^3/\mathscr{L}$ with the energy
injection rate into the turbulent cascade at scale $\mathscr{L}$. We obtain the
corresponding velocity and time-scales at the Kolmogorov length $\ell$,
\begin{align}
\label{eq:v_scaling}
\varv_\ell&=\left(\dot\epsilon\ell\right)^{1/3}
=\left(\dot\epsilon\nu_{\rmn{vis}}\right)^{1/4}
=\mathscr{V}\,\rmn{Re}^{-1/4},\\
\label{eq:tau_scaling}
\tau_\ell&=\frac{\ell}{\varv _\ell}
=\left(\frac{\nu_{\rmn{vis}}}{\dot\epsilon}\right)^{1/2}
=\tau\,\rmn{Re}^{-1/2},
\end{align}
where we have used the definition of the Reynolds number $\rmn{Re}$
(Eq.~\ref{eq:Reynolds}) and defined the eddy turnover time at the outer scale,
$\tau=\mathscr{L}/\mathscr{V}$.  Hence, we obtain the scaling of the outer to
inner eddy turnover timescale or -- equivalently -- the inverse growth rates,
\begin{align}
  \frac{\Gamma_\ell}{\mathscr{V}/\mathscr{L}}=\frac{\tau}{\tau_\ell}
  =\rmn{Re}^{1/2}.
\end{align}
In a small-scale dynamo, the magnetic field grows fastest at the resistive scale
\citep{2004ApJ...612..276S,2005PhR...417....1B}, which corresponds to the
Kolmogorov scale for a magnetic Prandtl number of unity. We identify this scale
with the local grid scale in ideal MHD (where the explicit kinematic viscosity
and resistivity are neglected, i.e.\ $\nu_{\rmn{vis}}=\eta_{\rmn{res}}=0$ and
are replaced by the numerical counterparts that operate on the grid
scale). Because our moving Voronoi mesh dynamically adjust the resolution in a
quasi-Lagrangian fashion (see Fig.~\ref{fig:d_cell}), we expect the small-scale
dynamo to grow fastest in the highest density regions in the galactic centre.

\subsection{Numerical resolution of the moving mesh}

\begin{figure*}
\centering
\includegraphics[width=0.495\textwidth]{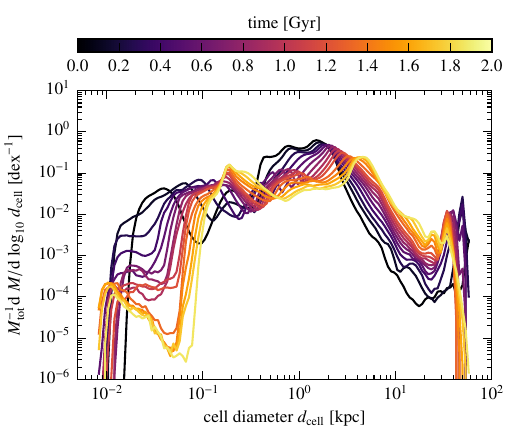}
\includegraphics[width=0.495\textwidth]{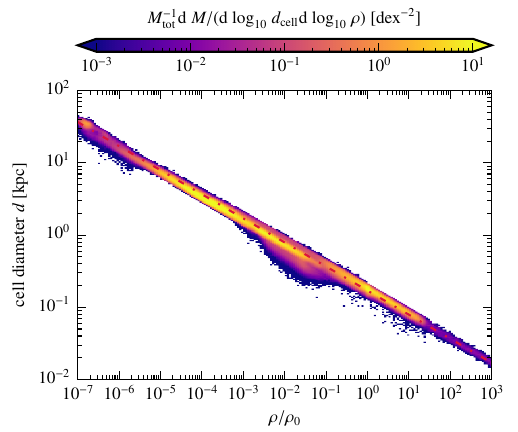}
\caption{Quantifying the numerical resolution in our simulation of a
  $10^{12}~\rmn{M}_\odot$ halo with concentration parameter $c_{200}=7$ and
  initially $10^7$ cells within the virial radius.  On the left-hand side, we
  show the mass-weighted probability density of cell diameters, $d_\rmn{cell}=
  (6V/\uppi)^{1/3}$ (assuming a spherical cell volume) for different times,
  initially spaced by 100~Myrs, and after 1~Gr, spaced by 200~Myr.  On the
  right-hand side, we show the mass-weighted probability density of resolution
  elements in the $d_\rmn{cell}$--$\rho$ plane (at $t=1$~Gyr), demonstrating the
  (almost) Lagrangian nature of the code that is manifested in the relation
  $d_\rmn{cell}\propto\rho^{-1/3}$ (the exact expression for $d_\rmn{cell}$ used
  in the simulation is shown with the red dash-dotted line). The density is
  scaled to the star formation threshold $\rho_0 = 4.05\times
  10^{-25}\,\rmn{g~cm}^{-3}$. The improved resolution at scaled densities
  $\gtrsim\rho/\rho_0=10^{-3}$ is due to the refinement criterion that ensures
  that the volume of adjacent Voronoi cells differs at most by a factor of ten.}
\label{fig:d_cell}
\end{figure*}

Recent adaptive mesh-refinement simulations of magnetic field growth in clusters
and galaxies have opted for a quasi-uniform maximum refinement level in the
high-density regions of interest
\citep{2016MNRAS.457.1722R,2018MNRAS.474.1672V,2022MNRAS.513.3326M}, which
facilitates the numerical dissipation properties and enables to quote a reliable
numerical Reynolds number of the flow. Because the numerical truncation error is
proportional to the sum of the absolute values of sound speed and gas velocity,
the schemes require an enormous resolution to resolve a small-scale dynamo
\citep{2022MNRAS.513.3326M}, which is otherwise quenched by numerical
dissipation. There are two alternatives to nevertheless resolve a small-scale
dynamo: (i) increasing the integral scales $\mathscr{L}$ and $\mathscr{V}$
through feedback \citep{2016MNRAS.457.1722R,2017MNRAS.471.2674R}, or (ii)
increasing the effective resolution by introducing a turbulent subgrid scheme,
which enables resolving the growth of a small-scale dynamo below the formal grid
resolution that would otherwise be numerically dissipated
\citep{2022MNRAS.513.6028L}.

On a quasi-Lagrangian moving mesh employed here, there is no explicit control of
the effective Reynolds number. Instead, the cell diameter $d_\rmn{cell}=
(6V/\uppi)^{1/3}= [6M/(\uppi\rho)]^{1/3}$ depends on the gas density $\rho$ and
target gas mass $M$ that the code keeps constant (up to a small tolerance
interval). As a result, at a given spatial resolution there is a smaller
truncation error in comparison to spatially-fixed adaptive mesh refinement codes
because the dominating advection error relative to the grid is significantly
reduced. Increasing the velocity of the mesh-generating points would cause
non-Lagrangian motion of the mesh and hence result in an increased truncation
error.

Figure~\ref{fig:d_cell} shows the numerical resolution in our high-resolution
Milky Way-like simulation. The mass-weighted probability density shows a wide
spectrum of cell sizes down to a minimum diameter of 10~pc. The two-dimensional
probability density of cell size and gas density demonstrates the (almost)
Lagrangian nature of the code that is manifested in the relation
$d_\rmn{cell}\propto\rho^{-1/3}$. Using our insight that the dynamo grows
fastest in the centre at the highest densities (Figs.~\ref{fig:B-K_evolution}
and \ref{fig:B-K_regions}), we conclude that the small-scale dynamo in the
centre grows with an effective numerical resolution of 10~pc, which should be
the cell size entering our estimate of the effective numerical Reynolds number
in Section~\ref{sec:growth_analysis}.

\subsection{Profile of the magnetic field strength at early times}
\label{sec:B_profile}

\begin{figure}
\centering
\includegraphics[width=\columnwidth]{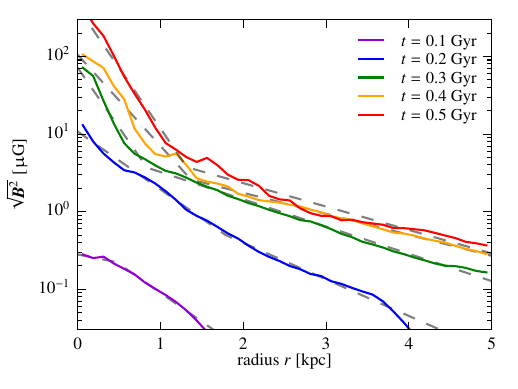}
\caption{Radial profiles of magnetic field strength during the initial stages of
  exponential growth for different time (colour-coded). The profiles are well
  fit by a joint exponential profile (dashed lines) of Eq.~\eqref{eq:Bfit2}.}
\label{fig:B_3D}
\end{figure}

\begin{table}
  \caption{Fitting parameters for the radial magnetic profiles $B(r)$ of the
    simulation of the $10^{12}~\msun$ halo with concentration $c_{200}=7$, and
    the `CR diff' model of Fig.~\ref{fig:B_3D}. The joint exponential profile in
    Eq.~\eqref{eq:Bfit2} is characterised by a normalisation $B_0$, a transition
    radius $r_{\rmn{tr}}$, and the scale radii of the inner and outer
    exponentials, $r_0$ and $r_1$.}
  \begin{center}
\begin{tabular}{c | cccc }
\hline
\hline
simulation time & $B_0$ & $r_{\rmn{tr}}$ & $r_0$ & $r_1$ \\
\phantom{\big|}%
[Gyr] & [$\upmu$G] & [kpc] & [kpc] & [kpc] \\
\hline
0.1 &~~~~0.3~ & 0.5 & 2.1 & 0.6 \\
0.2 & ~~10.8~ & 1.8 & 0.6 & 0.9 \\
0.3 & ~~71.6~ & 0.8 & 0.3 & 1.2 \\
0.4 &  105.4~ & 1.4 & 0.4 & 1.6 \\
0.5 &  430.2~ & 1.3 & 0.3 & 1.4 \\
\hline
\end{tabular}
\end{center}
\label{tab:fit2}
\end{table} 

Figure~\ref{fig:B_3D} shows the radial profiles of $B$ at different times
separated by 0.1~Gyr. We see the development of a steep inner and a more
shallower outer profile after $t\approx0.2$~Gyr. Hence, we fit the
three-dimensional radial profile of the magnetic field with two joint
exponentials that describe $B$ in two regions separated by a transition radius
$r_{\rmn{tr}}$, i.e., a four-parameter fit defined by
\begin{equation}
  \label{eq:Bfit2}
  B(r)=\left\{
  \begin{array}{ll}
    B_0\rmn{e}^{-r/r_0},&\rmn{if}~~r<r_{\rmn{tr}},\\
    B_0\rmn{e}^{-r_{\rmn{tr}}/r_0-(r-r_{\rmn{tr}})/r_1},&\rmn{if}~~r\geq{r_{\rmn{tr}}}.
  \end{array}\right.
\end{equation}
All fitting parameters are given in Table~\ref{tab:fit2}. Note that at
$t\gtrsim0.2$~Gyr the inner scale radius is rather constant, $r_0\sim0.3$ to
0.4~kpc.

Note that angular momentum conservation causes the disc to flatten and hence
challenges the simplified picture of a spherically symmetric magnetic field
profile (see Fig.~\ref{fig:turbulence}). While the central region exhibits
quasi-spherical density and magnetic profiles (especially at early times), after
the formation of a galactic disc, turbulence on scales larger than the disc
scale height becomes anisotropic. We postpone a more detailed analysis of the
effects of anisotropy on the small-scale dynamo in the disc and disc-halo
interface to future work and note that instead of using a power spectrum
analysis, this would likely require adopting azimuthally dependent velocity
structure functions.

\section{Comparing different magnetic initial conditions}
\label{sec:B_init}

\begin{figure*}
\centering
\begin{minipage}{0.415
    \textwidth}
  \vspace{.5em}
\includegraphics[width=\textwidth]{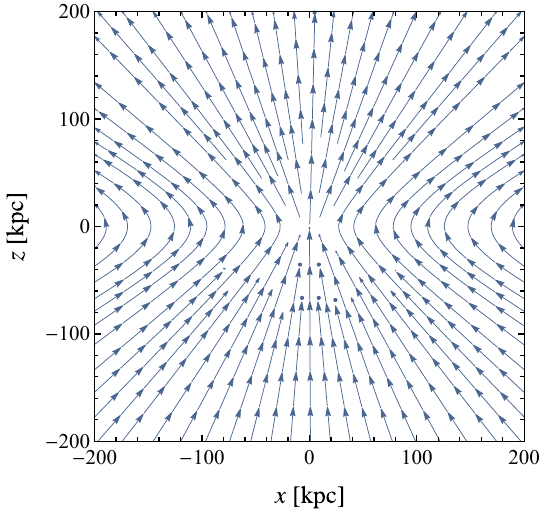}
\end{minipage}
\begin{minipage}{0.55\textwidth}
\includegraphics[width=\textwidth]{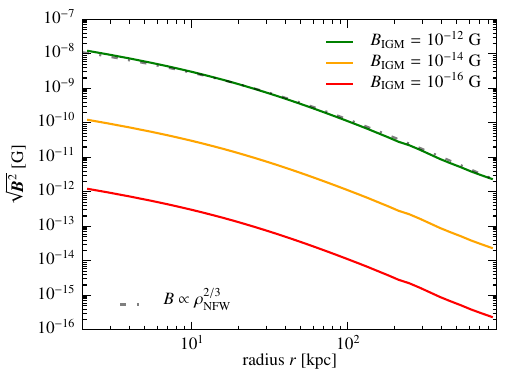}
\end{minipage}
\caption{Initial magnetic field distribution in our flux-freezing model with a
  dipole-like topology (left-hand panel). We superpose the individual magnetic
  dipoles so that the azimuthally averaged field strength is proportional to
  $\rho_\rmn{NFW}^{2/3}$ (right-hand panel). We simulate three models of our
  $10^{12}\,\rmn{M}_\odot$ halo (with virial radius $r_{200}\approx206$~kpc)
  that differ in their assumed magnetic field strength of the IGM,
  $B_\rmn{IGM}$.}
\label{fig:B_dipole}
\end{figure*}

\begin{figure}
\includegraphics[width=\columnwidth]{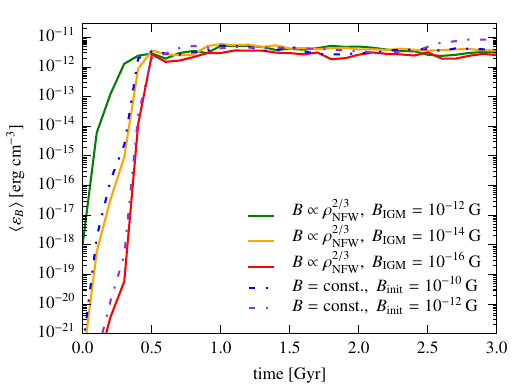}
\caption{Time evolution of volume-averaged magnetic energy densities in a disc
  of radius 10~kpc and total height 1~kpc of a $10^{12}~\msun$ halo with
  concentration $c_{200}=7$ and initially $10^6$ cells within the virial
  radius. We compare simulations with different initial magnetic field
  configurations: our flux-freezing models with a dipole-like magnetic topology,
  which scale with the gas density (solid lines), are confronted to simulations
  with an initially constant magnetic field (dash-dotted). This shows that the
  small-scale dynamo can more easily amplify the pre-compressed magnetic field
  so that the exponential dynamo growth rate of a constant initial magnetic
  field is equivalent to a pre-compressed field with a $10^4$ times smaller IGM
  field strength.}
\label{fig:B_evol}
\end{figure}

\begin{figure*}
\centering
\includegraphics[width=\textwidth]{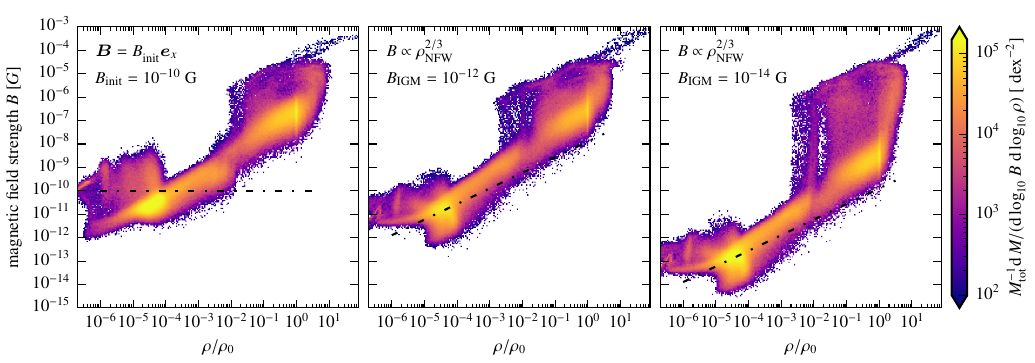}
\caption{Mass-weighted probability density of magnetic field strength $B$ and
  mass density, $\rho$, in the galactic halo for three different initial
  magnetic field models at $t=2$~Gyr. All gas densities are scaled to the star
  formation threshold $\rho_0 = 4.05\times 10^{-25}\,\rmn{g~cm}^{-3}$. From left
  to right, we show the constant field model with $B_\rmn{init}=10^{-10}$~G, and
  two flux-freezing models with $B_\rmn{IGM}=10^{-12}$ and $10^{-14}$~G. The
  initial distributions are shown with dash-dotted lines and we use simulations
  of a $10^{12}~\msun$ halo that initially have $10^6$ cells within the virial
  radius. Note that the distributions in the first two panels are very similar.}
\label{fig:B_rho}
\end{figure*}

In order to study the dependence of our conclusions on the choice of the initial
magnetic field distribution, we consider two types of configurations. First, we
adopt a constant initial magnetic field $\mathbfit{B} = B_\rmn{init}
\mathbfit{e}_x$, where $B_\rmn{init}=10^{-10}$ and $10^{-12}$~G. Second, if the
initial halo field results from isotropic collapse, magnetic flux freezing
implies a scaling $B\propto \rho^{2/3}$. We preserve the constraint
$\bnabla\bcdot\mathbfit{B}=0$ by deriving the magnetic field from its vector
potential $\mathbfit{A}$. Inspired by \citet{2010A&A...523A..72D}, our initial
magnetic field distribution is a superposition of small magnetic dipoles that
are aligned with the $z$ axis and have a strength proportional to
$\rho^{2/3}$ so that they add up to a global large scale dipole-like
topology on the halo scale albeit with a different radial behaviour.

For our flux-freezing initial conditions, we assume that gas and dark matter
trace the NFW profile \citep{1997ApJ...490..493N}, which we slightly soften at
the centre (below 0.1~kpc) to introduce a core into the gas, allowing for a
``quiet'' start of the simulations.  Outside, the gas mass density is given by
\begin{align}
  \label{eq:rho}
  \rho(r)=\rho_\rmn{NFW}(r)\frac{\Omega_\rmn{b}}{\Omega_\rmn{m}},
\end{align}
where $\Omega_\rmn{b}$ and $\Omega_\rmn{m}$ are the baryon and total mass
density in units of the critical density for geometrical closure of the
universe, $\rho_\rmn{cr}$, and the NFW profile is given by
\begin{align}
  \label{eq:NFW}
  \rho_\rmn{NFW}(r)&=\frac{\delta_\rmn{c}\rho_\rmn{cr}}{r/r_\rmn{s} ( 1 + r/r_\rmn{s} )^2},\\
  \delta_\rmn{c}&=\frac{200}{3}\,c_{200}^3
  \left[\ln\left(1+c_{200}\right) - \frac{c_{200}}{1+c_{200}}\right]^{-1},\\
  r_\rmn{s}&=\frac{r_{200}}{c_\rmn{200}}
  =\left(\frac{G M_{200}}{c_\rmn{200}^3 \,100 H^2(a)}\right)^{1/3},
\end{align}
where $r=\sqrt{x^2+y^2+z^2}$, and the Hubble function at the cosmic scale factor
$a$ is $H(a)=H_0\sqrt{\Omega_\Lambda+\Omega_\rmn{m} a^{-3}}$, where $H_0$ is the
current day value and $\Omega_\Lambda$ is the density parameter of the
cosmological constant. Using the mean density of the IGM,
$\rho_\rmn{IGM}=\rho_\rmn{cr}\Omega_\rmn{b}$, we can write down the vector
potential,
\begin{align}
  \label{eq:A}
  \mathbfit{A} = B_\rmn{IGM}\,
  \left[\frac{\rho(r)}{\rho_\rmn{IGM}}\right]^{2/3}
  \left(
  \begin{array}{l}
    -y \\
    \,~~x \\
    \,~~0
  \end{array}\right).
\end{align}
The solenoidal (divergence-free) magnetic field reads
\begin{align}
  \label{eq:B}
  \mathbfit{B}&=\bnabla\btimes\mathbfit{A}\\
  &= \frac{2B_\rmn{IGM}}{3 r^{8/3} (r+r_\rmn{s})}
  \left[\frac{r_\rmn{s}^3\,\delta_\rmn{c}\,\Omega_\rmn{m}^{-1}}{(r+r_\rmn{s})^2}\right]^{2/3}
  \left(
  \begin{array}{l}
    xz ( 3r+r_\rmn{s} ) \\
    yz ( 3r+r_\rmn{s} ) \\
    3rz^2 + (2 r^2 + z^2)r_\rmn{s}
  \end{array}\right),
\end{align}
which depends on the parameters $B_\rmn{IGM}$, $M_{200}$, and $c_{200}$ (see
Table~\ref{tab:simulations-overview} for our choices), and we adopt the
currently favoured concordance cosmological model at the present time with
$\Omega_\rmn{m}=0.315$, $\Omega_\Lambda=0.685$, and
$H_0=70~\rmn{km~s}^{-1}~\rmn{Mpc}^{-1}$.

Figure~\ref{fig:B_dipole} demonstrates that the initial magnetic field topology
resembles a dipole-like structure that is aligned with the $z$ axis (left-hand
panel) and that the azimuthally averaged field strength decreases with radius as
$\rho_\rmn{NFW}^{2/3}$ (right-hand panel). Defining the virial radius $r_{200}$
such that the average density contained in the spherical volume
$4/3{\upi}r_{200}^3$ is 200 times the critical density of the universe, we get
$r_{200}\approx206$~kpc for our $10^{12}\,\rmn{M}_\odot$ halo with a
concentration parameter $c_{200}=7$. Hence, the average magnetic profile
approaches $B_\rmn{IGM}$ only beyond $5 r_{200}$ and assumes average values of
approximately $30\,B_\rmn{IGM}$ at the virial radius and $10^4\,B_\rmn{IGM}$ at
$0.01 r_{200}$.

In Fig.~\ref{fig:B_evol}, we compare the time evolution of volume-averaged
magnetic energy densities in the gaseous galactic discs for different initial
magnetic configurations. While the models differ in their exponential growth
rate, they all saturate at the same level in equipartition with the turbulent
energy density, as discussed in Section~\ref{sec:mag_sat}. The growth rate of
the small-scale dynamo increases with a larger initial magnetic field strengths
and the dynamo can more easily amplify the pre-compressed magnetic field of our
flux-freezing model. A comparison of the different field configurations shows
that the exponential dynamo growth rate of a constant initial magnetic field is
equivalent to a pre-compressed field with a $10^4$ times smaller IGM field
strength. Thus, this empirical comparison suggests that the magnetic field
strength at around $0.01 r_{200}$ is responsible for setting the dynamo growth
rate in our moving-mesh finite-volume hydrodynamics. More work is needed to
explore the full dependencies of this finding on the numerical method and exact
magnetic configuration.

Figure~\ref{fig:B_rho} shows the distribution of the gas mass in the magnetic
field strength-mass density plane for three different initial magnetic field
models at $t=2$~Gyr, i.e., an epoch well into the saturated regime of the
dynamo. There are two branches visible: a lower branch that reflects the halo
field with approximately $B\propto \rho$ and an upper branch characterising the
disc field, which dominates the total magnetic energy. Those branches are
connected by a vertical bridge across which the dynamo exponentially amplifies
the magnetic field, which is even more clearly visible in the panel to the
right-hand side. As expected, the upper branch with the disc field is similar in
all three models, which is a consequence of the equipartition magnetic field in
the saturated dynamo state.

Interestingly, the distributions of the constant initial magnetic field with
$B_\rmn{init}=10^{-10}$~G and the flux-freezing model with
$B_\rmn{IGM}=10^{-12}$~G very closely resemble each other, while the model on
the right-hand side shows a halo magnetic field that is lower by a factor of
100. We have seen that the dynamo growth rates of the models on the left- and
right-hand panels of Fig.~\ref{fig:B_rho} are equal. This demonstrates while
there is no exact one-to-one mapping of constant initial field and flux-freezing
models, we can find close resemblances of specific properties in both models.
Most importantly, the mapping between the magnetic growth rates of the two
models (Fig.~\ref{fig:B_evol}) and our power spectrum analysis in
Section~\ref{sec:dynamo} show that most of the growth of the magnetic field is
caused by a small-scale dynamo and not by adiabatic compression as a result of
magnetic flux-freezing.

\section{Spectral resolution study}
\label{sec:resolution}

\begin{figure*}
  \includegraphics[width=\textwidth]{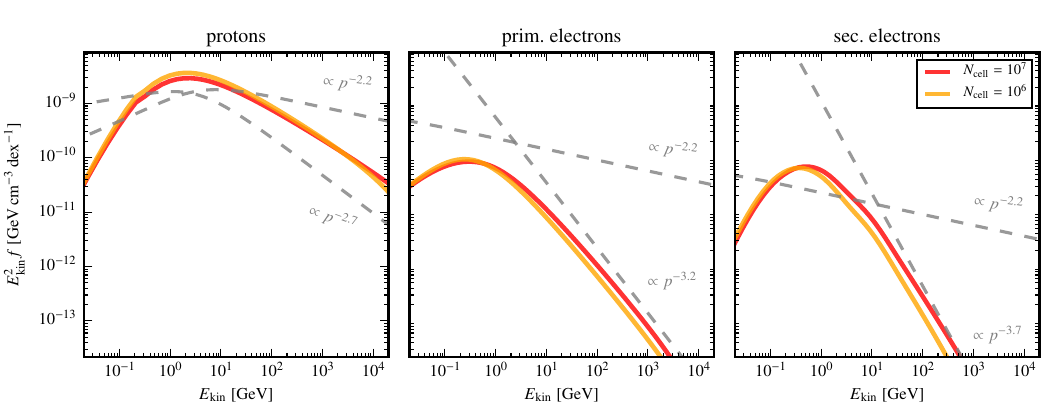}
\caption{Resolution study of steady-state CR spectra in the galactic disc of our
  simulation of a $10^{12}~\msun$ halo with concentration parameter $c_{200}=7$
  at $t=1$~Gyr.  The CR spectra $E_{\rmn{kin}}^2 f = (\ln 10)^{-1} E_{\rmn{kin}}
  \dd N / (\dd \log_{10} E_{\rmn{kin}}\,\dd V)$ are volume averaged over a
  cylinder of a radius that includes 99 per cent of the total radio luminosity
  and a height above and below the mid-plane that is equal to the scale-height
  of the gas density. We compare simulations with initially $10^6$ and $10^7$
  Voronoi cells within the virial radius and show from left to right the scaled
  kinetic energy spectra of CR protons, primary electrons and secondary
  electrons and positrons. The dashed lines represent pure power-law momentum
  spectra with the indicated momentum spectral indices for comparison.}
\label{fig:CR_spectra}
\end{figure*}

\begin{figure}
  \includegraphics[width=\columnwidth]{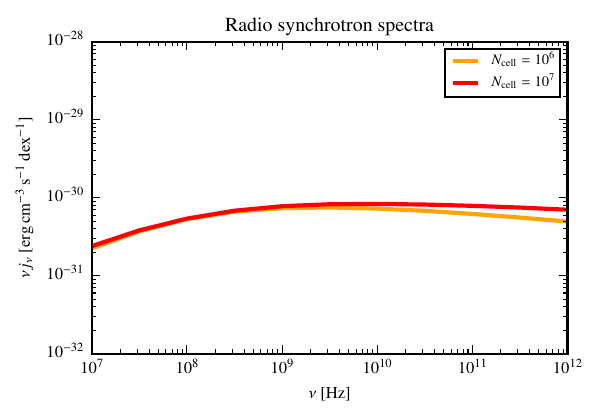}
\caption{Resolution study of total radio synchrotron spectra of our simulation
  of a $10^{12}~\msun$ halo with concentration parameter $c_{200}=7$ at
  $t=1$~Gyr. The total radio spectrum $\nu j_\nu = (\ln 10)^{-1} \dd N_\gamma /
  (\dd \log_{10} \nu\,\dd V\dd t)$ is dominated by primary CR electrons and is
  numerically fully converged at energies $\lesssim100$~GHz.}
\label{fig:syn_spectra}
\end{figure}

The radio luminosity depends on the magnetic field strengths and CR spectra. The
magnetic energy saturates early on in the galactic evolution in equilibrium with
the turbulent kinetic energy and is numerically well converged. In
Fig.~\ref{fig:CR_spectra}, wee assess the numerical convergence of our
steady-state spectra of CR protons, primary and secondary electrons. While the
low-energy part ($E_\rmn{kin}\lesssim10$~GeV) of the CR spectra are well
converged, our low-resolution spectra fall slightly short of the high-resolution
analogues at larger energies. Because the total radio spectra in our `CR~diff'
models are dominated by primary CR electrons \citep{2021WerhahnIII}, which are
well converged at kinetic energies $\lesssim10$~GeV, we find that our radio
luminosities at frequencies $\lesssim100$~GHz are also numerically well
converged (see Fig.~\ref{fig:syn_spectra}).

% Don't change these lines
\bsp	% typesetting comment
\label{lastpage}
\end{document}